\documentclass[12pt]{article}
\usepackage[utf8]{inputenc}
\usepackage[margin=1in,footskip=0.25in]{geometry}
\usepackage{mathtools}
\usepackage{amsfonts}
\usepackage{amsmath}
\usepackage{bm}
\usepackage{indentfirst}
\usepackage{graphicx}
\usepackage{float}
\usepackage{setspace}
\usepackage{textcomp}
\usepackage{makecell}
\usepackage{multirow}
\usepackage{amssymb}
\usepackage{color}
\usepackage{subcaption}
\usepackage{authblk}
\usepackage{booktabs}

\usepackage{placeins}

\usepackage[pdftex, colorlinks, urlcolor=black, linkcolor=black, anchorcolor=black, citecolor=black]{hyperref}

\providecommand{\keywords}[1]
{
  \small	
  \textbf{\textit{Keywords---}} #1
}

\usepackage[numbers,square,sort&compress]{natbib}
\title{B\'ezier-GP: B\'ezier-Surface-Based Gaussian Process for Chips-on-Wafer Warpage Modeling}

\author[]{Sangyup Lee} 
\author[]{Xiao Liu} 
\author[]{Jianjun Shi} 
\affil[]{H. Milton Stewart School of Industrial and Systems Engineering, \\Georgia Institute of Technology}

\begin{document}

\maketitle

\begin{abstract}
{\color{black}
As the semiconductor industry transitions to the post-Moore paradigm, wafer warpage has become a critical quality issue in advanced packaging; particularly for 3D stacking and high-bandwidth memory manufacturing where die-level deformation can affect bonding quality, interconnect reliability, and yield. However, direct high-throughput measurement of individual chip warpage remains difficult in mass-production environments, while existing wafer-level metrology systems routinely provide dense 3D shape data at the wafer scale. This paper proposes a B\'ezier-surface-based Gaussian Process---known as B\'ezier-GP---for chips-on-wafer warpage modeling. The proposed framework combines wafer-level point-cloud metrology data with wafer-map layout information to construct chip-region-level B\'ezier surface patches, imposes $\mathcal{C}^1$ continuity constraints between neighboring patches, and represents the resulting chips-on-wafer surface through a compact free-control-point vector. A GP is then introduced in this low-dimensional representation to model wafer-to-wafer variation, chip-region-level spatial dependence, and provide uncertainty quantification. Numerical experiments, using an ASC wafer-map layout,  demonstrate that the proposed framework can reconstruct chips-on-wafer surfaces, preserve chip-region-level geometric interpretability, and recover population-level mean and covariance structures under a controlled multi-wafer simulation setting. The results suggest that B\'ezier-GP provides a measurement-driven statistical modeling foundation for connecting wafer-level metrology data to chip-region-level warpage analysis in semiconductor manufacturing.
}
\end{abstract}

\doublespacing

\section{Introduction}
For decades, the semiconductor industry has achieved remarkable progress  largely observing the Moore’s Law. 
However, the Moore’s Law has encountered significant technical and economic limitations in recent years due to the physical constraints of advanced nodes, quantum tunneling, rapidly increasing process costs, heat dissipation and power consumption, and reduced yield associated with process miniaturization \citep{ref1}. 
To overcome these limitations, the semiconductor industry is seeking a transition to the ``More-than-Moore'' or ``Post-Moore'' paradigm. Such a new paradigm aims for functionality-centric innovation by integrating heterogeneous functions, maximizing system-level performance, and enhancing space efficiency, rather than relying solely on transistor scaling \citep{ref2}. Against this backdrop, advanced packaging technology---which improves performance at the package level rather than within individual chips---has emerged as a key technology, attracting active research and investment from both academia and industry \citep{ref3,ref4}.

\begin{figure}[!h]
    \centering
    \includegraphics[width=0.65\linewidth]{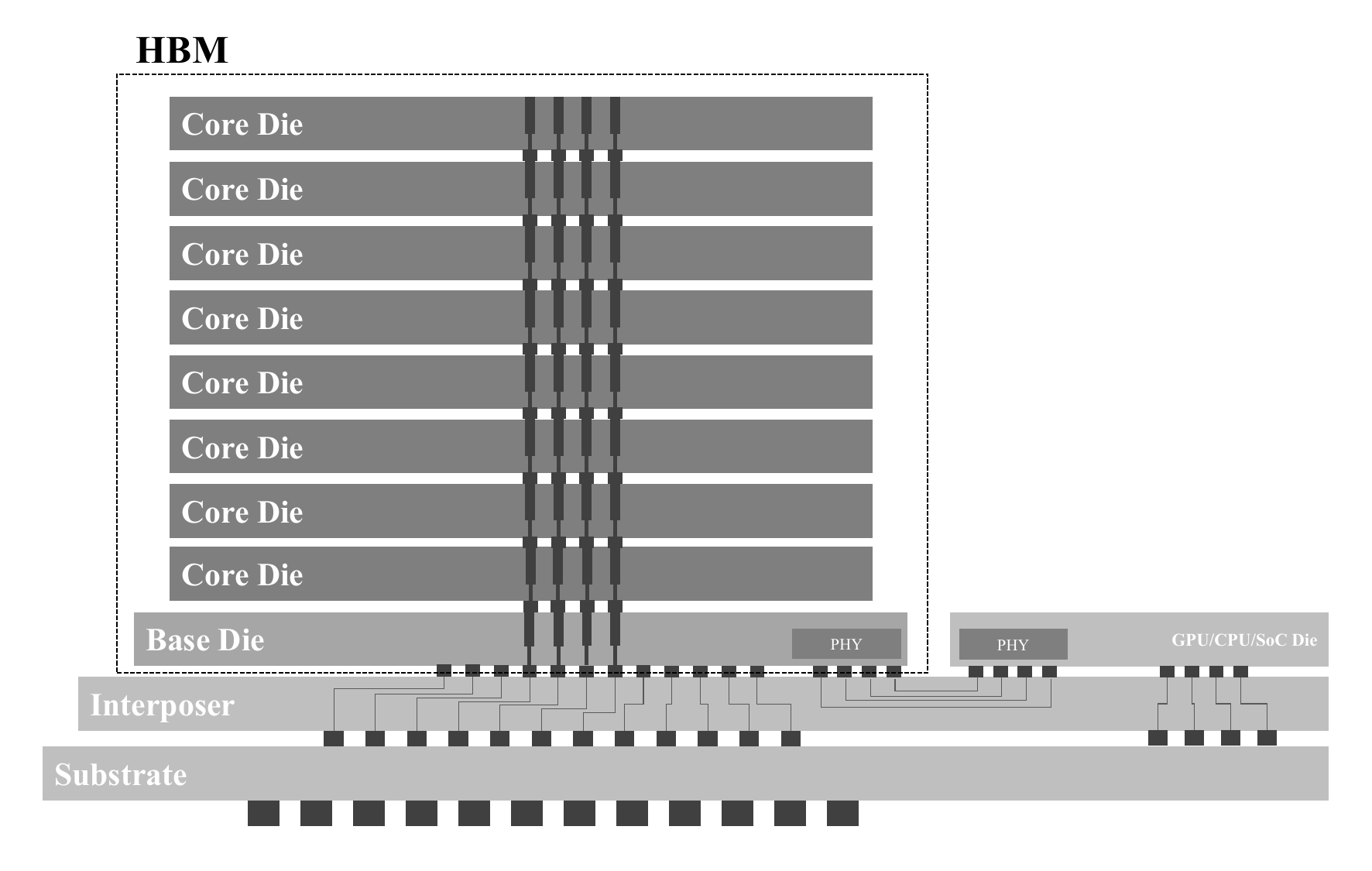}
    \caption{An Illustration of the structure of High Bandwidth Memory (HBM)}
    \label{fig:hbm}
\end{figure}
New technologies have emerged in the field of advanced packaging, including 2.5D integration, Fan-Out Wafer-Level Packaging (FOWLP), chiplet, and 3D stacking \citep{ref5}. Among them, 3D chip stacking is a cutting-edge semiconductor packaging technology that overcomes area limitations and enhances high-speed interconnects and integration density by vertically stacking chips. It enables the integration of heterogeneous or homogeneous chips, such as memory, logic, and sensors. Compared to 2D integration, 3D stacking allows for the same functionality to be implemented in a smaller area, contributing to better power efficiency and system performance. 
Techniques used to realize such stacking include Through-Silicon Via (TSV), which forms electrical connections between chips via metal vias penetrating the silicon, micro-bump interconnects, and the latest hybrid bonding technology, which aligns and directly bonds both metal and dielectric layers with high precision. To manufacture products based on 3D chip stacking, precise stacking processes that achieve both electrical and mechanical interconnects between chips are required.

Since ultra-thin chips are vertically stacked in multiple layers, the warpage phenomenon occurring during this process significantly affects the stability and yield of the process. \textit{Warpage refers to the bending of thin chips away from their original flat shape due to exposure to various thermal and mechanical stresses}. This phenomenon  causes quality issues such as non-wet bonding failures, interconnect misalignment, and crack propagation during processing especially in fine interconnect technologies such as TSV and Hybrid Bonding \citep{ref6, ref7, ref8, ref9}. Since chip warpage occurs at the individual-chip level, conventional wafer-level flatness measurements are insufficient to accurately capture this issue. Nevertheless, the industry currently lacks technologies that can measure the warpage of individual chips in a non-destructive and high-speed manner immediately before 3D stacking. In the automation metrology equipment industry for inline semiconductor mass production, chip-level warpage measurement capability remains largely unavailable. At the same time, the need for measurement and analysis of chip and wafer warpage is rapidly increasing.

The most successful and commercialized product based on 3D stacking to date is High Bandwidth Memory (HBM), a high-performance memory dedicated to GPUs. HBM features vertically stacked core DRAM chips using TSV and implements I/O and interface functionality through a base chip; see Figure \ref{fig:hbm}. In particular, HBM requires ultra-thin DRAM chips to be vertically stacked with precise TSV and micro-bump bonding. These requirements make warpage-related quality issues even more critical. Warping that occurs in each DRAM chip becomes a major source of various process defects such as misalignment, non-wet bonding failure, and TSV opening during stacking \citep{ref8, ref9}. Furthermore, the high-speed interconnect architecture of HBM---which operates at several tens of gigabits per second (Gbps)---means that even the slightest bonding failure can directly lead to degraded performance or device failure. Yield degradation caused by chip-level warpage in HBM has emerged as a critical issue and is also responsible for defects in many other advanced packaging products. Therefore, warpage analysis at the chip-level is becoming an urgent research priority to secure yield and reliability in 3D stacking processes.

{\color{black}
\textbf{Literature Review.} 
Existing wafer warpage modeling can be divided into two categories: physics-based and data-driven approaches. The former investigates deformation mechanisms based on material properties, process conditions, structural geometry, and boundary conditions, while the latter approach learns the surface geometry and spatial variation from wafer metrology data. 
In the remaining part of the literature review, we discuss the potentials and limitations of both physics-based and data-driven approaches in advanced packaging mass-production environments, and then connect the discussion to the chip-region-level modeling gap to be addressed in this research.

$\bullet$ The first approach---physics-based warpage modeling---has largely been developed around finite-element analysis (FEA). This approach models warpage as a thermo-mechanical response to changes in material, process, structure, and boundary conditions \citep{ref12, ref13, ref14}.
In the context of advanced packaging processes, 
\cite{ref15} analyzed wafer warpage generated during a wafer-to-wafer bonding fabrication process, and  \cite{ref16} addressed both warpage measurement and FE analysis in wafer-level molding. \cite{ref17} analyzed reconstructed wafer warpage during an encapsulation process through numerical simulation. \cite{ref18} and \cite{ref19} proposed simulation-based strategies for optimizing process-induced warpage in advanced packaging processes, and  \cite{ref20} proposed a modeling framework for predicting wafer-to-die warpage transfer. These studies are useful for explaining the effects of process window, material selection, and stack design on warpage, are are fundamentally close to constructing physics models for specific product or process scenarios.
In general, physics-based warpage modeling is powerful in explaining warpage mechanisms and performing what-if analysis for a specific reference wafer, product stack structure, material system, or process condition. 
However, in a high-volume manufacturing environment, wafer-to-wafer, lot-to-lot, and chip-to-chip variation continuously occur, making it difficult to specify residual stress, material properties, layer-level process history, chucking and handling conditions for every production wafer. Solving full physics-based models for every wafer or chip is  unrealistic in terms of computational cost and inline throughput. Therefore, FEA is highly suitable for mechanism-oriented analysis, but an additional modeling layer is needed for the purpose of directly utilizing metrology data accumulated in mass-production environments to continuously learn and quantify chip-region-level geometric variation.

$\bullet$ The second approach---data-driven warpage modeling---involves wafer geometry measurement, surface reconstruction, spatial sampling, spatial variation modeling, and geometric quality monitoring based on wafer metrology data. \cite{ref21} presented a method for interpolating scattered 3D point data using tensor-product B-spline surfaces, thereby providing a basis for transforming irregular point data into a smooth surface representation. \cite{ref22} developed a wafer geometry measurement system based on double-sided stitching interferometry and proposed a method for generating a full wafer shape map by stitching multiple interferometry shots.
Data-driven reconstruction has also been investigated to efficiently recover wafer profiles and reduce measurement cost. \cite{ref23} proposed a dynamic spatial sampling and reconstruction methodology to reduce wafer profile measurement cost while maintaining reconstruction accuracy and detecting localized anomalies. \cite{ref24} combined B-spline surface reconstruction with hierarchical local refinement to complement the limitations of a uniform-sampling-based global model. Furthermore, statistical modeling, kriging, and generalized-least-squares-based estimation for analyzing spatially correlated surface data provide a methodological basis for interpreting wafer-level geometric variation \citep{ref25}. \cite{ref26} represented wafer spatial variation with a hierarchical model to analyze process variation characteristics. \cite{ref27} proposed a method for monitoring wafer geometric quality using an additive Gaussian process model, and  \cite{ref28} presented a sequential measurement strategy that uses GP-based profile prediction and prediction uncertainty to select subsequent measurement locations. From the perspective of statistical process monitoring for manufacturing surface data, \cite{ref29} proposed an SPC framework for monitoring surface/manifold geometry obtained from 3D sensor data using intrinsic geometric features. 
This stream of studies provides an important basis for reconstructing wafer-level surfaces from sparse or irregular measurement data, interpreting wafer-level geometric variation and prediction uncertainty, and addressing the trade-off between metrology cost and reconstruction accuracy.


\textbf{Research Gaps and Contributions.} Although the prior investigations have provided important foundations for the current work, 
existing data-driven and statistical approaches mostly focus on wafer-level surface reconstruction or wafer-level spatial variation modeling. The geometry/warpage metrology infrastructure that is routinely deployed in actual semiconductor manufacturing lines is also concentrated on full-wafer shape maps, wafer-level shape/flatness/thickness measurement, and patterned-wafer geometry measurement \citep{ref22,ref31,ref32,ref33,ref34}, while free-form, patterned-wafer, and large-area optical shape/strain metrology has likewise developed mainly toward wafer-scale distortion or in-plane distortion prediction \citep{ref35,ref36,ref37}. In contrast, direct chip-level warpage measurement is an emerging approach that requires special process treatment and hardware modification \citep{ref9}. Hence, such direct measurement approaches require separate validation depending on product type, process flow, carrier/support condition, and calibration protocol, and cannot yet be regarded as directly replacing the current general mass-production metrology infrastructure. In this context, wafer map information, including commonly used ASCII/ASC wafer map files, defines chip layout and die coordinates and provides a practical data structure for defining chip-region boundaries. SEMI E5 specifies equipment--host message content \citep{ref38}, while an example of an ASCII-based wafer-map format is available \citep{ref39}. The wafer-level 3D measurement data can be converted into chip-region-level analysis while maintaining compatibility with existing semiconductor manufacturing data infrastructure. Hence, \textit{the core unresolved problem is to use wafer-scale 3D metrology data, together with wafer map layout, to transform the data into a chip-region-level geometric representation on wafer and to statistically connect the gap between wafer-level measurement and chip-level analysis}.

This observation leads to the following gaps between existing warpage modeling  and the needs of advanced packaging mass-production environments: (i) Physics-based modeling has strengths in analyzing warpage mechanisms and optimizing process/material design for a specific product stack structure or reference wafer, but it has limitations as a framework for directly learning and updating wafer-level metrology data and multi-wafer variation that are repeatedly accumulated in high-volume manufacturing; (ii) Measurement-driven surface reconstruction studies are effective for reconstructing wafer-level geometry, but they have not been sufficiently extended to explicitly treat individual chip regions on a wafer map as geometric units and to represent each chip-region surface as a continuous surface patch; and (iii) Existing statistical wafer modeling studies address wafer-level spatial variation and uncertainty, but they have not connected wafer-level point-cloud measurement, chip layout, neighboring-chip continuity, chip-region-level surface representation, and multi-wafer random variation within a unified modeling framework.

To address these gaps, this research proposes a B\'ezier-Surface-Based Gaussian Process (B\'ezier-GP) for chips-on-wafer surface modeling that transforms wafer-scale 3D metrology data into a chip-region-level geometric representation according to the die coordinates of the wafer map. Here, chip-level warpage representation does not directly refer to free-standing die warpage that appears after singulation due to changes in boundary conditions and stress release; rather, it refers to the pre-singulation chips-on-wafer surface geometry defined for each chip region existing on the wafer. This definition focuses on formalizing chip-region geometry on wafer as a consistent low-dimensional representation as a preceding step for predicting post-singulation die warpage. Hence, the proposed chips-on-wafer representation provides a modeling foundation that can later be extended to a free-standing die warpage prediction model that jointly considers singulation-induced stress release, boundary condition change, and die-level deformation behavior. The proposed framework is not dependent on a specific metrology modality, and it is designed to be applicable to various measurement systems that can provide wafer surfaces in point-cloud form, such as interferometry, 3D laser scanning, and optical shape measurement. 

The proposed B\'ezier-GP framework integrates the B\'ezier surface representation and Gaussian Process. 
Measured wafer-level point-cloud data are transformed into chip-region-level B\'ezier control points, and the entire chips-on-wafer surface is represented as a compact free-control-point representation through continuity constraints. This low-dimensional representation serves to transform irregular 3D metrology data into a structured chip-region-level shape-parameter space. Then, by modeling the control-point vector as a GP, the model performs continuous chip-region-level surface reconstruction for a single wafer and statistically captures the wafer-to-wafer variation, chip-region-level spatial variability, and prediction uncertainty in a multi-wafer setting.
In particular, the B\'ezier-GP  framework constructs individual chip-region B\'ezier patch representations from wafer-scale point-cloud measurements. Then,   neighboring chip patches are stitched in a $\mathcal{C}^1$ continuity-constrained manner to generate a continuous chips-on-wafer surface over the wafer domain. By representing chip-region-level surface geometry as a compact free-control-point vector, we transform high-dimensional 3D surface data into a low-dimensional statistical modeling problem, and through the B\'ezier-GP model, we provide a probabilistic framework capable of quantifying multi-wafer warpage variation and prediction uncertainty.

From the perspective of engineering practice, the proposed method can function as a manufacturing oriented solution that connects existing production metrology data streams to chip-region-level statistical modeling. As more production wafers and process information are accumulated, the proposed representation can be extended to downstream decision support, such as learning dominant chip-region-level warpage modes, predicting wafer-to-wafer and chip-region-to-chip-region variation, quantifying discrepancies between FEA prediction and measurement data, bonding-risk assessment, and stacking-oriented chip selection or chip pairing strategy. In other words, the B\'ezier-GP can be viewed as a measurement-driven statistical modeling layer that connects physics-based understanding and production metrology data, and it presents a foundational framework for enabling chips-on-wafer warpage analysis in a mass-production setting.

\vspace{4pt}
The remainder of this paper is organized as follows. Section \ref{sec:preliminaries} describes the wafer map, chip layout, and wafer-level point-cloud measurement data structure. Section \ref{sec:model} presents the proposed B\'ezier-GP and describes chip-region-level B\'ezier patch construction, continuity constraints, free-control-point representation, and statistical inference. Section \ref{sec:experiments} validates the reconstruction capability and multi-wafer statistical modeling performance of the proposed framework through numerical experiments. Section \ref{sec:conclusion} summarizes the research results and discusses future research. 
}

\section{Wafer Map, Chip Layout and Point Cloud Data} 
\begin{figure}[!ht] 
    \centering
\includegraphics[width=1\linewidth]{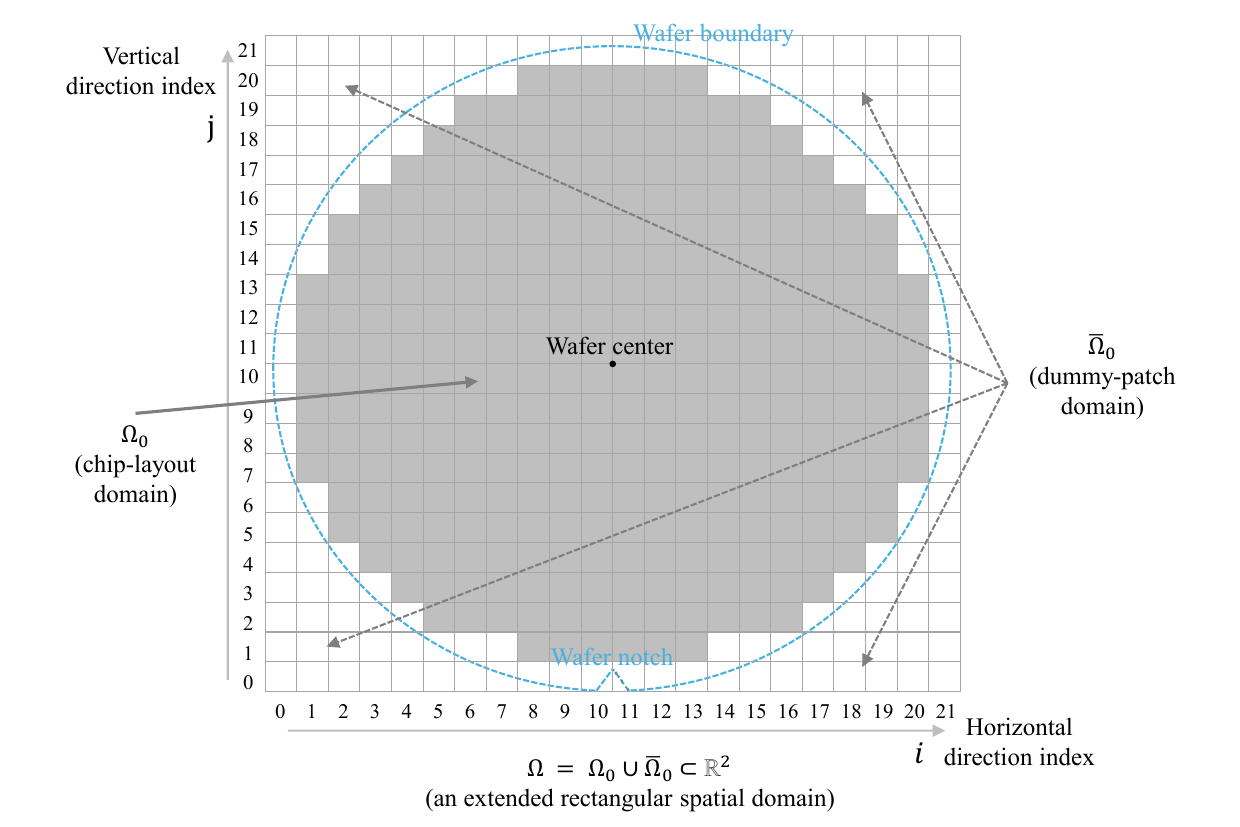}
\vspace{-6pt}
\caption{Illustration of a wafer map embedded in an extended rectangular spatial domain.} \label{fig:map}
\end{figure}

\vspace{-8pt}
\label{sec:preliminaries}
\textbf{Geometric Settings.} This section describes the fundamental geometry of a wafer map, chip layout, and the point cloud measurement data. A wafer is a thin slice of silicon used for the formation of integrated circuits (i.e., chips). {\color{black}
As illustrated in Figure \ref{fig:map}, the chip-layout domain is denoted by $\Omega_0$ (the shaded area).
} Each  small shaded square represents a chip location on the wafer map.
{\color{black}
Here, $n_x$ and $n_y$ denote, respectively, the numbers of columns and rows in the rectangular patch grid used for modeling, i.e., the numbers of patches in the horizontal and vertical directions, including dummy chips. The layout information is used to identify the cells corresponding to actual chips within this grid.}
Horizontal indices, $i=0,1,\cdots,n_x-1$, and vertical indices, $j=0,1,\cdots,n_y-1$, are used to specify chip locations in a given wafer map. 

To make the subsequent wafer warpage modeling tractable, we also introduce the patched domain, $\bar{\Omega}_0$, 
such that the extended domain, $\Omega = \Omega_0  \cup  \bar{\Omega}_0 \subset \mathbb{R}^2$, becomes a rectangular spatial domain. The introduction of the extended domain avoids the irregular geometric shape of the spatial domain in our modeling. 
Figure \ref{fig:map} provides an illustration of a wafer map embedded in a rectangular spatial domain. 
Throughout this paper, the in-plane coordinate unit is defined as the patch size, {\color{black} which is the sum of the die size and the scribe-lane width between adjacent dies.} We normalize the in-plane coordinates by the patch size so that each chip location corresponds to a unit square in $\Omega$. Thus, $\Omega=[0,n_x]\times[0,n_y]\subset\mathbb{R}^2$ in this paper.
{\color{black}
Here, $\Omega$ is expressed in normalized coordinates.
The physical registration of the chip layout grid is performed before normalization.
}

Over the extended spatial domain $\Omega$, let $c_{i,j}$ denote the small square indexed by $(i,j)$, where $i = 0, 1, \ldots, n_x - 1$ and $j = 0, 1, \ldots, n_y - 1$. If $c_{i,j}$ lies in $\Omega_0$, it represents an actual chip. {\color{black} If $c_{i,j}$ lies in $\bar{\Omega}_0$, i.e., the dummy-patch domain outside the chip-layout domain, it represents a dummy chip.}
After the wafer warpage model is completed, the dummy-chip region is excluded and only the surface over the actual-chip domain $\Omega_0$ is used for chip-warpage evaluation.

For any $c_{i,j}$, we let $\mathcal{N}^{t}_{i,j}$, $\mathcal{N}^{b}_{i,j}$, $\mathcal{N}^{r}_{i,j}$, and $\mathcal{N}^{l}_{i,j}$ respectively be the sets that include the indices of the top, bottom, right and left neighbors of $c_{i,j}$. Some of these neighbor sets become empty for chips  located on the wafer map boundary, and they are used to determine whether an adjacent patch exists so that the continuity constraint can be imposed in Section \ref{sec:model}.
\begin{equation}
\begin{split}
 & \mathcal{N}^{t}_{i,j} = 
\begin{cases*}
    \{(i,j+1)\} & $i \in I,\; j=0,\ldots,n_y-2$\\
    \emptyset   & $i \in I,\; j=n_y-1$
  \end{cases*}, \quad
  \mathcal{N}^{b}_{i,j} = 
\begin{cases*}
    \{(i,j-1)\} & $i \in I,\; j=1,\ldots,n_y-1$\\
    \emptyset   & $i \in I,\; j=0$
  \end{cases*}, \\
& \mathcal{N}^{l}_{i,j} = 
\begin{cases*}
    \{(i-1,j)\} & $i=1,\ldots,n_x-1,\; j \in J$\\
    \emptyset   & $i=0,\; j \in J$
  \end{cases*}, \quad
\mathcal{N}^{r}_{i,j} = 
\begin{cases*}
    \{(i+1,j)\} & $i=0,\ldots,n_x-2,\; j \in J$\\
    \emptyset   & $i=n_x-1,\; j \in J$
  \end{cases*}
\end{split}
\end{equation}
\noindent where $I = \{0,1,\ldots,n_x-1\}$ and $J = \{0,1,\ldots,n_y-1\}$.

\vspace{4pt}
\textbf{Point Cloud Measurement and Data Processing.} Wafer-level metrology systems obtain the 3D point-cloud data, which are further processed to extract chip-level warpage information for subsequent modeling. To improve the readability of the paper, we leave the details of point cloud data measurement and processing to the \textcolor{blue}{Appendix}.

\section{B\'ezier-Surface-Based Gaussian Process}
\label{sec:model}


In this section, we present the B\'ezier-Surface-Based Gaussian Process (B\'ezier-GP) for chips-on-wafer warpage.  The use of B\'ezier surfaces is naturally motivated by how such surfaces are constructed: to construct a B\'ezier surface, one starts with a number of patches with rich geometric shapes, and the patches are then stitched together to form a surface. For the modeling of wafer warpage, each patch naturally corresponds to a chip, and a wafer surface is formed by connecting the chips together. 
Based on the B\'ezier surface (Section \ref{sec:Bezier}), the GP process is employed to capture the wafer-to-wafer warpage variation and within-wafer warpage correlation (Section \ref{sec:Bezier-GP}). 

\subsection{B\'ezier Surface for Wafer Warpage}
\label{sec:Bezier}

Let $(u,v)\in[0,1]^2$ denote the local in-plane coordinates within chip $c_{i,j}$. In this work we focus on warpage height modeling and thus treat the chip-level surface, $S_{i,j}(u,v)$, as a scalar-valued height field (known as the out-of-plane displacement). 
For a chosen B\'ezier order $(K,L)$, the chip-level out-of-plane displacement $S_{i,j}(u,v)$ is represented as
\begin{equation} \label{eq:S}
    S_{i,j}(u,v)=\sum_{k=0}^K\sum_{l=0}^L B_{k,K}(u)\,B_{l,L}(v)\,p_{i,j}^{k,l},
\end{equation}
\noindent where $B_{k,K}(u)$ and $B_{l,L}(v)$ are the Bernstein basis functions of degrees $K$ and $L$ on $[0,1]$, 
\begin{equation}
   B_{k,K}(u) =\binom{K}{k}u^{k}(1-u)^{K-k}, \quad
   B_{l,L}(v) =\binom{L}{l}(v)^{l}(1-v)^{L-l},
\end{equation} 
\noindent and $\{p_{i,j}^{k,l}\}_{k=0,\cdots,K, l=0,\cdots,L}$ are the B\'ezier control points associated with chip $c_{i,j}$ to be determined from the measurement data. For convenience, we collect the control points into a matrix
\begin{equation}
\label{eq:P}
\bm{P}_{i,j} = \begin{bmatrix}
p_{i,j}^{0,0} & p_{i,j}^{0,1} & \cdots & p_{i,j}^{0,L} \\
p_{i,j}^{1,0} & p_{i,j}^{1,1} & \cdots & p_{i,j}^{1,L} \\
\vdots  & \vdots  & \ddots   & \vdots  \\
p_{i,j}^{K,0} & p_{i,j}^{K,1} & \cdots & p_{i,j}^{K,L}  \\
\end{bmatrix},
\end{equation}

\noindent and (\ref{eq:S}) can be written in a compact tensor-product form as 
$S_{i,j}(u,v)=\bm{b}_K(u)^{\top} \bm{P}_{i,j} \bm{b}_L(v)$, where $\bm{b}_K(u)=[B_{0,K}(u),\ldots,B_{K,K}(u)]^{\top}$ and $\bm{b}_L(v)=[B_{0,L}(v),\ldots,B_{L,L}(v)]^{\top}$. 

\begin{figure}[htbp] 
    \centering
\includegraphics[width=0.7\linewidth]{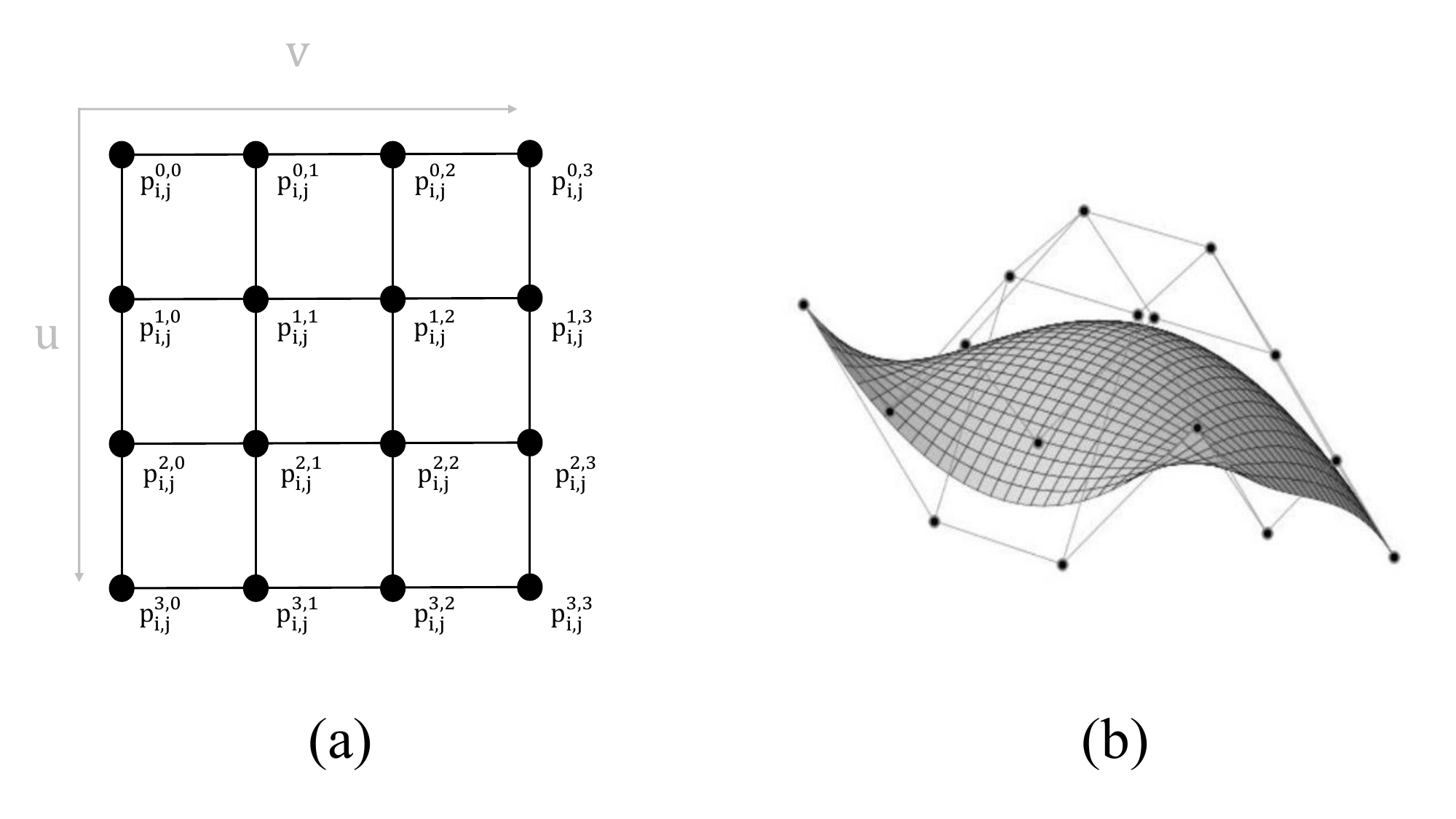}
\vspace{-8pt}
\caption{Illustration of control points (left) and a B\'ezier patch that defines a continuous mapping from the 2D parameter domain to the chip surface (right).} \label{fig:patch}
\end{figure}
Figure~\ref{fig:patch} gives an illustrative example of the control-point layout for the cubic case $K=L=3$. The B\'ezier surface (\ref{eq:S}) defines a continuous mapping from the 2D parameter domain $[0,1]\times[0,1]$ to the chip surface (which is effective for modeling out-of-plane displacement). The Bernstein basis functions act as nonnegative blending weights that smoothly combine the control points over the chip domain, yielding an implicit and compact geometric representation of die-level warpage.

Next, stitching together all chip patches \(c_{i,j}\) over \(i=0,\ldots,n_x-1\) and \(j=0,\ldots,n_y-1\) yields a wafer-level B\'ezier surface defined on the spatial domain \(\Omega=[0,n_x]\times[0,n_y]\). {\color{black} In particular, let $(x,y)\in[0,n_x]\times[0,n_y]$
be a point in the normalized modeling domain, where
$(x,y)$ denote the normalized coordinates
$(\tilde{x},\tilde{y})$ rather than millimeter coordinates.
Following the patch-assignment convention in Section 2.2,
for $0\le x<n_x$, define
$i=\lfloor x\rfloor$ and
$u=\operatorname{frac}(x)\equiv x-\lfloor x\rfloor$.
Similarly, for $0\le y<n_y$, define
$j=\lfloor y\rfloor$ and
$v=\operatorname{frac}(y)\equiv y-\lfloor y\rfloor$.
At $x=n_x$, set $i=n_x-1$ and $u=1$;
at $y=n_y$, set $j=n_y-1$ and $v=1$.} The wafer-level B\'ezier surface (representing wafer warpage) is given by a compact tensor product:
\begin{equation}
\label{eq:surface}
Z(x,y)
=
\sum_{k=0}^K\sum_{l=0}^L B_{k,K}(u)\,B_{l,L}(v)\,p_{i,j}^{k,l}
=
\bm{b}_K(u)^{\top}\bm{P}_{i,j}\bm{b}_L(v).
\end{equation}






Note that \textit{continuity} must be imposed when stitching together B\'ezier patches to form a wafer surface. In this work, we consider $\mathcal{C}^1$ continuity in B\'ezier surfaces, also known as tangent continuity, which requires neighboring B\'ezier patches to share compatible tangent planes (or normal vectors) along their shared boundary. 
In particular, for any patch $c_{i,j}$, the $\mathcal{C}^1$ continuity can be imposed by imposing the following constraints to the control points of neighboring patches:

\begin{itemize}
    \item if $\mathcal{N}_{i,j}^t \neq \emptyset$
    \begin{equation}
        p_{i,j}^{k,L} = \frac{p_{i,j}^{k,L-1}+p_{i,j+1}^{k,1}}{2},\quad\quad\text{for }k=0,\cdots,K
    \end{equation}
    \item if $\mathcal{N}_{i,j}^b \neq \emptyset$
    \begin{equation}
        p_{i,j}^{k,0} = \frac{p_{i,j}^{k,1}+p_{i,j-1}^{k,L-1}}{2},\quad\quad\text{for }k=0,\cdots,K
    \end{equation}
    \item if $\mathcal{N}_{i,j}^l \neq \emptyset$
    \begin{equation}
        p_{i,j}^{0,l} = \frac{p_{i,j}^{1,l}+p_{i-1,j}^{K-1,l}}{2},\quad\quad\text{for }l=0,\cdots,L
    \end{equation}
    \item if $\mathcal{N}_{i,j}^r \neq \emptyset$
    \begin{equation}
        p_{i,j}^{K,l} = \frac{p_{i,j}^{K-1,l}+p_{i+1,j}^{1,l}}{2},\quad\quad\text{for }l=0,\cdots,L
    \end{equation}
\end{itemize}

\begin{figure}[htbp] 
    \centering
\includegraphics[width=0.85\linewidth]{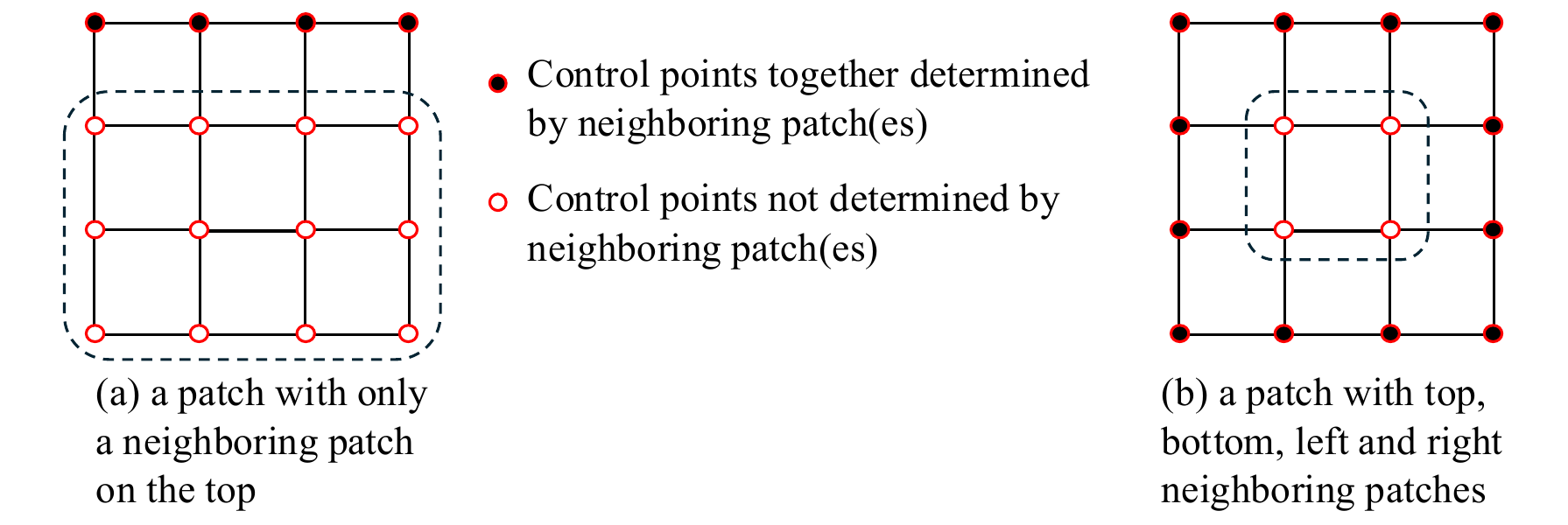}
\caption{Illustration of conditions on control points that ensure the $\mathcal{C}^1$ continuity} \label{fig:continuity}
\end{figure}

The conditions above impose constraints among control points between neighboring patches. Given the position of a patch on the surface, only a subset of the $(K+1)\times(L+1)$ control points need to be determined while the remaining control points are directly computed from the $\mathcal{C}^1$ continuity. Figure \ref{fig:continuity} presents two examples. In Figure \ref{fig:continuity}(a), it is only necessary to determine 12 control points (rows 2$\sim$4) for a patch with a neighboring patch on its top: the top 4 control points are directly computed by averaging the 4 control points in the second row and the 4 control points in second row from the bottom of the patch on top. Figure \ref{fig:continuity}(b) shows another patch with neighboring patches on all four sides: one only needs to determine 4 control points while the remaining 12 control points are directly computed from the 4 control points and the control points from the neighboring patches. 

Hence, similar to the $(K+1)\times(L+1)$ matrix $\bm{P}_{i,j}$ in (\ref{eq:P}), it is possible to define a smaller $(K'+1)\times(L'+1)$ matrix $\bm{Q}_{i,j}$, for patch $c_{i,j}$, that only contains the control points which can be freely chosen (i.e., control points not involved in the continuity constraints). In particular, the matrix $\bm{Q}_{i,j}$ is obtained by removing the rows and/or columns of the matrix $\bm{P}_{i,j}$, and is given by
\begin{equation}
\label{eq:Q_select}
    \bm{Q}_{i,j} = \bm{H}_{i,j}^{R}\bm{P}_{i,j}\bm{H}_{i,j}^{C}
\end{equation}
\noindent where $\bm{H}_{i,j}^{R}$ is a $(K'+1)\times(K+1)$ sparse matrix with only one $``1"$ in each row at the $k$-th position if the $k$-th row of $\bm{P}_{i,j}$ is kept, and $\bm{H}_{i,j}^{C}$ is a $(L+1)\times(L'+1)$ sparse matrix with only one $``1"$ in each column at the $k$-th position if the $k$-th column of $\bm{P}_{i,j}$ is kept. The structure of $\bm{H}_{i,j}^{R}$ and $\bm{H}_{i,j}^{C}$ is further exploited in the next subsection when constructing the B\'ezier-GP model. 


\subsection{The B\'ezier-GP Model}
\label{sec:Bezier-GP}
Building on the B\'ezier surface for wafer warpage above, this section describes the B\'ezier-GP that captures both the within- and between wafer variation. 
Suppose that a number of $D_{i,j}$ measurements are taken from a chip (patch) $c_{i,j}$. The measurements are indexed by
$d=1,\ldots,D_{i,j}$, with local coordinates $(\tilde{u}_{i,j}(d),\tilde{v}_{i,j}(d))\in[0,1]^2$ and measured heights $z_{i,j}(d)$. Hence, the measurements from a chip $c_{i,j}$ can be collected in a chip-level measurement vector $\bm{z}_{i,j}
=
\big[z_{i,j}(1),\,z_{i,j}(2),\,\ldots,\,z_{i,j}(D_{i,j})\big]^{\top}\in\mathbb{R}^{D_{i,j}}$.
Motivated by the B\'ezier surface (\ref{eq:surface}), one can envision a (chip-level) model that takes a generic form as follows:
\begin{equation}
\label{eq:chip_linear}
\bm{z}_{i,j}=\bm{\beta}_{i,j}\,\vec{\bm{P}}_{i,j}+ \bm{\epsilon}_{i,j},
\end{equation}
\noindent where $\vec{\bm{P}}_{i,j}
=
\big[p^{0,0}_{i,j},\,p^{1,0}_{i,j},\,\ldots,\,p^{K,0}_{i,j},\,
p^{0,1}_{i,j},\,\ldots,\,p^{K,1}_{i,j},\,\ldots,\,
p^{0,L}_{i,j},\,\ldots,\,p^{K,L}_{i,j}\big]^{\top}
\in\mathbb{R}^{(K+1)(L+1)}$
denotes the column-wise vectorization of $\bm{P}_{i,j}$, 
\begin{equation}
\label{eq:beta_ij}
\bm{\beta}_{i,j}
=
\begin{bmatrix}
\bm{b}_L(\tilde{v}_{i,j}(1))^{\top}\otimes \bm{b}_K(\tilde{u}_{i,j}(1))^{\top}\\
\bm{b}_L(\tilde{v}_{i,j}(2))^{\top}\otimes \bm{b}_K(\tilde{u}_{i,j}(2))^{\top}\\
\vdots\\
\bm{b}_L(\tilde{v}_{i,j}(D_{i,j}))^{\top}\otimes \bm{b}_K(\tilde{u}_{i,j}(D_{i,j}))^{\top}
\end{bmatrix}\in\mathbb{R}^{D_{i,j}\times (K+1)(L+1)}, 
\end{equation}
\noindent with $\bm{b}_K(u)=[B_{0,K}(u),\ldots,B_{K,K}(u)]^{\top}$ and $\bm{b}_L(v)=[B_{0,L}(v),\ldots,B_{L,L}(v)]^{\top}$, and $\bm{\epsilon}_{i,j}$ represents the modeling error. Note that in deriving the model (\ref{eq:chip_linear}), we utilize the identity
$
\bm{b}_K(u)^{\top}\bm{P}_{i,j}\bm{b}_L(v)
=
\big(\bm{b}_L(v)^{\top}\otimes \bm{b}_K(u)^{\top}\big)\,\vec{\bm{P}}_{i,j}$. 
Next, by stacking all chip-level measurements, control-point vectors and measurement errors---from all chips over a wafer---into the following vectors
\begin{equation}
\label{eq:Z_global}
\bm{z}
=
\begin{bmatrix}
\bm{z}_{0,0}\\
\bm{z}_{0,1}\\
\vdots\\
\bm{z}_{n_x-1,n_y-1}
\end{bmatrix}
\in\mathbb{R}^{\sum_{(i,j)\in I\times J} D_{i,j}}, \quad
\vec{\bm{P}}
=
\begin{bmatrix}
\vec{\bm{P}}_{0,0}\\
\vec{\bm{P}}_{0,1}\\
\vdots\\
\vec{\bm{P}}_{n_x-1,n_y-1}
\end{bmatrix}, \quad
\bm{\epsilon} = 
\begin{bmatrix}
\bm{\epsilon}_{0,0}\\
\bm{\epsilon}_{0,1}\\
\vdots\\
\bm{\epsilon}_{n_x-1,n_y-1}
\end{bmatrix},
\end{equation}
we extend (\ref{eq:chip_linear}) to obtain the model for wafer-level warpage
\begin{equation}
\label{eq:wafer_linear}
\bm{z}=\bm{\beta}\vec{\bm{P}}+ \bm{\epsilon}
\end{equation}
\noindent with $\bm{\beta}=\mathrm{diag}\{\bm{\beta}_{i,j}\}_{(i,j)\in I\times J}$. 
Although (\ref{eq:wafer_linear}) suggests that one may model the wafer warpage through a model with a generic linear form, the model itself cannot be directly applied. This is because model (\ref{eq:wafer_linear}) does not yet account for the $\mathcal{C}^1$ continuity constraints (6)-(9). In other words, not all elements in the long vector $\vec{\bm{P}}$ can be independently determined. As shown in (\ref{eq:Q_select}), one only needs to determine the elements in $\vec{\bm{Q}}$ (a subset of $\vec{\bm{P}}$), and compute the remaining elements in $\vec{\bm{P}}$ based on the $\mathcal{C}^1$ continuity. 
Hence, the model envisioned needs to take the following form
\begin{equation}
\label{eq:wafer_linear_Q}
\bm{z}=\bm{\beta}\mathcal{L}_{\vec{\bm{Q}}}+ \bm{\epsilon},
\end{equation}
\noindent where $\mathcal{L}$ is an operator, to be found, that maps $\vec{\bm{Q}}$ back to $\vec{\bm{P}}$. 

We show that the operator $\mathcal{L}$ needed is a matrix operator in the sense that $\vec{\bm{P}}$ can be obtained by multiplying the matrix operator from the left to $\vec{\bm{Q}}$. 
From (\ref{eq:Q_select}), it is known that $\bm{Q}_{i,j} = \bm{H}_{i,j}^{R}\bm{P}_{i,j}\bm{H}_{i,j}^{C}$. Therefore, a naive idea of recovering $\bm{P}_{i,j}$ from $\bm{Q}_{i,j}$ involves finding the left inverse of $\bm{H}_{i,j}^{R}$ and the right inverse of $\bm{H}_{i,j}^{C}$, and writing $\bm{P}_{i,j} =  \tilde{\bm{H}}^{R}_{i,j}\,\bm{Q}_{i,j}\,\tilde{\bm{H}}^{C}_{i,j}$ with $\tilde{\bm{H}}^{R}_{i,j}$ and $\tilde{\bm{H}}^{C}_{i,j}$ respectively being the left and right inverse. However, a closer examination of the structures of $\bm{H}_{i,j}^{R}$ and $\bm{H}_{i,j}^{C}$ immediately shows that neither the left nor the right inverse matrices exist. This is not surprising as it is impossible to create a matrix with a high dimension from a smaller matrix without additional information. 
Fortunately, the $\mathcal{C}^1$ continuity conditions provide the additional information needed to construct the operator $\mathcal{L}$. Let $\bm{0}$ be a $(K+1)\times (L+1)$ zero matrix. 
Then, from (\ref{eq:Q_select}) and by leveraging the structure of $\bm{H}_{i,j}^{R}$ and $\bm{H}_{i,j}^{C}$, it can be seen that $\bm{P}_{i,j}^{\text{embed}} = \bm{0}+(\bm{H}_{i,j}^{R})^{\top}\bm{Q}_{i,j}(\bm{H}_{i,j}^{C})^{\top}$ embeds the $(K'+1)\times (L'+1)$ matrix $\bm{Q}_{i,j}$ into a bigger $(K+1)\times (L+1)$ matrix $\bm{P}_{i,j}^{\text{embed}}$, such that $\bm{Q}_{i,j} = \bm{H}_{i,j}^{R}\bm{P}_{i,j}^{\text{embed}}\bm{H}_{i,j}^{C}$.
The remaining positions with zeros in $\bm{P}_{i,j}^{\text{embed}}$ are computed from the $\mathcal{C}^1$ continuity conditions. 
Let
\begin{equation}
    \vec{\bm{P}}^{\text{embed}}
=
\begin{bmatrix}
\vec{\bm{P}}^{\text{embed}}_{0,0}\\
\vec{\bm{P}}^{\text{embed}}_{0,1}\\
\vdots\\
\vec{\bm{P}}^{\text{embed}}_{n_x-1,n_y-1}
\end{bmatrix}, \quad
\vec{\bm{Q}}
=
\begin{bmatrix}
\vec{\bm{Q}}_{0,0}\\
\vec{\bm{Q}}_{0,1}\\
\vdots\\
\vec{\bm{Q}}_{n_x-1,n_y-1}
\end{bmatrix}, \quad
\tilde{\bm{H}}
=
\begin{bmatrix}
\tilde{\bm{H}}_{0,0} &  & & \\
 &  \tilde{\bm{H}}_{0,1} & &  \\
& & \ddots &\\
&   &  & \tilde{\bm{H}}_{n_x-1,n_y-1}\\
\end{bmatrix}
\end{equation}
\noindent with $\tilde{\bm{H}}_{i,j}={\color{black}\bm{H}_{i,j}^{C}\otimes(\bm{H}_{i,j}^{R})^{\top}}$, 
 we show that it is possible to define a sparse matrix $\bm{A}$ so that
\begin{equation}
\label{eq:A_operator}
\vec{\bm{P}} \;=\; \bm{A}\,\vec{\bm{P}}_{\mathrm{embed}}  \;=\; \bm{A}\,\tilde{\bm{H}}\,\vec{\bm{Q}}.
\end{equation}
\noindent where the second equality above is due to the identity that $\vec{\bm{P}}^{\mathrm{embed}}_{i,j}=\tilde{\bm{H}}_{i,j}\vec{\bm{Q}}_{i,j}$ if $\bm{P}_{i,j}^{\text{embed}} = {\color{black}(\bm{H}_{i,j}^{R})^{\top}\bm{Q}_{i,j}(\bm{H}_{i,j}^{C})^{\top}}$. The matrix $\bm{A}$ is found as follows. 

Define the full control-point index set $\Omega_P=\big\{0,1,2,\ldots, n_x n_y (K+1)(L+1)-1\big\}$. 
We use the following mapping from a 4-tuple $(i,j,k,l)$ to a global index $g(i,j,k,l)\in\Omega_P$:
\begin{equation}
\label{eq:global_indexing}
g(i,j,k,l) \;=\; i\,n_y\,(K+1)(L+1)\;+\; j\,(K+1)(L+1)\;+\; l\,(K+1)\;+\; k
\end{equation}
\noindent for $i=0,\ldots,n_x-1$, $j=0,\ldots,n_y-1$, $k=0,\ldots,K$, and $l=0,\ldots,L$.
This indexing is consistent with the vectorization order used in $\vec{\bm{P}}_{i,j}$.
Based on the predefined chip layout and the stitching pattern, we partition $\Omega_P$ into three disjoint subsets:
\begin{equation}
\label{eq:Omega_partition}
\Omega_P = \Omega_{\mathrm{corner}} \;\cup\; \Omega_{\mathrm{edge}} \;\cup\; \Omega_{\mathrm{int}},
\qquad
\Omega_{\mathrm{corner}}\cap\Omega_{\mathrm{edge}}=\Omega_{\mathrm{edge}}\cap\Omega_{\mathrm{int}}
=\Omega_{\mathrm{int}}\cap\Omega_{\mathrm{corner}}=\emptyset.
\end{equation}
Intuitively, $\Omega_{\mathrm{int}}$ corresponds to interior control points (unique to a single chip),
$\Omega_{\mathrm{edge}}$ corresponds to edge-shared control points (coupled between two neighboring chips),
and $\Omega_{\mathrm{corner}}$ corresponds to corner-shared control points (coupled among four neighboring chips).
The operator $\bm{A}$ is constructed row-by-row according to the partition in \eqref{eq:Omega_partition}.
For each global index $a\in\Omega_P$, define a small neighbor-index set $\mathcal{N}(a)\subset\Omega_P$ such that $|\mathcal{N}(a)|=1$ if $a\in\Omega_{\mathrm{int}}$,  $|\mathcal{N}(a)|=2$ if $a\in\Omega_{\mathrm{edge}}$, and $|\mathcal{N}(a)|=4$ if $a\in\Omega_{\mathrm{corner}}$.
Here, $\mathcal{N}(a)$ is the index set of nonzero columns in row $a$ of $\bm{A}$.
Then, $\bm{A}$ enforces the averaging rule
\begin{equation}
\label{eq:A_rule}
\vec{\bm{P}}(a)
=
\frac{1}{|\mathcal{N}(a)|}\sum_{b\in\mathcal{N}(a)} \vec{\bm{P}}_{\mathrm{embed}} (b),
\qquad a\in\Omega_P.
\end{equation}

Equivalently, each row $a$ of $\bm{A}$ has nonzero entries only in columns indexed by $\mathcal{N}(a)$, with weights
$1$ (int), $1/2$ (edge), or $1/4$ (corner), respectively:
\begin{equation}
\bm{A}_{a,b}=
\begin{cases}
1, & a\in\Omega_{\mathrm{int}},\; b\in\mathcal{N}(a),\\
\frac{1}{2}, & a\in\Omega_{\mathrm{edge}},\; b\in\mathcal{N}(a),\\
\frac{1}{4}, & a\in\Omega_{\mathrm{corner}},\; b\in\mathcal{N}(a),\\
0, & \text{otherwise}.
\end{cases}
\end{equation}

This construction follows the proposed patch-stitching rule:
a corner control point equals the average of its four adjacent representatives,
an edge control point equals the average of its two adjacent representatives,
and interior control points remain unchanged.
Combining \eqref{eq:wafer_linear} and \eqref{eq:A_operator}, the model (\ref{eq:wafer_linear_Q}) becomes
\begin{equation}
\label{eq:global_final}
\bm{z} \;=\; \bm{J}\vec{\bm{Q}} + \bm{\epsilon}, \quad \bm{J}:=\bm{\beta}\,\bm{A}\,\tilde{\bm{H}}, \quad \bm{\epsilon} \sim \mathcal{N}(\bm{0},\,\bm{\Sigma}_{\epsilon})
\end{equation}

Equation~\eqref{eq:global_final} provides a single global linear system that links all wafer-level point-cloud measurements to the free control points across the wafer, while enforcing chip-to-chip continuity via the sparse coupling operator $\bm{A}$. Once all chip-wise free control points are stacked and the coupling operator $\bm{A}$ is introduced, the boundary control points can be determined consistently using neighboring-patch information across the wafer. 

Since different wafers must have distinct shapes (i.e., the wafer-to-wafer variation), we further introduce the Gaussian random effect on the free control-point vector:
\begin{equation} \label{eq:distribution_Q}
\vec{\bm{Q}} \sim \mathcal{N}\!(\bm{\mu}_{\vec{\bm{Q}}},\,\bm{\Sigma}_{\vec{\bm{Q}}}),
\end{equation}
\noindent where $\bm{\mu}_{\vec{\bm{Q}}}$ represents the mean wafer shape in the free-control-point space and $\bm{\Sigma}_{\vec{\bm{Q}}}$ captures spatial dependence of chip-level shape variation.
To properly construct $\bm{\Sigma}_{\vec{\bm{Q}}}$ in a structured and computationally convenient manner, we first define a covariance $\bm{\Sigma}_{\vec{\bm{P}}}$ on the vectorized full control-point space. Then, by removing the rows and columns according to the $\mathcal{C}^1$ continuity, $\bm{\Sigma}_{\vec{\bm{Q}}}$ is obtained and is guaranteed to be positive semi-definite. 
Let $\vec{\bm{P}}$ denote the vectorized full control points over the wafer (prior to reduction by the $\mathcal{C}^1$ constraints), we adopt a separable covariance structure, $\bm{\Sigma}_{\vec{\bm{P}}}
=
\bm{\Sigma}_{\text{between}} \otimes \bm{\Sigma}_{\text{within}}$, 
where $\bm{\Sigma}_{\text{between}}$ captures \emph{between-chip} (chip-to-chip) dependence across the wafer, and $\bm{\Sigma}_{\text{within}}$ captures \emph{within-chip} dependence among B\'ezier control points inside a chip.  
In this paper, we let
\begin{equation}
[\bm{\Sigma}_{\text{between}}]_{m,n}
=
\xi_1^2
\exp\!\left(
-\frac{\mathrm{dist}_{\text{chip}}(m,n)}{\xi_4^2}
\right),
\end{equation}
\noindent and
\begin{equation}
\big[\bm{\Sigma}_{\text{within}}\big]_{(k,l),(k',l')}
=
\exp\!\left(
-\frac{
\bm{d}_{(k,l),(k',l')}^{\top}
\bm{\Lambda}_{\xi}
\bm{d}_{(k,l),(k',l')}
}{2}
\right),
\qquad
\bm{\Lambda}_{\xi}=
\begin{bmatrix}
\xi_2^{-2} & 0\\
0 & \xi_3^{-2}
\end{bmatrix}.
\end{equation}
\noindent where $\mathrm{dist}_{\text{chip}}(m,n)$ denotes the distance between the centers of chips $m$ and $n$, $(k,l)$ and $(k',l')$ index B\'ezier control points on the $(K{+}1)\times(L{+}1)$ control-point grid inside a chip, and $\bm{d}_{(k,l),(k',l')}$ denotes the corresponding in-plane displacement vector (e.g., in normalized local $(u,v)$ coordinates) between two control-point locations. The hyperparameters $\bm{\xi}=(\xi_1,\xi_2,\xi_3,\xi_4)$ control the overall strength of correlation, the within-chip length scales along the local $u$ and $v$ directions, and the between-chip correlation decay, respectively. Under the between-chip kernel parameterization \(\exp\{-\mathrm{dist}_{\text{chip}}(m,n)/\xi_4^2\}\), a larger \(\xi_4\) implies slower correlation decay across chip locations. It is also noted that this parameterization avoids scale non-identifiability in the separable covariance. If separate amplitude parameters are assigned to both $\bm{\Sigma}_{\text{between}}$ and $\bm{\Sigma}_{\text{within}}$, only their product is identifiable through the Kronecker product. Therefore, we use a single global scale parameter $\xi_1$ and absorb it into $\bm{\Sigma}_{\text{between}}$, while $\bm{\Sigma}_{\text{within}}$ is modeled as a correlation matrix.

Finally, let $\bm{S} = \text{blockdiag}(\bm{S}_{0,0},\bm{S}_{0,1},\cdots,\bm{S}_{n_x-1,n_y-1} )$
with $\bm{S}_{i,j}=(\bm{H}_{i,j}^{C})^{\top}\otimes \bm{H}_{i,j}^{R}$, we have $\vec{\bm{Q}} = \bm{S}\vec{\bm{P}}$, and the covariance of the latent free control-point vector is obtained by restricting $\bm{\Sigma}_{\vec{\bm{P}}}$ to the free set:
\begin{equation} \label{eq:Sigma_Q}
\bm{\Sigma}_{\vec{\bm{Q}}} = \bm{S}\,\bm{\Sigma}_{\vec{\bm{P}}}\,\bm{S}^{\top}.
\end{equation}

\noindent Hence, it follows from (\ref{eq:global_final}), (\ref{eq:distribution_Q}) and (\ref{eq:Sigma_Q}) that the wafer surface warpage is modeled by
\begin{equation} \label{eq:model_final}
\bm{z} = \bm{J}\bm{\mu}_{\vec{\bm{Q}}} + \bm{\eta},
\qquad
\bm{\eta}\sim \mathcal{N}(\bm{0},\bm{V}(\bm{\theta})), \qquad
\bm{V}(\bm{\theta})
=
\bm{J} \bm{\Sigma}_{\vec{\bm{Q}}}(\bm{\theta}_Q)\bm{J}^\top
+
\bm{\Sigma}_{\epsilon}(\bm{\theta}_\epsilon),
\end{equation}
\noindent where $\bm{\theta}_Q$ and $\bm{\theta}_\epsilon$ denote the hyperparameters associated with the covariance matrices $\bm{\Sigma}_{\vec{\bm{Q}}}$ and $\bm{\Sigma}_{\epsilon}$, respectively, and 
$\bm{\theta}=(\bm{\theta}_Q,\bm{\theta}_\epsilon)$. 
Because the matrix $\bm{J}$ is obtained from the B\'ezier surface, the model (\ref{eq:model_final}) is referred to as the B\'ezier-GP for which $\bm{z}\sim
\mathcal N\!(
\bm{J}\bm{\mu}_{\vec{\bm{Q}}},
\bm{J}\bm{\Sigma}_{\vec{\bm{Q}}}\bm{J}^{\top}
+
\bm{\Sigma}_{\epsilon})$.

\vspace{2pt}
Given measurement data, parameters to be estimated include the mean vector and the hyperparameters associated with the covariance matrices. In particular, if we consider \textit{i.i.d.} measurement error, i.e., $\bm{\Sigma}_{\epsilon}(\bm{\theta}_\epsilon)=\sigma_\epsilon^2 \bm{I}$, the model involves unknown parameters $\bm{\mu}_{\vec{\bm{Q}}}$, $\bm{\theta}_Q = (\xi_1,\xi_2,\xi_3,\xi_4)$
and $\sigma_\epsilon$. 
The Iteratively Re-Weighted Generalized Least Squares (IRWGLS) can be used for estimating the model parameters, which involves the following 3 steps \citep{ref25}:

Step 1: Set initial $\hat{\bm{V}}$ to an identity matrix.

Step 2: Estimate the mean vector $\bm{\mu}$ using the Estimated Generalized Least Squares (EGLS)
\begin{equation}
\hat{\bm{\mu}}
=
(
\bm{J}^\top
\hat{\bm{V}}^{-1}
\bm{J})^{-1}
\bm{J}^\top
\hat{\bm{V}}^{-1}
\bm{z}.
\end{equation}

In the numerical implementation, singular mean normal equations are solved using the Moore--Penrose pseudoinverse via singular value decomposition (SVD) and selecting the minimum-norm solution without a ridge term.

Step 3: Compute the residuals, $\bm{r}
=
\bm{z}-\bm{J}\hat{\bm{\mu}}$, and estimate $\bm{\theta}$ associated with the covariance matrix $\bm{V}(\bm{\theta})$ by maximizing the log-likelihood function:
\begin{equation}
\hat{\bm{\theta}}
=
\arg\max_{\bm{\theta}}
[
-\frac12
(
\log|\bm{V}(\bm{\theta})|
+
\bm{r}^{\top}
\bm{V}(\bm{\theta})^{-1}
\bm{r}
)
]
\end{equation}
\noindent and update the estimate $\hat{\bm{V}}$. 

Iterate steps 2 and 3 until the relative changes of {\color{black}$\hat{\bm{\mu}}$} and $\hat{\bm{\theta}}$ are sufficiently small. Asymptotically, the
EGLS estimator in Step 2 possesses the asymptotic properties of the Maximum Likelihood estimator, 
and is equivalent to the Generalized Least Squares (GLS) estimator under regularity conditions; see Chapter 6.2 of  \cite{ref25}.



\vspace{-6pt}
\section{Numerical Experiments}
\label{sec:experiments}
\vspace{-6pt}
\subsection{Experiment Setup}

\begin{figure}[!ht]
    \centering
    \begin{subfigure}{0.48\linewidth}
        \centering
        \includegraphics[
            width=\linewidth,
            height=0.35\textheight
        ]{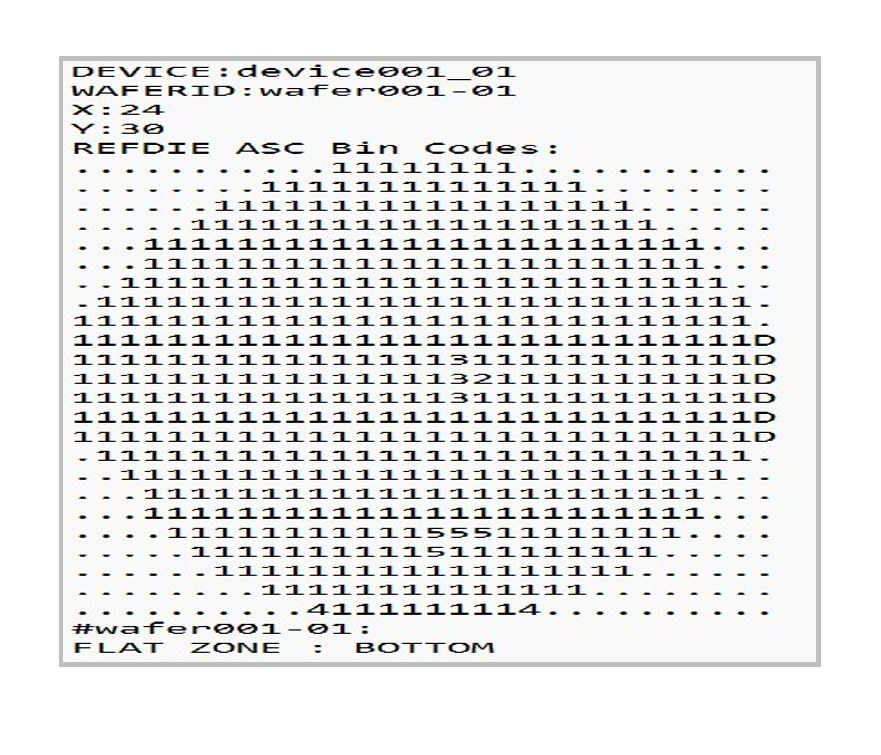}
        \caption{ASC bin-map file}
    \end{subfigure}
    \hfill
    \begin{subfigure}{0.48\linewidth}
        \centering
        \includegraphics[
            width=\linewidth,
            height=0.35\textheight
        ]{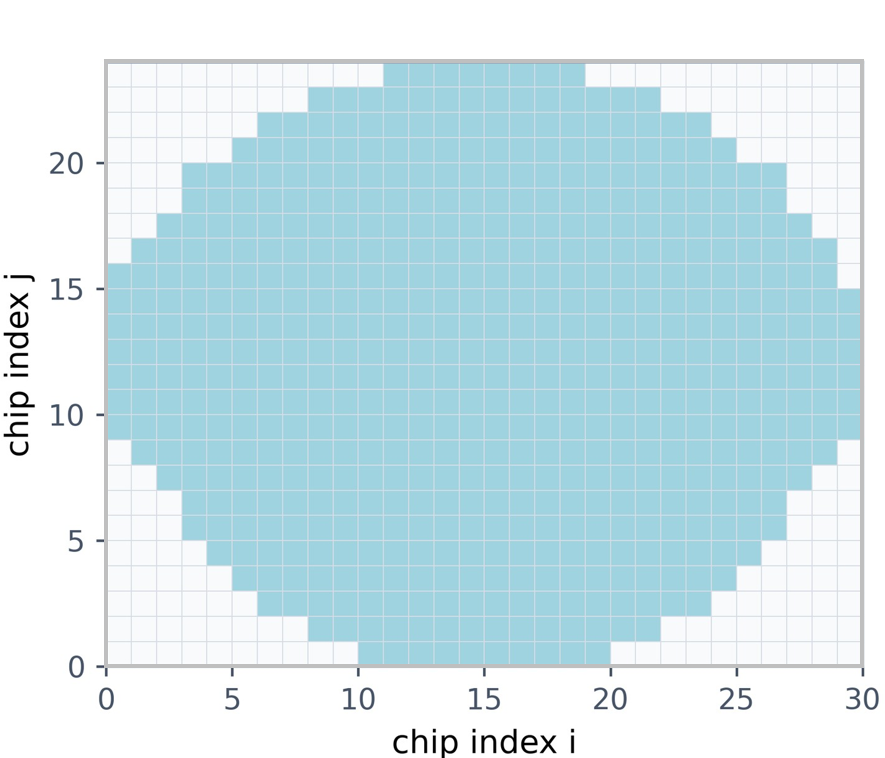}
        \caption{Parsed real-chip mask}
    \end{subfigure}
    \caption{ASC wafer-map input and parsed chip-layout mask.}
    \label{fig:asc_layout}
\end{figure}

This section investigates the performance of the proposed B\'ezier-GP through numerical experiments. We consider synthetic data based on the chip layout defined from an ASC bin wafer-map file. This profile is a manufacturing-compatible text representation of die locations and bin codes, 
and provides metadata such as device ID, wafer ID, horizontal and vertical die counts, flat-zone information, and an ASCII bin-code matrix. Product-bin entries identify actual chip locations, while non-product entries and outside-wafer cells are treated as dummy or empty cells. This layout information is used to build the chips-on-wafer patch grid in our numerical experiment and define which fitted patches are retained for final real-chip analysis.
As an illustration, Figure~\ref{fig:asc_layout} shows the ASC text input and the parsed chip mask. The parsed chip-layout index matrix has size $30\times 24$ after applying the coordinate convention used by the modeling code, with 554 actual-chip cells. This mask determines the real-chip domain.
{\color{black} For computation, the parsed $30\times24$ chip layout is embedded in a $32\times26$ rectangular patch grid covering the full wafer disk. The additional patches are treated as dummy cells. Coordinate transformations use the dimensions and origin of this extended grid, while preserving the physical locations and bin identities of the original cells.
}

\begin{figure}[!h]
    \centering
    \begin{subfigure}{0.49\linewidth}
        \centering
        \includegraphics[
            width=\linewidth,
            height=0.30\textheight
        ]{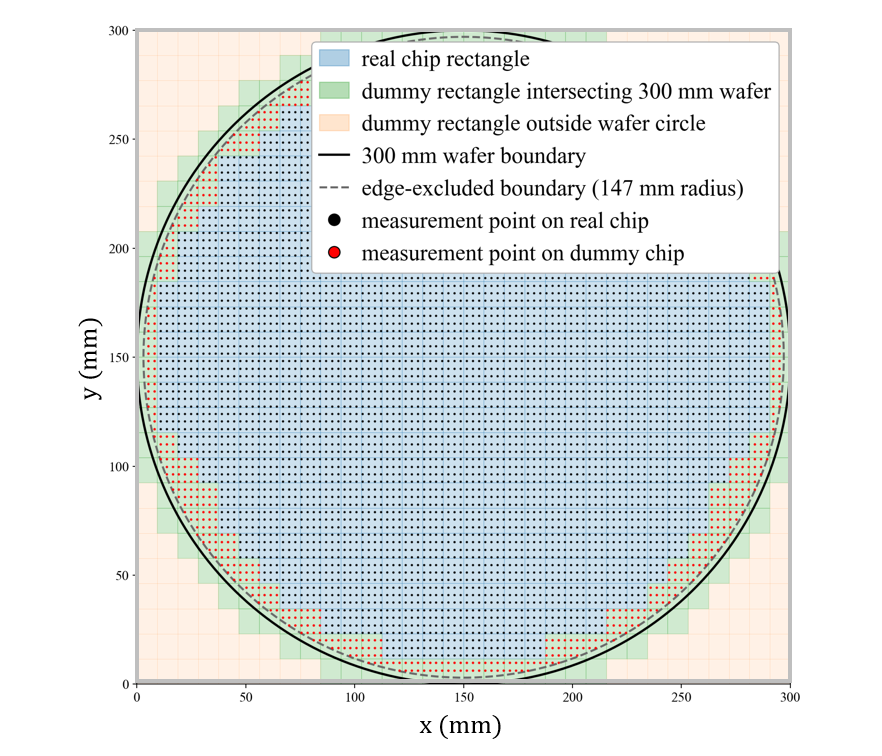}
        \caption{Chip layout area and measurement locations}
    \end{subfigure}
    \hfill
    \begin{subfigure}{0.49\linewidth}
        \centering
        \includegraphics[
            width=\linewidth,
            height=0.30\textheight
        ]{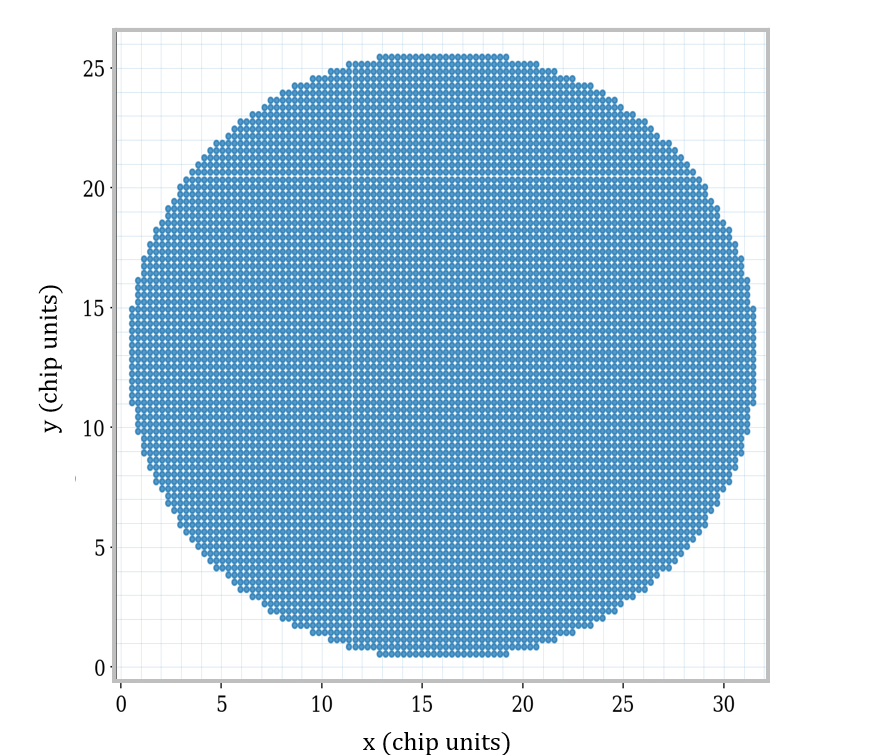}
        \caption{Common-$\bm{J}$ measurement locations}
    \end{subfigure}
    \caption{Measurement geometry over the chip layout and 300 mm wafer domain.}
    \label{fig:measurement_geometry}
\end{figure}

After the ASC map is parsed, its physical placement is defined relative to the 300 mm wafer disk centered at $(150,150)$ mm, using the registration and pitch normalization described above. In this experiment, we set $\Delta_x=\Delta_y=0$, aligning the center of the chip layout grid with the wafer center. Measurement locations specified by the sampling recipe are assigned to chip patches through the same physical-to-local coordinate transformation. A 3mm edge exclusion is applied, resulting in an inner sampling radius of 147 mm. The chip pitch is 9.3750 mm in the horizontal direction and 11.5385 mm in the vertical direction. Therefore, the 0.30-chip-unit spacing corresponds to 2.8125 mm in the \(x\) direction and 3.4615 mm in the \(y\) direction. Figure~\ref{fig:measurement_geometry} shows the single-wafer measurement geometry over the ASC layout and the measurement locations  in our experiments. 

\begin{figure}[!ht]
    \centering
    \begin{subfigure}{0.45\linewidth}
        \centering
        \includegraphics[
            width=\linewidth,
            height=0.30\textheight
        ]{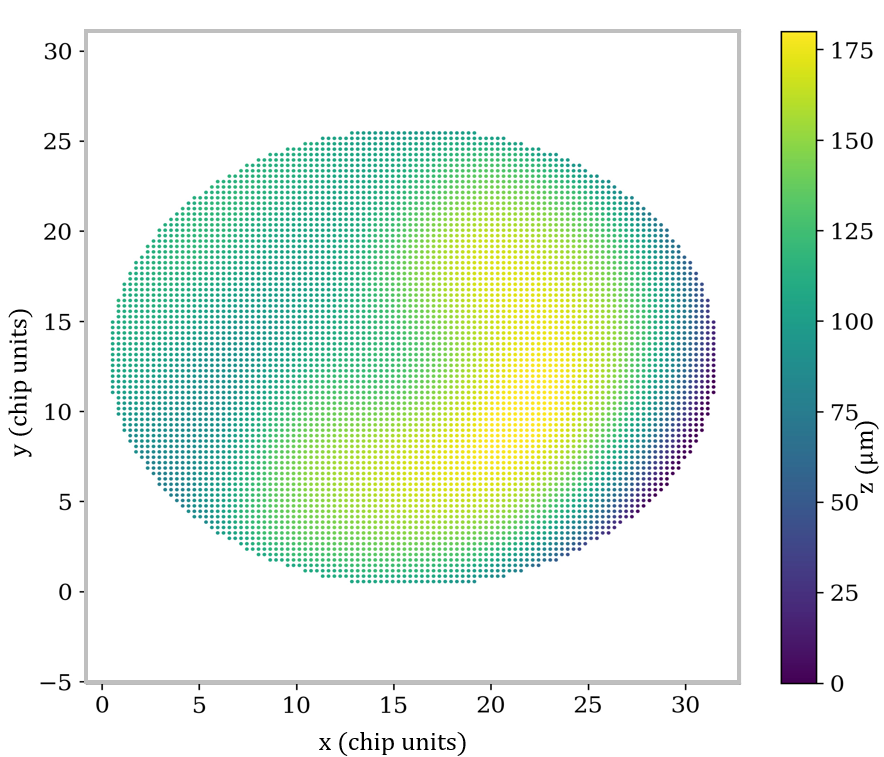}
        \caption{Input point-cloud heights}
    \end{subfigure}
    \hfill
    \begin{subfigure}{0.54\linewidth}
        \centering
        \includegraphics[
            width=\linewidth,
            height=0.30\textheight
        ]{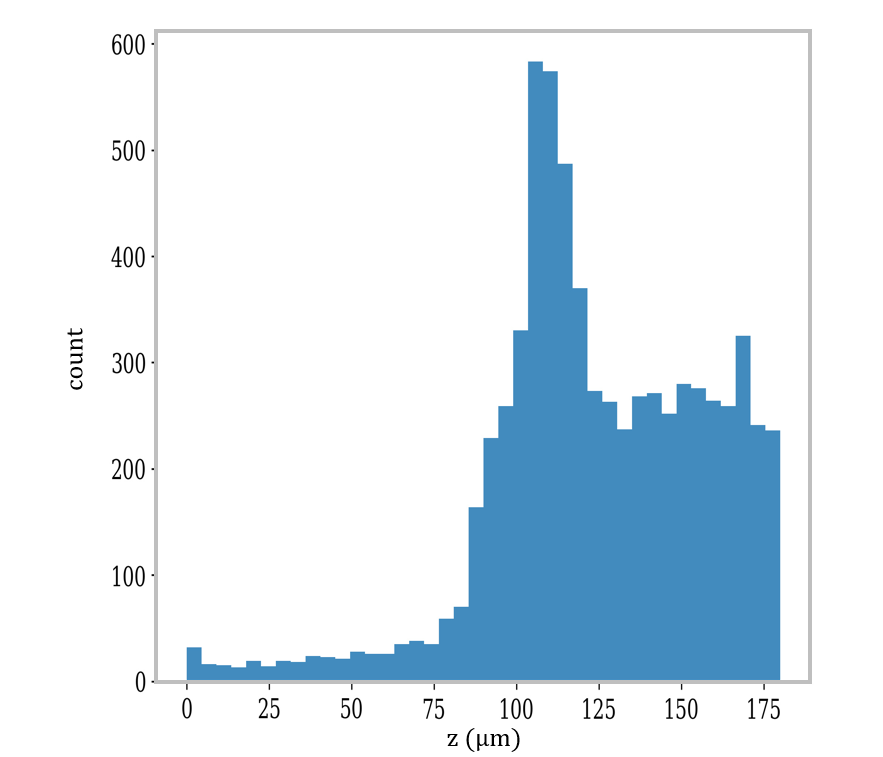}
        \caption{Height histogram}
    \end{subfigure}
    \caption{The wafer-shape point cloud and the histogram of the height measurements.}
    \label{fig:single_input}
\end{figure}

The input wafer shape is obtained by digitizing the wafer shape plot reported in \citep{ref35} and mapping the normalized image intensity to the height range \([0,180]\). Measurement locations are generated over the inner wafer disk using the geometry described above. As an illustration, Figure~\ref{fig:single_input} shows the input height measurements and the corresponding histogram.

{\color{black}
The mean is initialized using the minimum-norm OLS solution. Mean updates and single-wafer covariance/noise parameter updates are applied in full, whereas multi-wafer covariance/noise parameter updates use 50\% relaxation in log space to moderate abrupt changes and improve the stability of iterative estimation. To identify sufficiently small changes in the mean vector and the covariance/noise parameter vector, the change-based stopping criterion requires the relative $L_2$ changes of both vectors to be below $10^{-5}$.
}
The main settings  are summarized in Table~\ref{tab:experiment_settings}.

\vspace{0.3\baselineskip}
\begin{table}[htbp]
\centering
\caption{Main Experimental Settings}
\label{tab:experiment_settings}
\begin{tabular}{p{0.36\linewidth}p{0.54\linewidth}}
\toprule
Setting & Value \\
\midrule
Wafer diameter & 300 mm \\
Edge exclusion & 3 mm \\
Parsed ASC grid & $30\times 24$ chip cells \\
Actual chip cells & 554 \\
B\'ezier degree & $K=L=3$ \\
Full control-point dimension & 13,312 \\
Free control-point dimension & 3,564 \\
Measurement spacing & 0.30 chip units \\
Physical spacing (Resolution) & 2.8125 mm in $x$, 3.4615 mm in $y$ \\
Measurements per wafer & 6,972 \\
Multi-wafer training wafers & 10 \\
Computing environment & \makecell[l]{Google Colab GPU runtime ({\color{black} A100})\\CUDA 13.0; CuPy 14.0.1} \\
\bottomrule
\end{tabular}
\end{table}

\vspace{-6pt}
\subsection{Experiment I: Single-Wafer Warpage Modeling}
\begin{figure}[!ht]
    \centering
    \begin{minipage}[t]{0.49\linewidth}
        \centering
        \includegraphics[width=\linewidth]{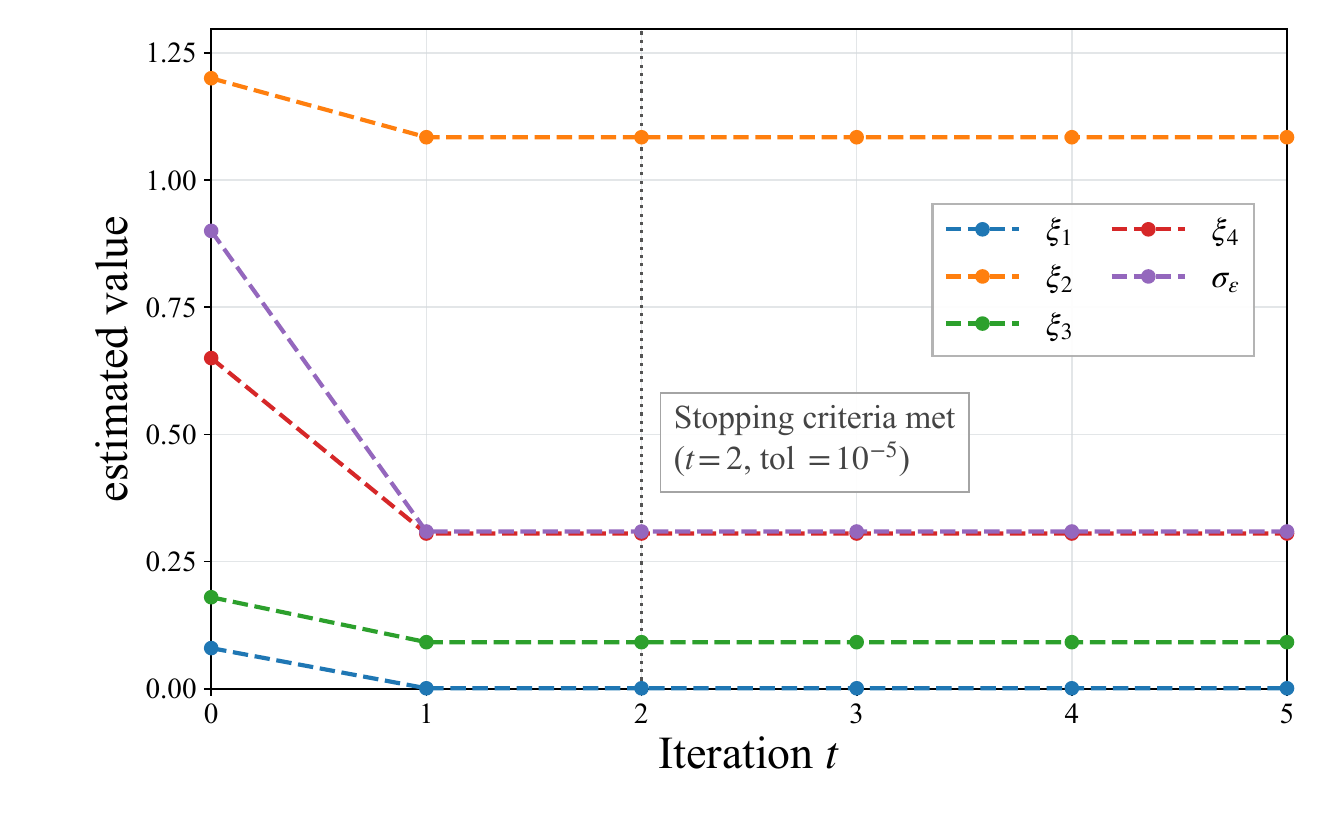}
        \vspace{-6pt}
        \caption{Covariance hyperparameter convergence}
        \label{fig:single_convergence}
    \end{minipage}
    \hfill
    \begin{minipage}[t]{0.49\linewidth}
        \centering
        \includegraphics[width=\linewidth]{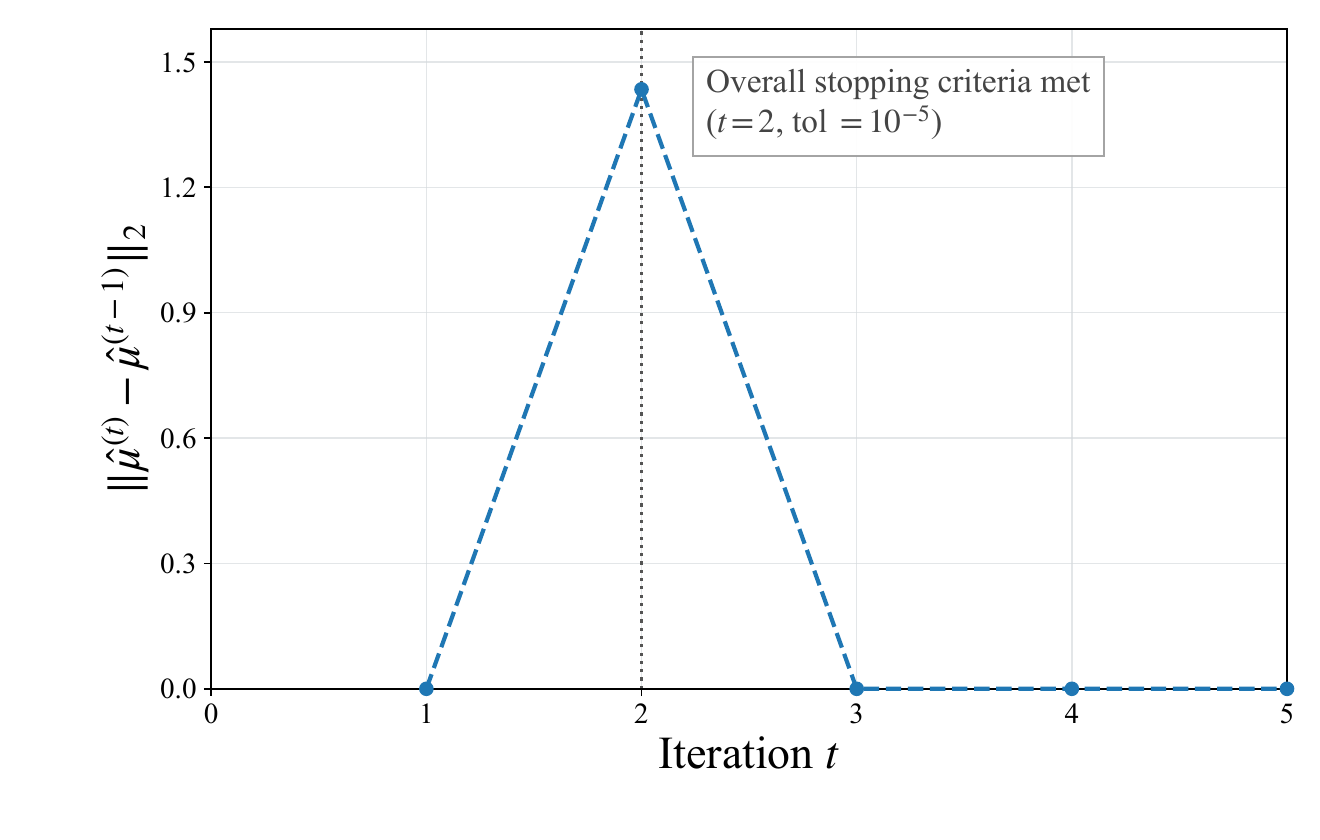}
        \vspace{-6pt}
        \caption{Absolute mean-vector update norm}
        \label{fig:single_parameter_history_combined}
    \end{minipage}
\end{figure}

\vspace{-4pt}
\textbf{Parameter Estimation.} 
To investigate the modeling capability of the proposed B\'ezier-GP model, we first fit the model to data collected from a single wafer (before the model is applied to data from multiple wafers) using the IRWGLS procedure. Figure~\ref{fig:single_convergence} shows the trajectories of the estimated covariance and noise hyperparameters over the IRWGLS iterations, and Figure~\ref{fig:single_parameter_history_combined} shows the update,  $\|\hat{\bm{\mu}}^{(t)}-\hat{\bm{\mu}}^{(t-1)}\|_2$, of the estimated mean vector. 

The stabilization of these traces indicates the convergence of the IRWGLS procedure, and the estimated parameters are given in Table~\ref{tab:single_theta_hat}. {\color{black} Because the single-wafer experiment does not include repeated wafer observations, these covariance/noise estimates are interpreted as model-fitting and uncertainty-diagnostic parameters rather than reflecting the wafer-to-wafer warpage variation.}

\vspace{0.3\baselineskip}
\begin{table}[!h]
\centering
\caption{Estimated hyperparameter values for the single-wafer experiment.}
\label{tab:single_theta_hat}
\begin{tabular}{llr}
\toprule
Parameter & Parameter description & Estimates \\
\midrule
\(\xi_1\) & Global scale & 0.0010 \\
\(\xi_2\) & Within-chip \(u\) length scale & 1.0841 \\
\(\xi_3\) & Within-chip \(v\) length scale & 0.0916 \\
\(\xi_4\) & Between-chip correlation decay & 0.3052 \\
\(\sigma_\epsilon\) & Observation residual std. & 0.3091 \\
\bottomrule
\end{tabular}
\end{table}

\begin{figure}[htbp]
    \centering
    \includegraphics[width=\linewidth]{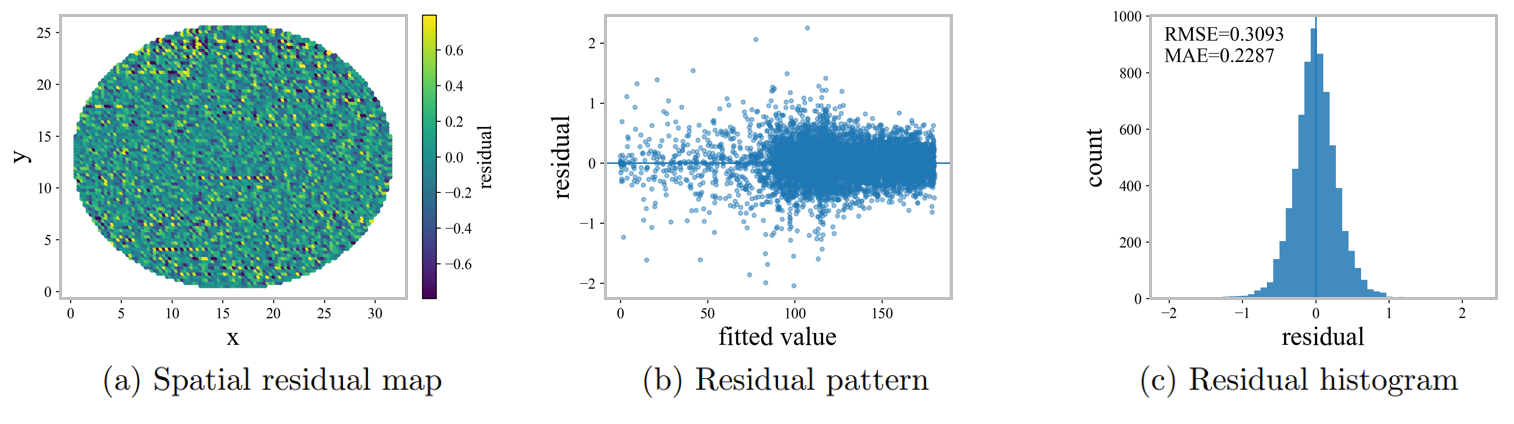}
    \caption{Centered residual diagnostics after IRWGLS estimation for the single-wafer experiment.}
    \label{fig:single_runtime_residuals}
\end{figure}

To further validate the model, Figure~\ref{fig:single_runtime_residuals} shows the centered residuals. It is seen that neither the spatial plot of the residuals nor the plot of residuals against fitted values shows any noticeable patterns. The residuals are approximately normally distributed with RMSE and MAE being 0.3093 and 0.2287, respectively. 

\begin{figure}[htbp]
    \centering
    \begin{subfigure}{0.49\linewidth}
        \centering
        \includegraphics[width=\linewidth]{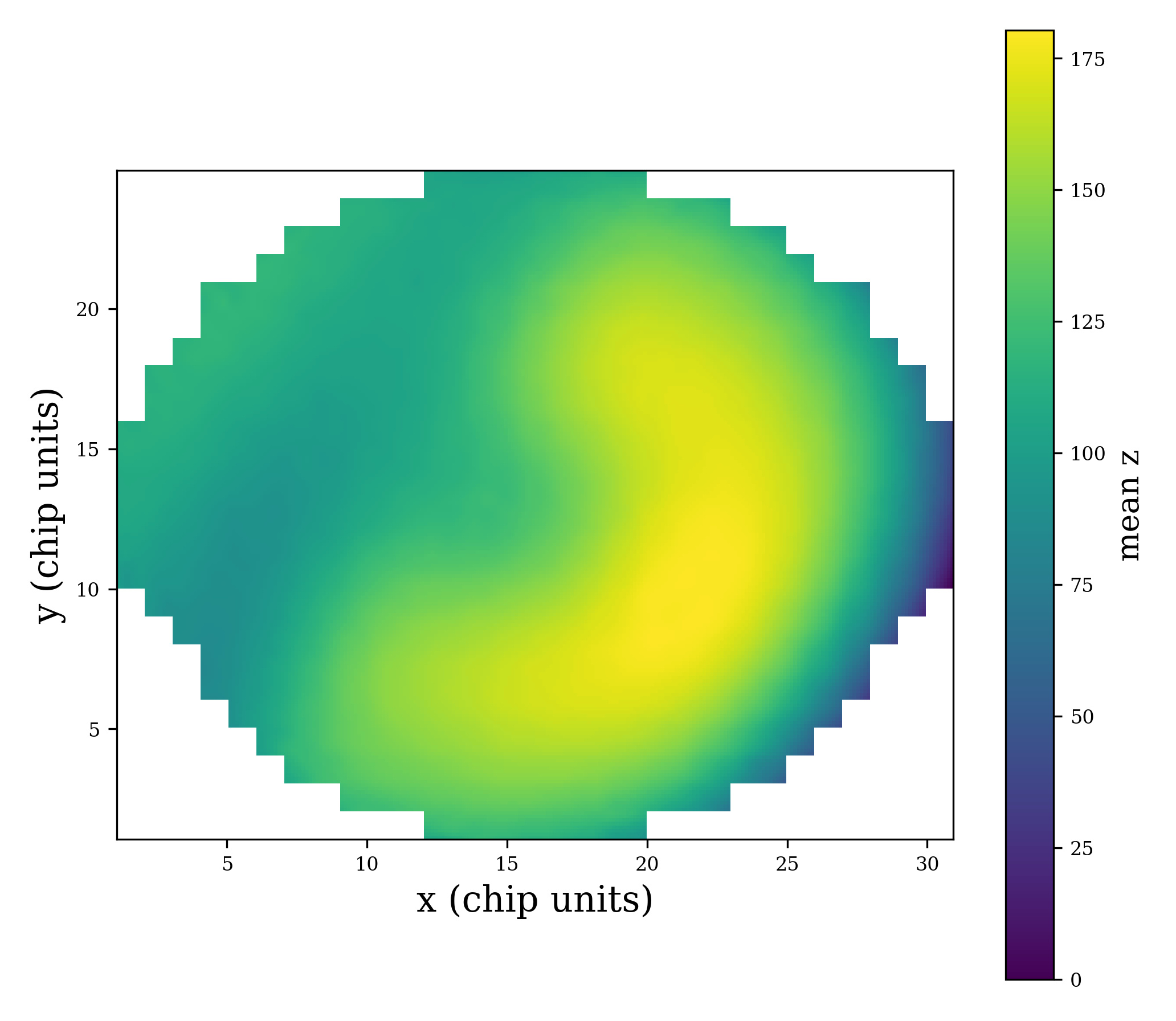}
        \caption{Estimated mean map}
    \end{subfigure}
    \hfill
    \begin{subfigure}{0.49\linewidth}
        \centering
        \includegraphics[width=\linewidth]{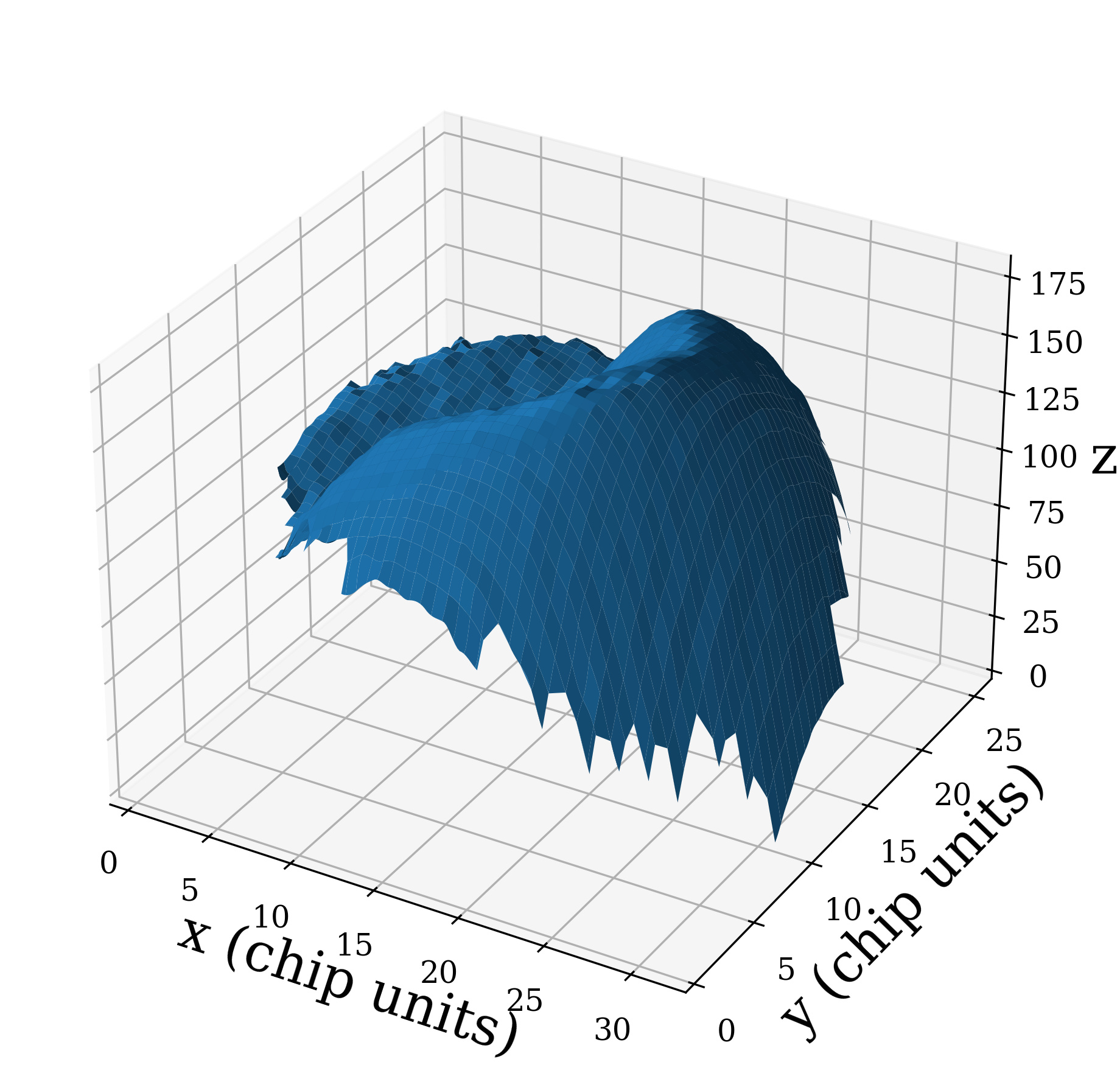}
        \caption{Estimated mean surface}
    \end{subfigure}
    \caption{Estimated mean surface over real-chip locations.}
    \label{fig:single_mean_surface}
\end{figure}

\textbf{Main Results.} 
Figure~\ref{fig:single_mean_surface} presents the 2D and 3D views of the estimated mean surface over the real-chip-only domain, $\Omega_0$, excluding the dummy chips. Comparing the fitted surface in Figure~\ref{fig:single_mean_surface} and the input point-cloud heights data shown in Figure \ref{fig:single_input}, the fitted B\'ezier surface preserves the large-scale wafer-shape pattern while producing a continuous chips-on-wafer representation.

\begin{figure}[!h]
    \centering
    \begin{subfigure}{0.49\linewidth}
        \centering
        \includegraphics[width=\linewidth]{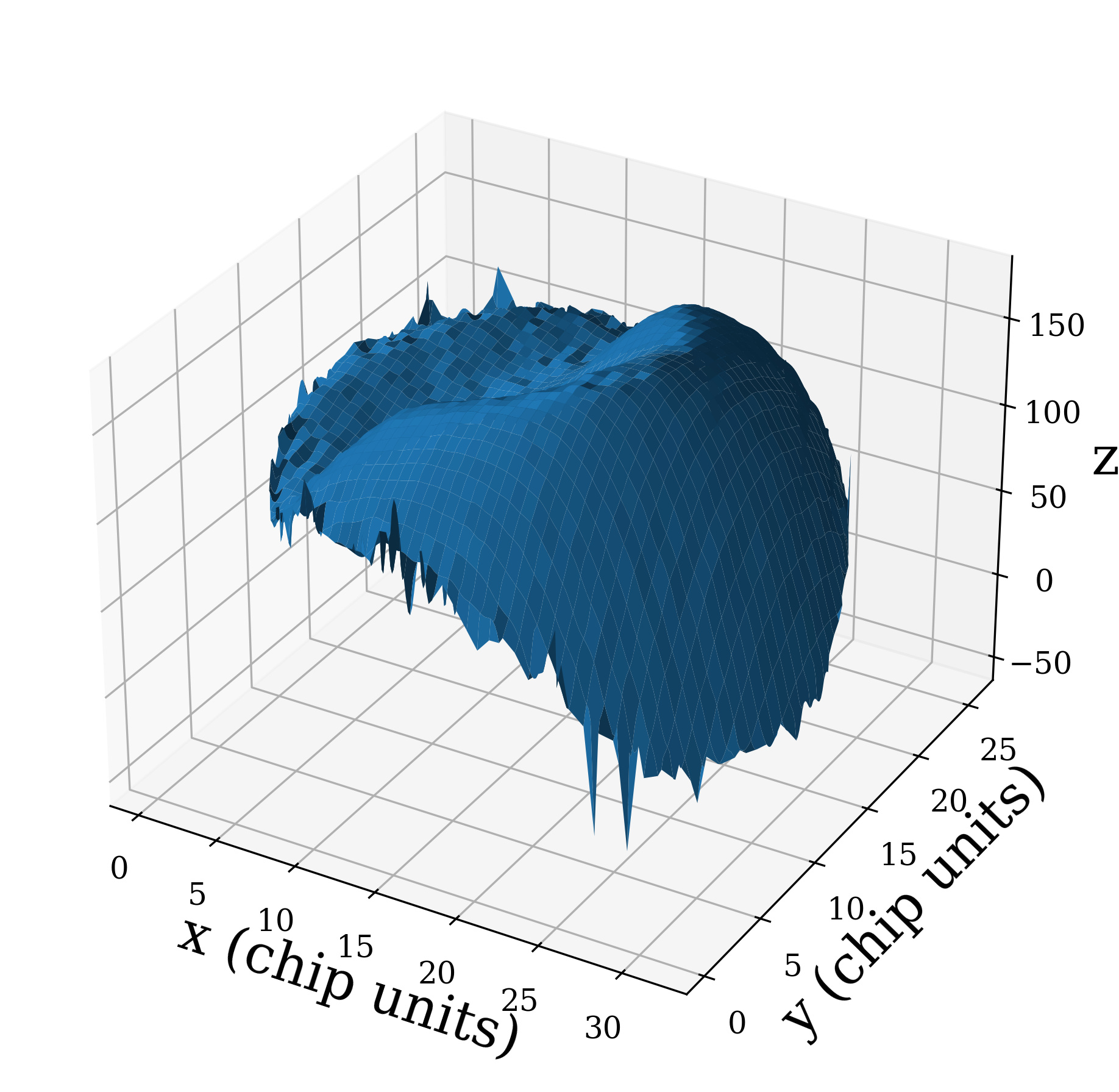}
        \caption{Full fitted rectangular domain}
    \end{subfigure}
    \hfill
    \begin{subfigure}{0.49\linewidth}
        \centering
        \includegraphics[width=\linewidth]{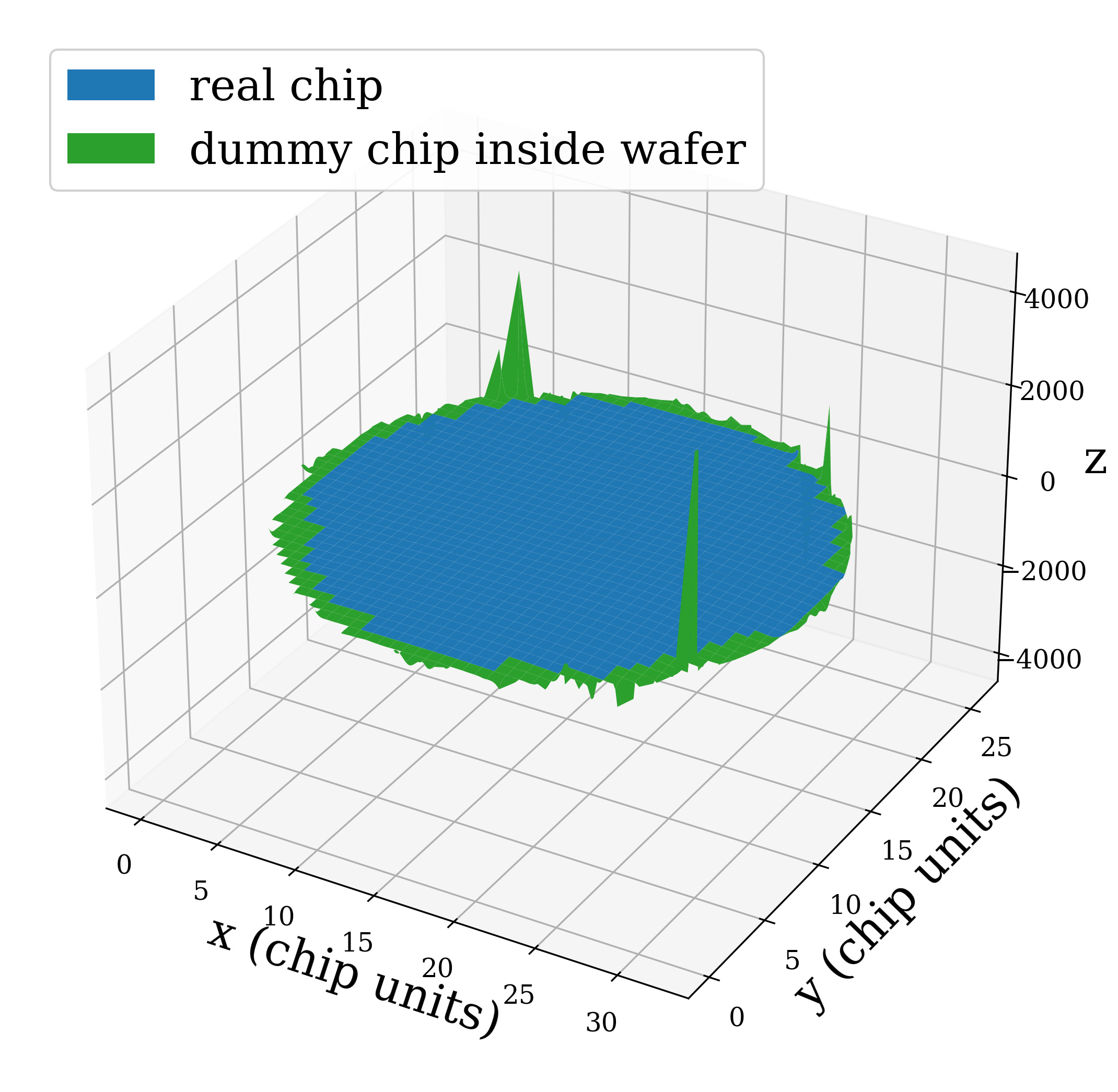}
        \caption{Real chips + inside-wafer dummy chips}
    \end{subfigure}
    \caption{Intermediate fitted surfaces before removing the regions with dummy chips.}
    \label{fig:single_crop_stage_a}
\end{figure}

\begin{figure}[!h]
    \centering
    \begin{subfigure}{0.49\linewidth}
        \centering
        \includegraphics[width=\linewidth]{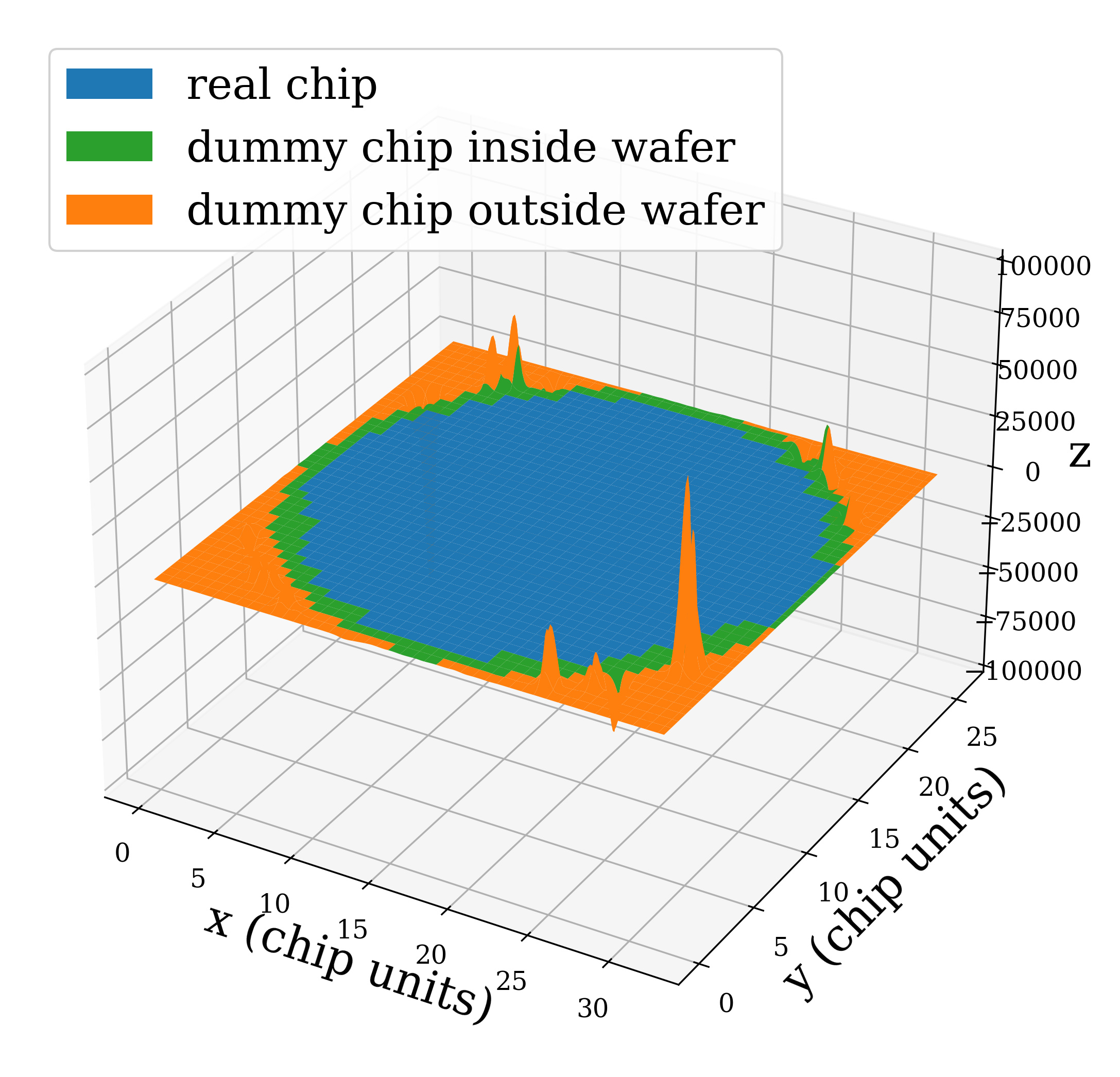}
        \caption{Real chips + all dummy chips}
    \end{subfigure}
    \hfill
    \begin{subfigure}{0.49\linewidth}
        \centering
        \includegraphics[width=\linewidth]{figure/fig_02h_estimated_mean_shape_3d_real_chip_only.jpg}
        \caption{Final real-chip-only surface}
    \end{subfigure}
    \caption{Final restriction from the extended fitted surface to the real-chip domain.}
    \label{fig:single_crop_stage_b}
\end{figure}

Note that the proposed model is fitted on an extended rectangular domain so that B\'ezier patches can be assembled and coupled through sparse matrix operations. This extended domain contains actual chips and dummy cells. Dummy cells are useful during geometric construction because they allow the continuity operators to be applied on a regular chip grid, but they are not part of the final chip-warpage interpretation. Hence, the surface is restricted to the real-chip domain for analysis by excluding the region with dummy chips. 
Figures~\ref{fig:single_crop_stage_a} and~\ref{fig:single_crop_stage_b} show this post-fitting restriction. The full fitted rectangular domain may include unsupported extrapolated values in dummy regions. Visualizing inside-wafer and outside-wafer dummy chips makes it clear which portions are geometric support regions. The final real-chip-only result removes these unsupported regions and is the surface used for chip-level evaluation.


\begin{figure}[htbp]
    \centering
    \begin{subfigure}{0.65\linewidth}
        \centering
        \includegraphics[width=\linewidth]{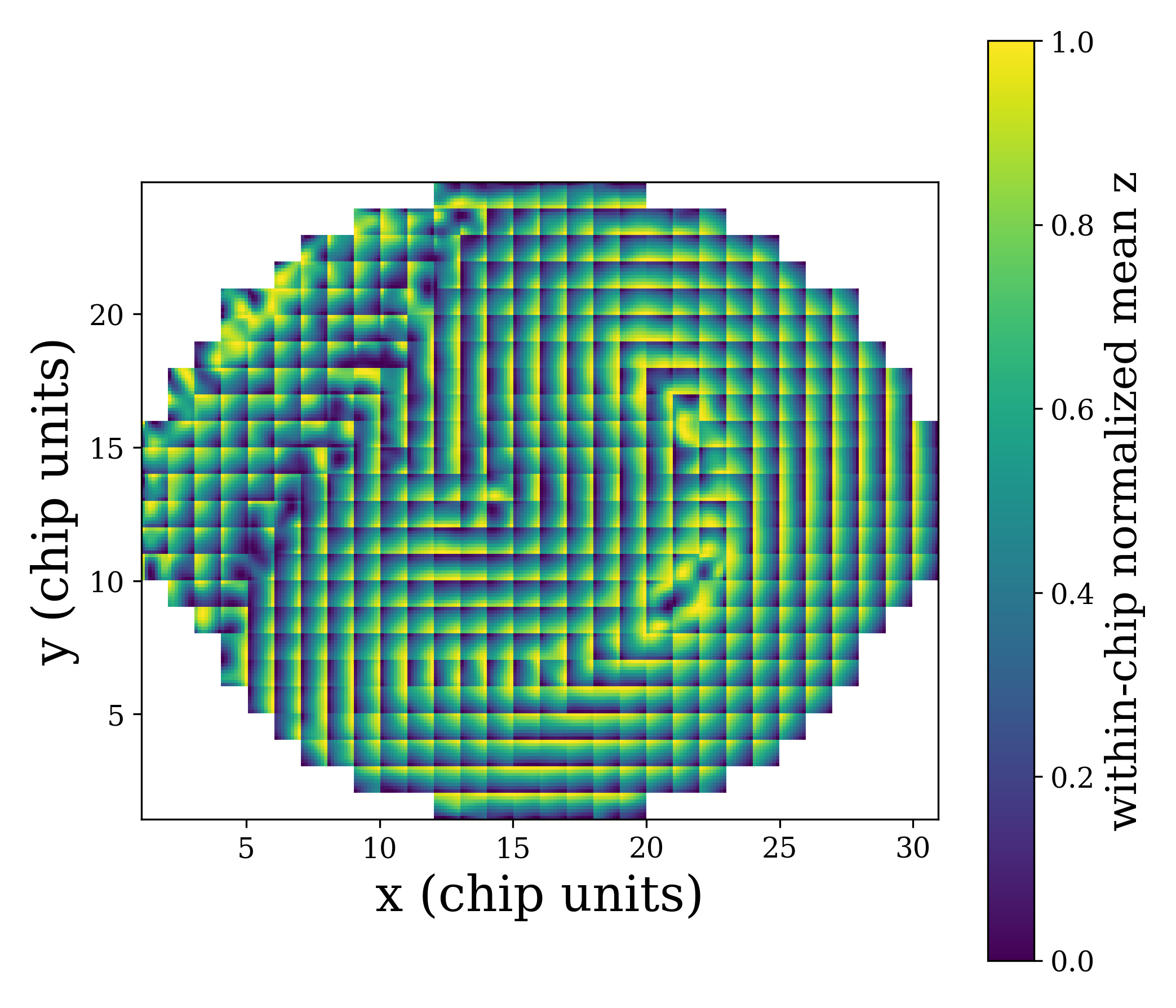}
        \caption{Within-chip normalized mean map}
    \end{subfigure}
    \par\medskip
    \begin{subfigure}{1\linewidth}
        \centering
        \includegraphics[width=\linewidth]{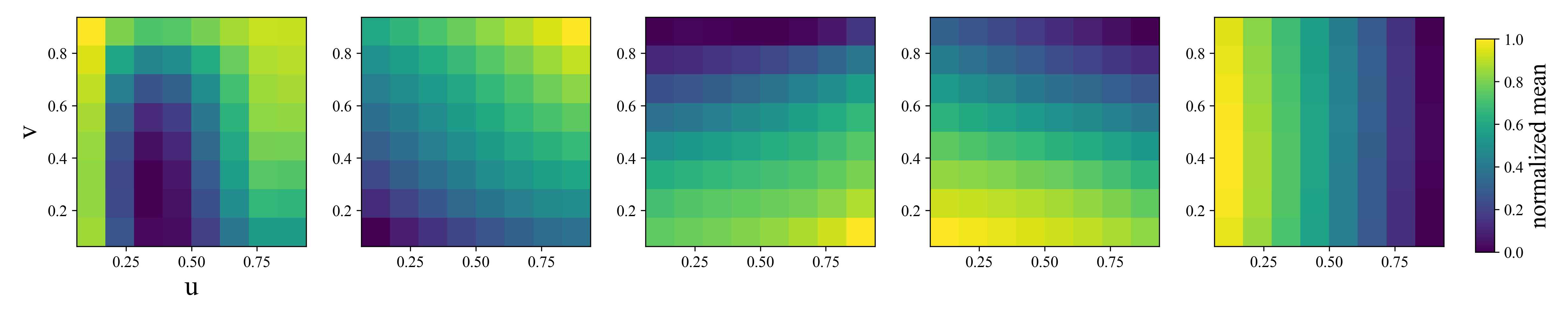}
        \caption{Representative chip-level patches}
    \end{subfigure}
    \caption{Within-chip normalized map and representative chip patches, demonstrating the capability of the proposed B\'ezier-GP model in capturing the chip-level warpage rather than fitting a global smooth surface to the entire wafer surface which lacks the chip-level shape details.}
    \label{fig:single_chip_level}
\end{figure}

Recall that one of the key advantages of using the B\'ezier surface is that it allows us to start with a number of patches with rich geometric shapes before stitching the patches together to form a surface (this is different from modeling the entire wafer surface by a global smooth surface which lacks the chip-level shape details).
Hence, to clearly visualize the capability of modeling chip-level warpage,  Figure~\ref{fig:single_chip_level} shows the within-chip normalized maps and representative chip patches. The chipwise-scale map normalizes each chip separately, making local intra-chip shape patterns visible even when global wafer-level variation dominates the color scale. The representative patches show that each chip has its own fitted surface while remaining embedded in the globally continuous wafer surface---a key advantage of the proposed B\'ezier-GP model. 

\begin{figure}[!ht]
    \centering
    \begin{subfigure}{0.48\linewidth}
        \centering
        \includegraphics[width=\linewidth]{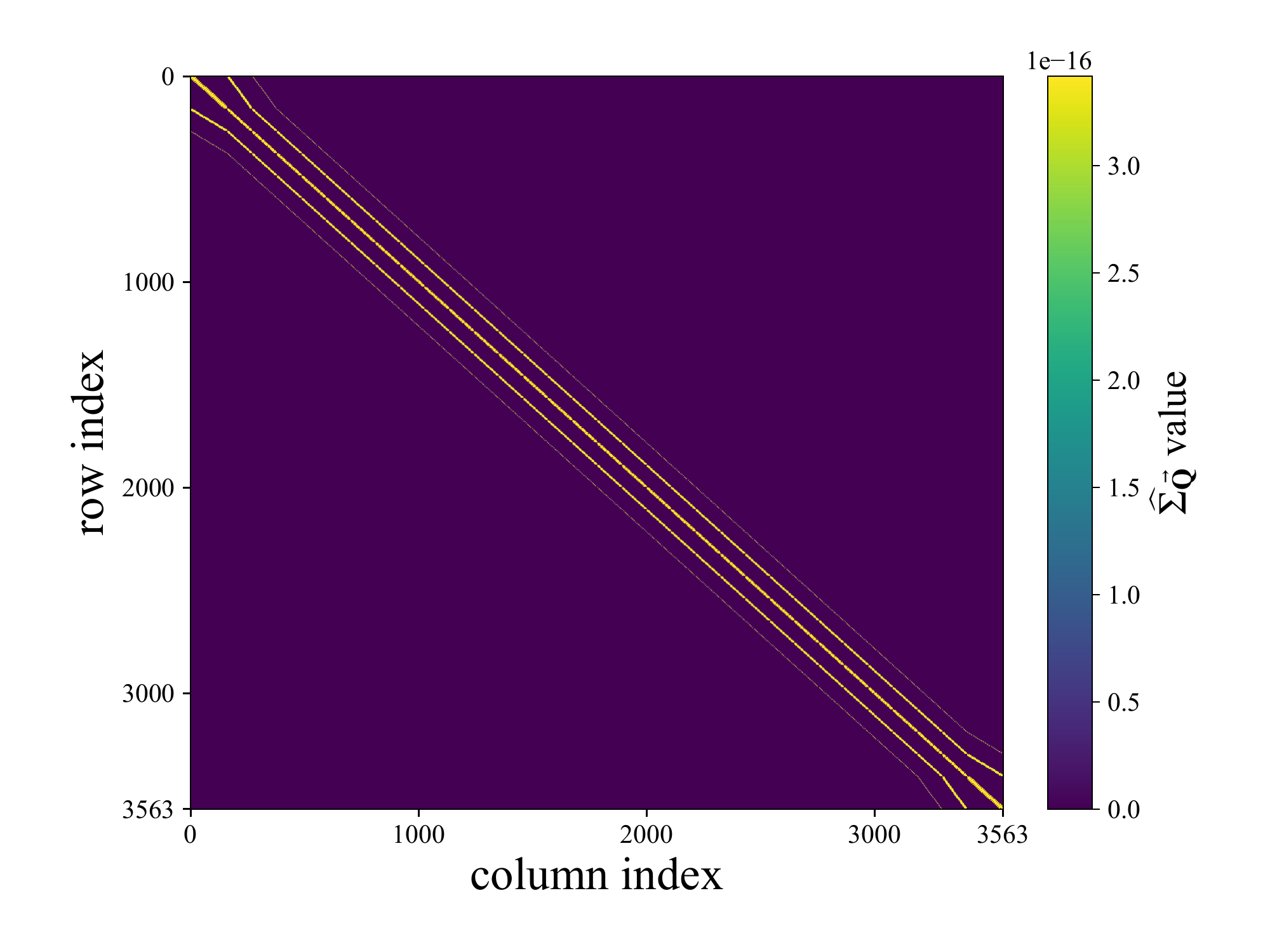}
        \caption{Estimated $\widehat{\bm{\Sigma}}_{\vec{\bm{Q}}}$}
    \end{subfigure}
    \hfill
    \begin{subfigure}{0.48\linewidth}
        \centering
        \includegraphics[width=\linewidth]{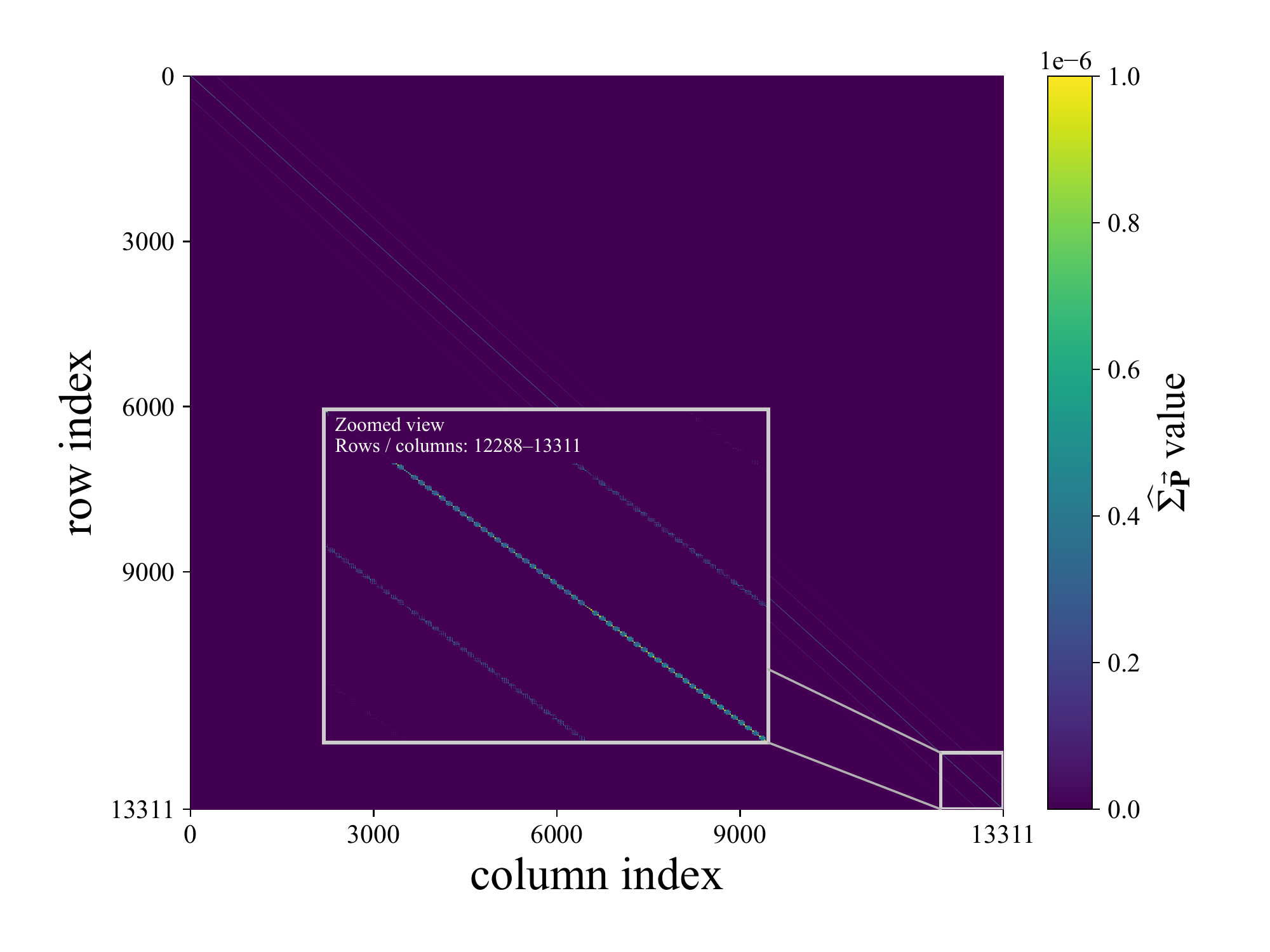}
        \caption{Estimated $\widehat{\bm{\Sigma}}_{\vec{\bm{P}}}$}
    \end{subfigure}
    \caption{Estimated covariance matrices $\hat{\bm{\Sigma}}_{\vec{\bm{Q}}}$ and $\hat{\bm{\Sigma}}_{\vec{\bm{P}}}$.}
    \label{fig:single_sigma}
\end{figure}

\begin{figure}[!ht]
    \centering
    \begin{subfigure}{0.32\linewidth}
        \centering
        \includegraphics[width=\linewidth]{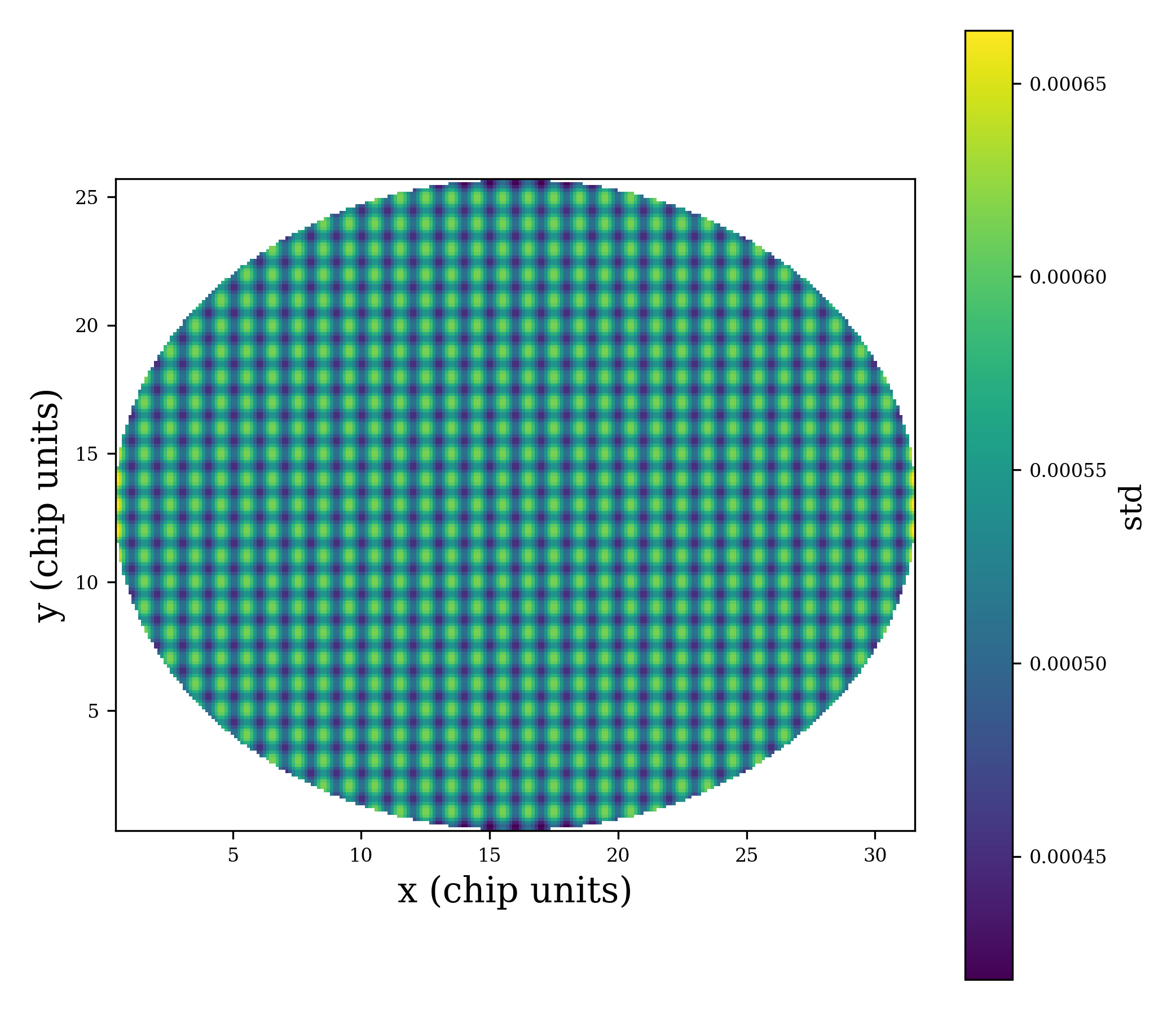}
        \caption{Computed standard deviation}
    \end{subfigure}
    \hfill
    \begin{subfigure}{0.32\linewidth}
        \centering
        \includegraphics[width=\linewidth]{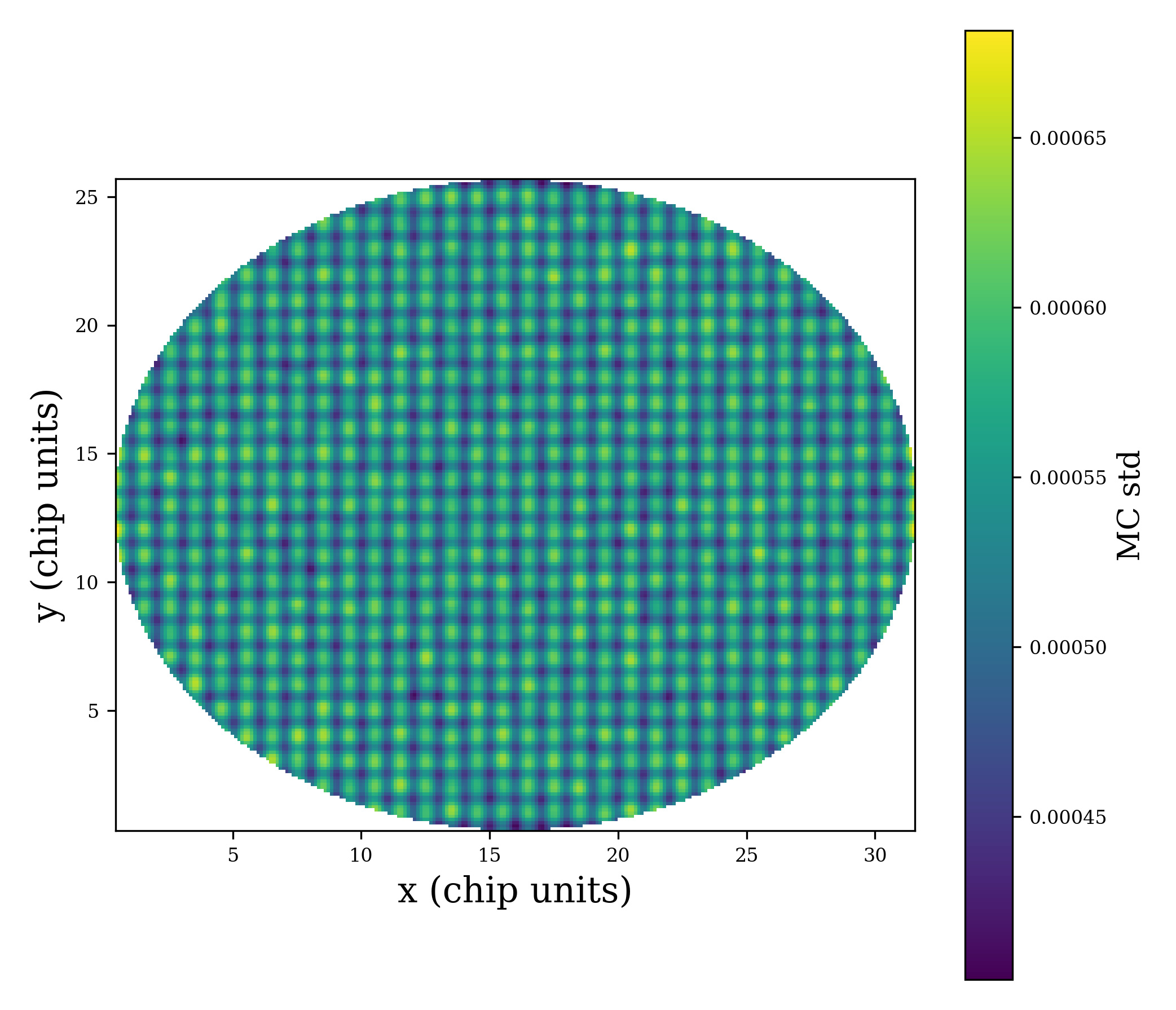}
        \caption{Sample standard deviation}
    \end{subfigure}
    \hfill
    \begin{subfigure}{0.32\linewidth}
        \centering
        \includegraphics[width=\linewidth]{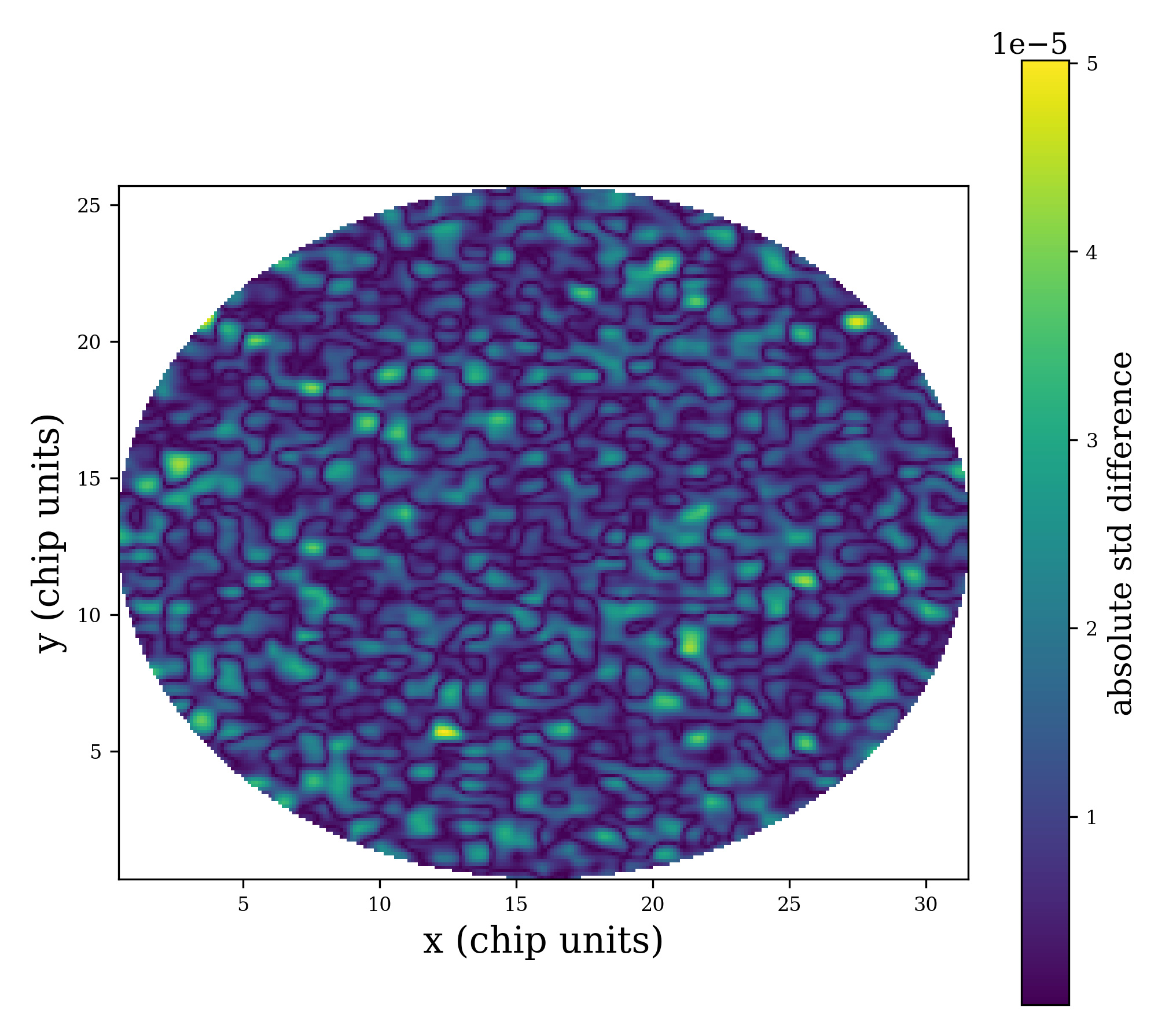}
        \caption{Absolute difference}
    \end{subfigure}
    \caption{Pointwise standard-deviation diagnostics for the single-wafer fitted model.}
    \label{fig:single_std_maps}
\end{figure}

\textbf{Uncertainty Quantification.} 
Figure~\ref{fig:single_sigma} shows the estimated covariance matrices $\hat{\bm{\Sigma}}_{\vec{\bm{Q}}}$ and $\hat{\bm{\Sigma}}_{\vec{\bm{P}}}$. Given the estimated covariance matrix $\hat{\bm{\Sigma}}_{\vec{\bm{Q}}}$, it follows from (\ref{eq:model_final}) that the wafer surface is a spatial GP. 
Figure~\ref{fig:single_std_maps}(a) shows the  pointwise standard deviation at each chip location directly computed from the estimated model. For comparison purposes, Figure~\ref{fig:single_std_maps}(b) also shows the pointwise sample standard deviation computed from 1000 Monte Carlo simulations, and Figure~\ref{fig:single_std_maps}(c) shows the absolute difference between the directly computed and sample standard deviation. This comparison is used as a consistency check. 




\vspace{-26pt}
\subsection{Experiment II: Multi-Wafer Warpage Modeling}
\begin{figure}[!ht]
    \centering
    \includegraphics[width=0.95\linewidth]{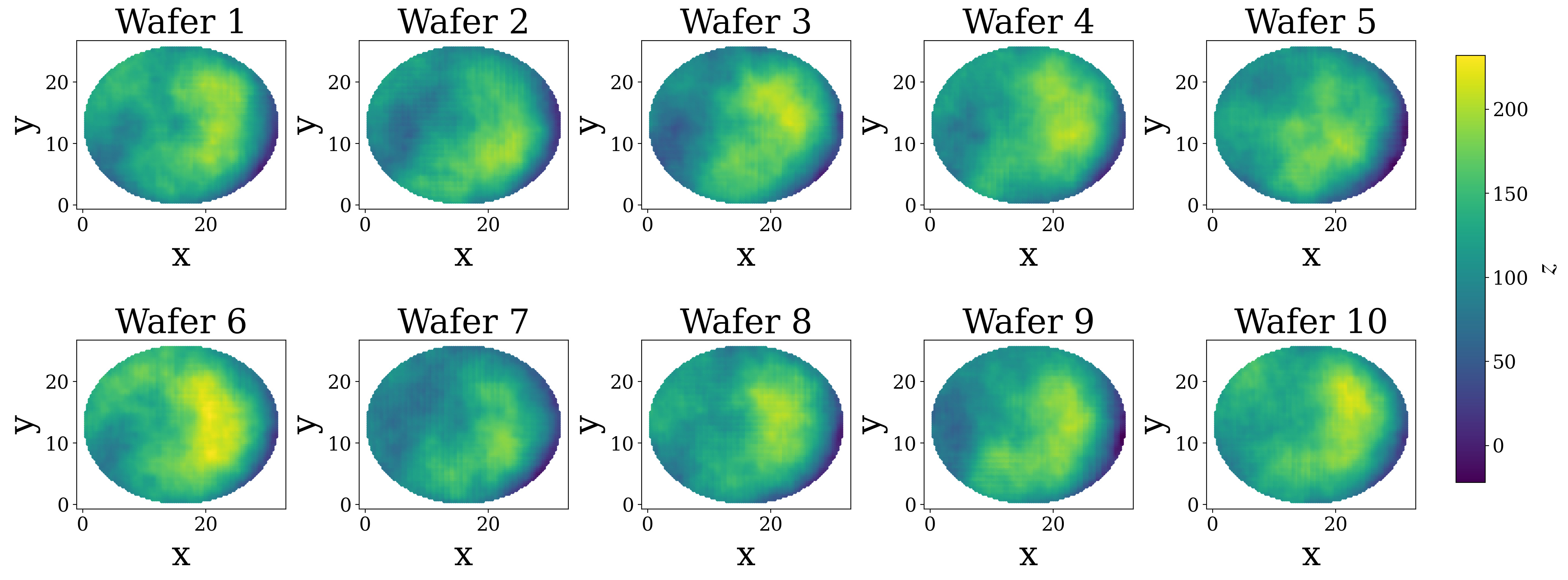}
    \caption{Synthetic wafer observations generated using a common mean and covariance.}
    \label{fig:multi_wafer_montage}
\end{figure}

\begin{figure}[!ht]
    \centering
    \includegraphics[width=0.95\linewidth]{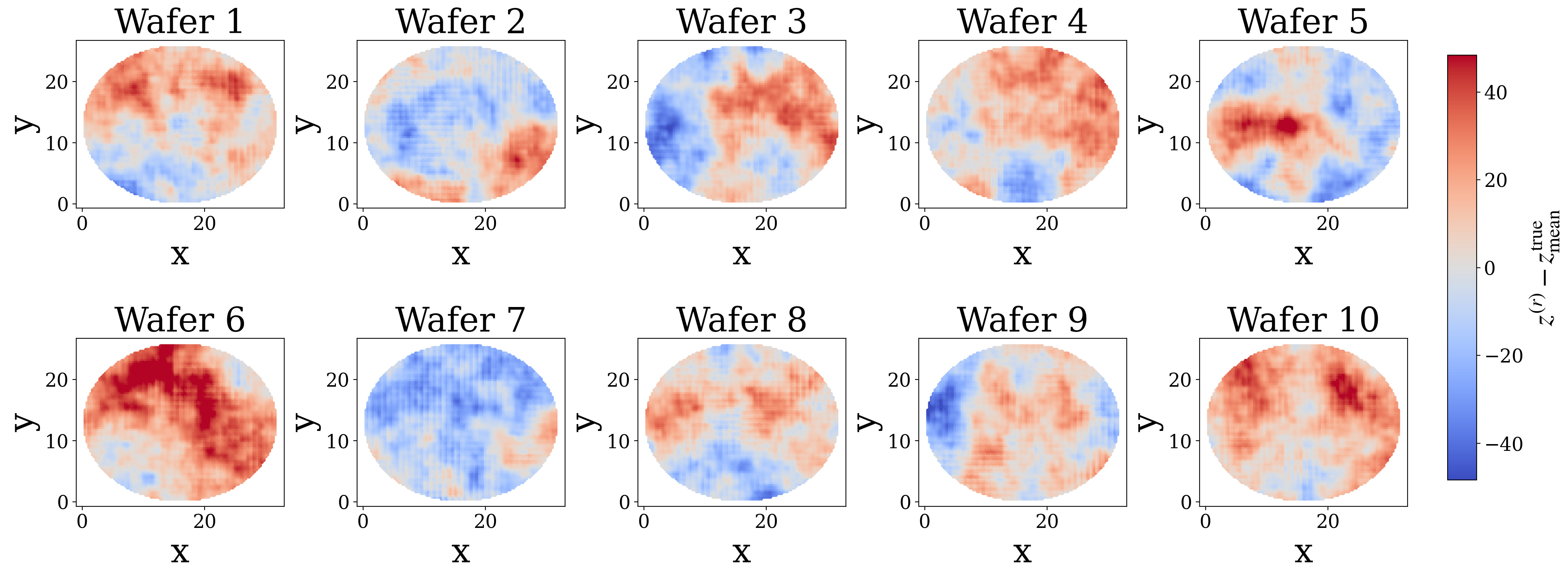}
    \caption{Generated synthetic wafer deviations from the true population mean.}
\label{fig:multi_wafer_deviation_montage}
\end{figure}

\vspace{-6pt}
\textbf{Parameter Estimation.} In Experiment I, we focus on the modeling of wafer warpage using data from a single wafer, while Experiment II investigates the modeling using data from multiple wafers. The key difference lies in the interpretation of the estimated $\hat{\bm{\Sigma}}_{\vec{\bm{Q}}}$. In the former case, $\hat{\bm{\Sigma}}_{\vec{\bm{Q}}}$ is used to check the internal consistency of covariance propagation, while in the latter case, it estimates wafer-to-wafer variation.
Figure~\ref{fig:multi_wafer_montage} shows ten generated wafer observations, and 
Figure~\ref{fig:multi_wafer_deviation_montage} shows the deviation of each wafer observation from the true population mean. 
By removing the dominant mean shape from the raw height maps, this figure clearly visualizes between-wafer warpage variation.

Data from the ten synthetic wafers are used to estimate the population mean vector $\bm{\mu}$ and covariance/noise hyperparameters $\bm{\theta}$ in (\ref{eq:model_final}) using IRWGLS. Table~\ref{tab:parameter_recovery} compares the true (used in data generation) and estimated parameter values. The global scale parameter $\xi_1$ has the largest relative error because it controls the overall wafer-to-wafer variation level and is more sensitive to the finite number of training wafers. The within-chip length scales, between-chip correlation-decay parameter, and measurement-noise standard deviation are recovered with small relative errors.

\vspace{0.3\baselineskip}
\begin{table}[htbp]
\centering
\caption{Comparison between the true and estimated parameters}
\label{tab:parameter_recovery}
\begin{tabular}{llrrrr}
\toprule
Parameter & Symbol & True & Estimate & Abs. error & Rel. error \\
\midrule
Global scale & $\xi_1$ & 20.0000 & 18.5054 & 1.4946 & 7.4730\% \\
Within-chip $u$ length scale & $\xi_2$ & 0.4500 & 0.4420 & 0.0080 & 1.7711\% \\
Within-chip $v$ length scale & $\xi_3$ & 0.6500 & 0.6599 & 0.0099 & 1.5186\% \\
Between-chip correlation decay & $\xi_4$ & 3.0000 & 2.9272 & 0.0728 & 2.4253\% \\
Measurement noise std. & $\sigma_\epsilon$ & 0.0300 & 0.0299 & 0.0001 & 0.4539\% \\
\bottomrule
\end{tabular}
\end{table}

\begin{figure}[htbp]
    \centering
    \begin{subfigure}[t]{0.49\linewidth}
        \centering
        \includegraphics[width=\linewidth]{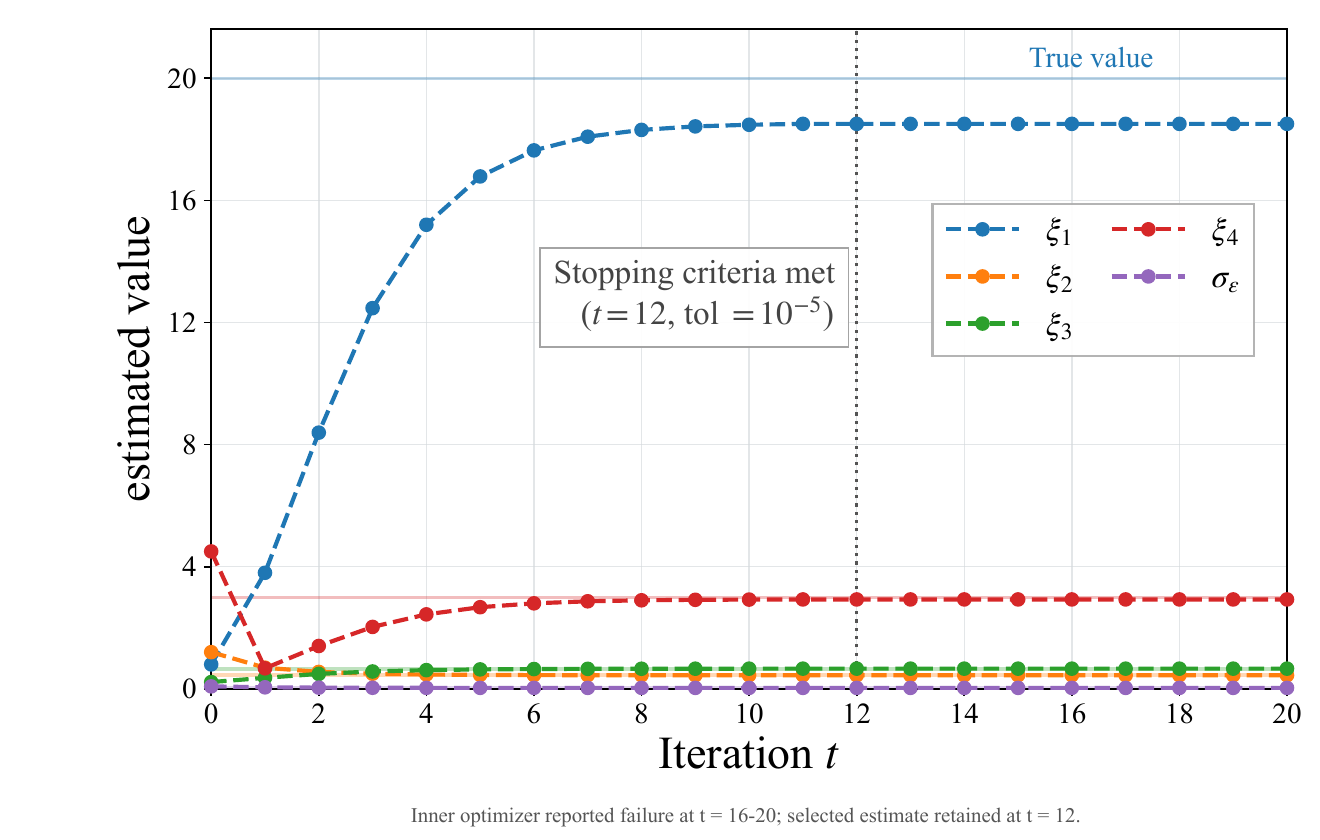}
        \caption{y-axis on the original scale}
    \end{subfigure}
    \hfill
    \begin{subfigure}[t]{0.49\linewidth}
        \centering
        \includegraphics[width=\linewidth]{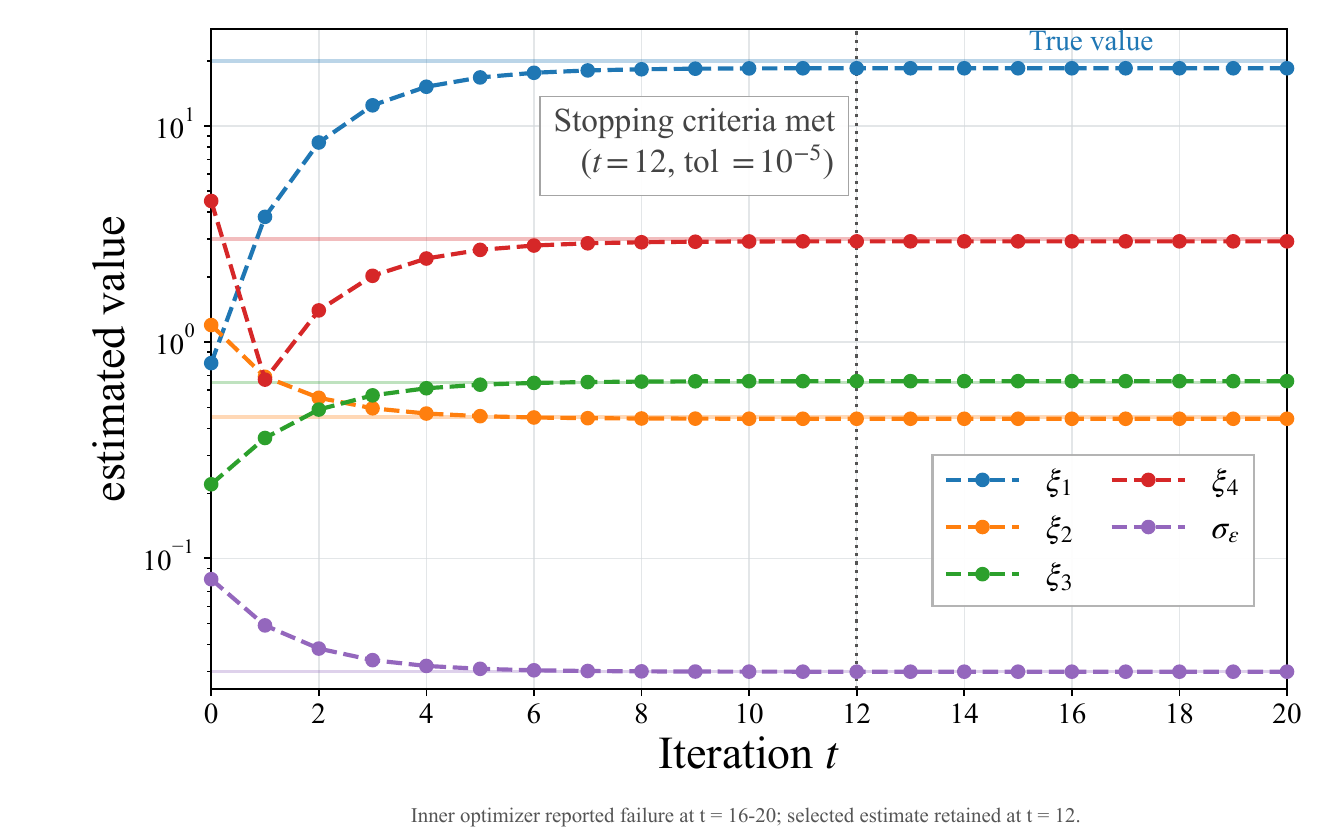}
        \caption{y-axis on the logarithmic scale}
    \end{subfigure}
    \caption{Covariance/noise hyperparameter convergence in the multi-wafer experiment.}
    \label{fig:multi_nll_history}
\end{figure}

\begin{figure}[!ht]
    \centering
    \includegraphics[width=0.50\linewidth]{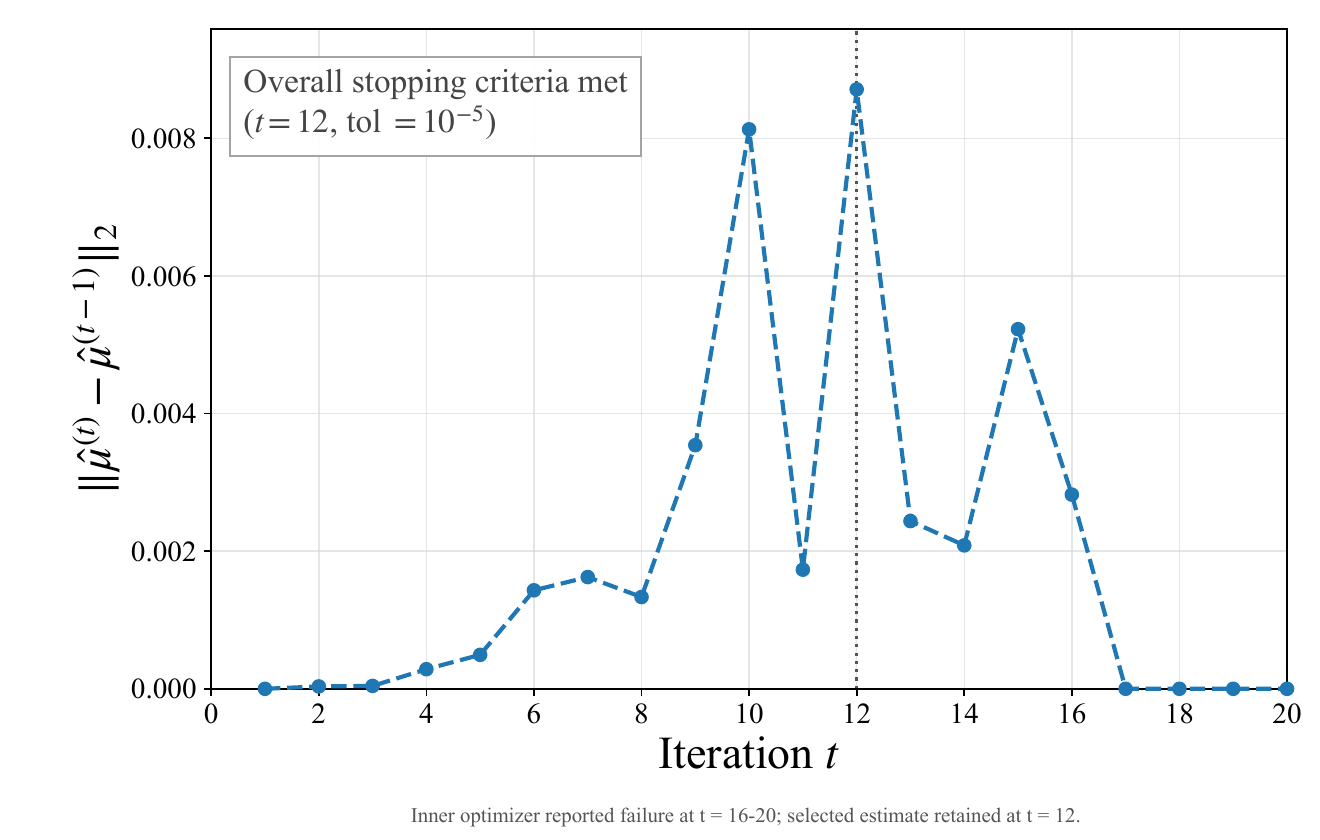}
    \caption{Absolute mean-vector update norm over iterations for the multi-wafer experiment.}
    \label{fig:multi_parameter_history}
\end{figure}

{\color{black} Similar to Experiment I, Figure~\ref{fig:multi_nll_history} shows the trajectories of the covariance/noise hyperparameters over the outer IRWGLS iterations. Figure~\ref{fig:multi_parameter_history} shows the update,  $\|\hat{\bm{\mu}}^{(t)}-\hat{\bm{\mu}}^{(t-1)}\|_2$, of the estimated mean vector over the iterations. The stabilization of these traces indicates the convergence of the IRWGLS procedure.}
{\color{black} The overall stopping conditions are first met at $t=12$, although the mean relative change is already below $10^{-5}$ from $t=2$ onward. The nonmonotonic absolute updates in Figure~\ref{fig:multi_parameter_history} remain small relative to the mean-vector norm. The faint horizontal lines in Figure~\ref{fig:multi_nll_history} indicate the corresponding true parameter values.}

\begin{figure}[!ht]
    \centering
    \includegraphics[width=\linewidth]{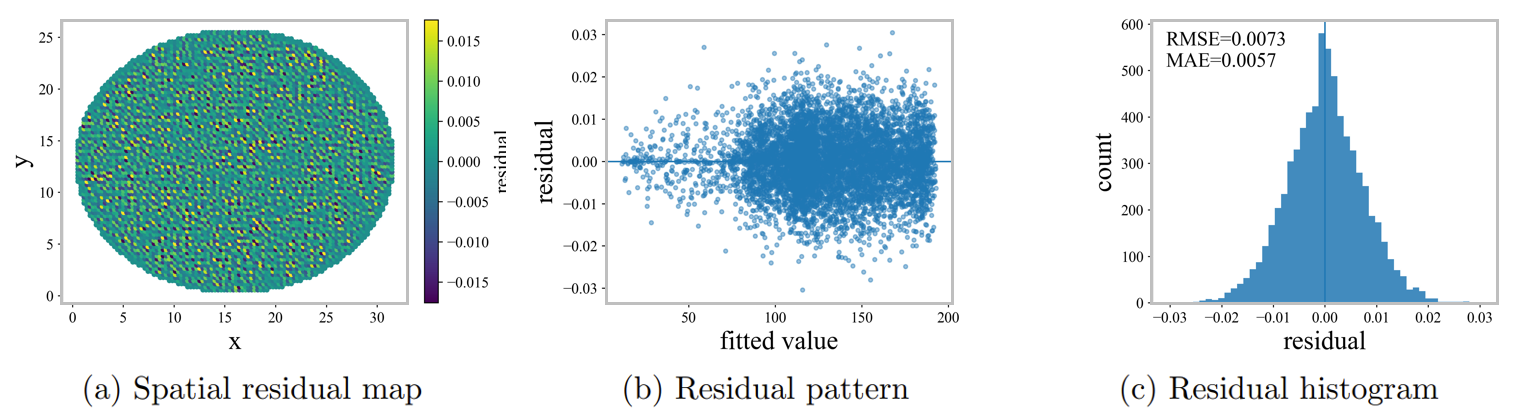}
    \caption{Centered residual diagnostics after IRWGLS estimation for the multi-wafer experiment.}
    \label{fig:multi_runtime_residuals}
\end{figure}

Table~\ref{tab:timing} reports the computational cost (for both Experiments I and II), and Figure~\ref{fig:multi_runtime_residuals} shows the final centered residual diagnostics. The computational time is separated into covariance hyperparameter estimation and GLS mean update within the  IRWGLS iterations. In both experiments, covariance hyperparameter estimation dominates the runtime. The GLS update is comparatively fast because the implementation carries out the main linear solve in the free-control-point space using Woodbury-type algebra together with GPU acceleration. The runtime of the multi-wafer experiment does not increase linearly with the number of wafers because the implementation uses a common-\(\bm{J}\) q-space/Woodbury formulation, where matrix construction and linear algebra associated with the shared design matrix and latent covariance are reused across wafers.

{\color{black} For model updates using ten wafers, the total runtime is approximately 229s (Table~\ref{tab:timing}). Assuming an effective handling and measurement time of 30--600s per wafer for inline warpage metrology, acquiring ten wafers takes 300--6,000s. Under these operating assumptions and the computational conditions of this experiment, model updates can be completed concurrently within the acquisition time of the next ten wafers.}

\vspace{0.3\baselineskip}
\begin{table}[!ht]
\centering
\caption{Computational time decomposition.}
\label{tab:timing}
\begin{tabular}{lrr}
\toprule
Runtime component & Single wafer (s) & Multi wafer (s) \\
\midrule
System/data generation & 2.5407 & 17.7220 \\
Covariance-parameter MLE & 22.2539 & 183.6984 \\
Mean update & 11.7846 & 22.2730 \\
Other runtime & 4.4032 & 5.3669 \\
Total runtime & 40.9824 & 229.0603 \\
\bottomrule
\end{tabular}
\end{table}

\FloatBarrier

\begin{figure}[!h]
    \centering
    \begin{subfigure}{0.32\linewidth}
        \centering
        \includegraphics[width=\linewidth]{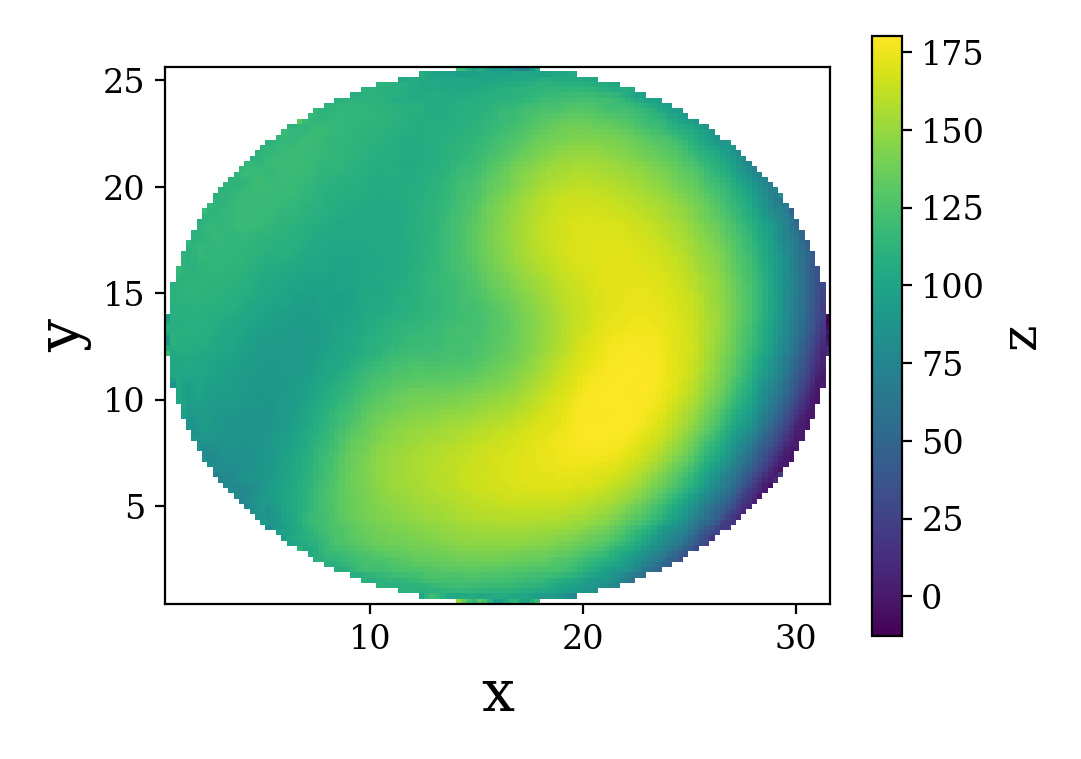}
        \caption{True mean surface}
    \end{subfigure}
    \hfill
    \begin{subfigure}{0.32\linewidth}
        \centering
        \includegraphics[width=\linewidth]{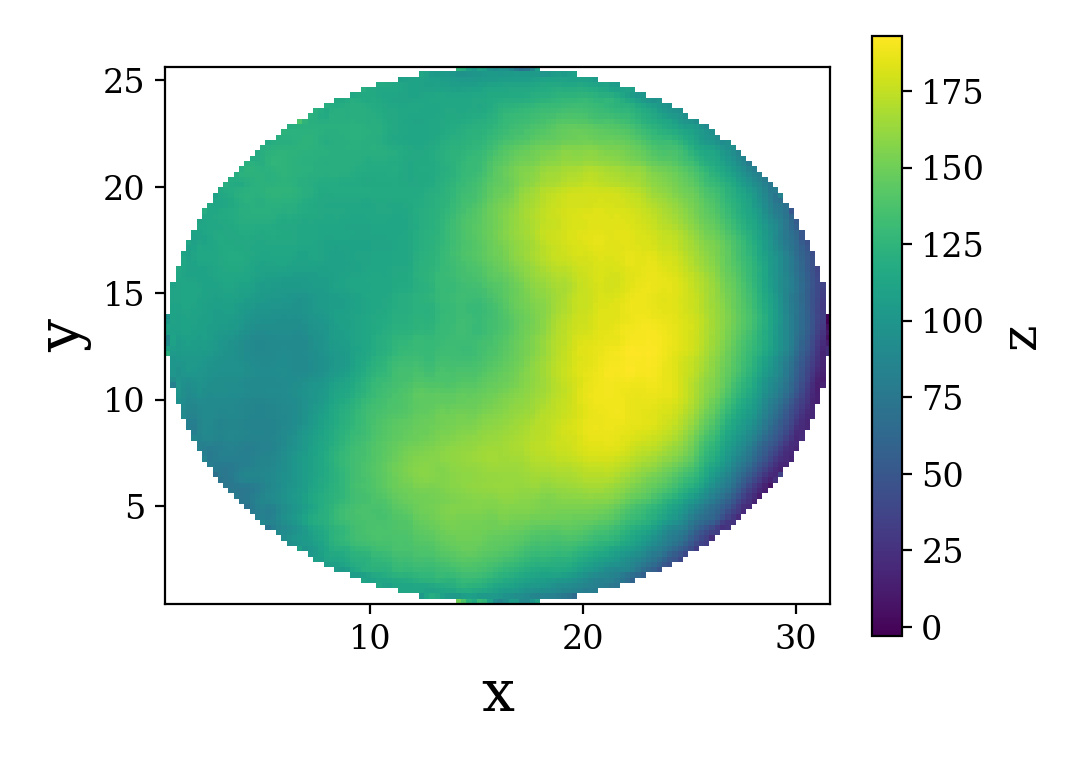}
        \caption{Estimated mean surface}
    \end{subfigure}
    \hfill
    \begin{subfigure}{0.32\linewidth}
        \centering
        \includegraphics[width=\linewidth]{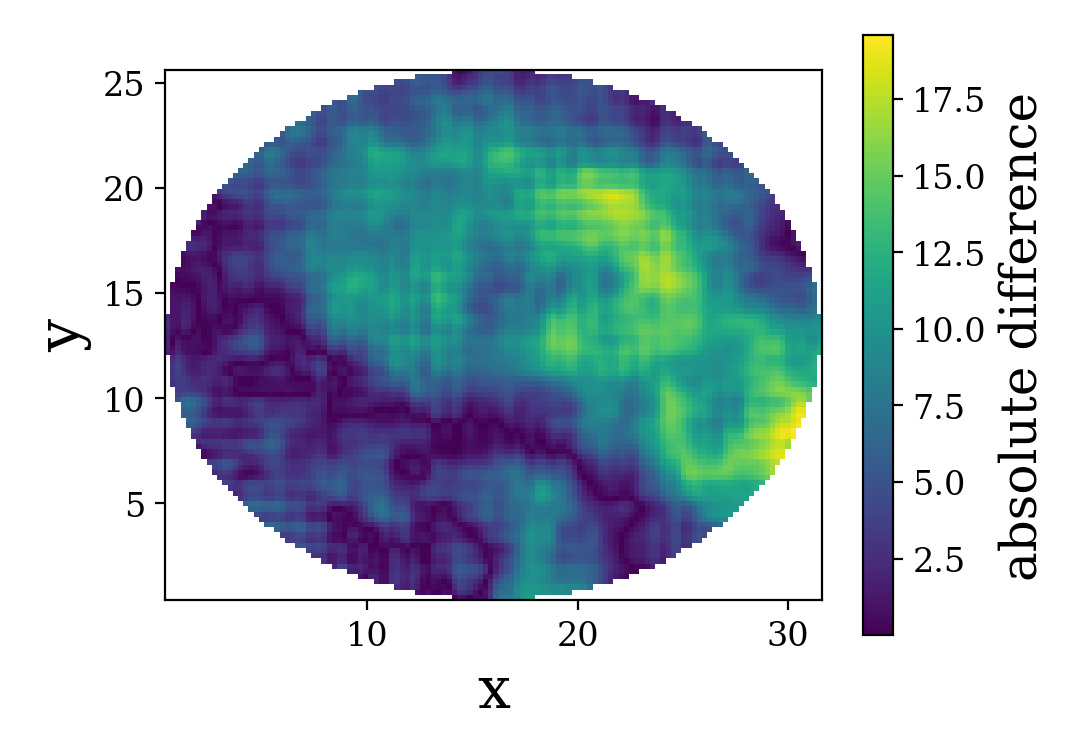}
        \caption{Absolute difference}
    \end{subfigure}
    \caption{True and estimated population mean surfaces.}
    \label{fig:multi_mean_recovery}
\end{figure}

\textbf{Main Results.} Based on the estimated model, Figure~\ref{fig:multi_mean_recovery} compares the true and estimated mean surfaces. The estimated mean surface closely follows the true image-derived population mean. The mean-surface RMSE is 8.1061 and the MAE is 6.8265.
{\color{black} The population-mean surface estimated from ten wafers has a peak-to-valley (P--V) value of 195.8685, compared with the true value of 192.8679 (relative error: 1.5558\%), indicating close agreement in overall warpage amplitude in this simulation. For context, a previous study reported simulated peak warpage within 5\% of experimental measurements~\cite{ref40}, although differences in evaluation targets and metrics preclude a direct performance comparison. In a separate preliminary study evaluated after 50 iterations, increasing the wafer sample size from ten to 100 reduced the mean-surface RMSE from approximately 8.1 to 2.2, suggesting potential gains from accumulating measurements under stable process conditions.}

\begin{figure}[!ht]
    \centering
    \includegraphics[width=\linewidth]{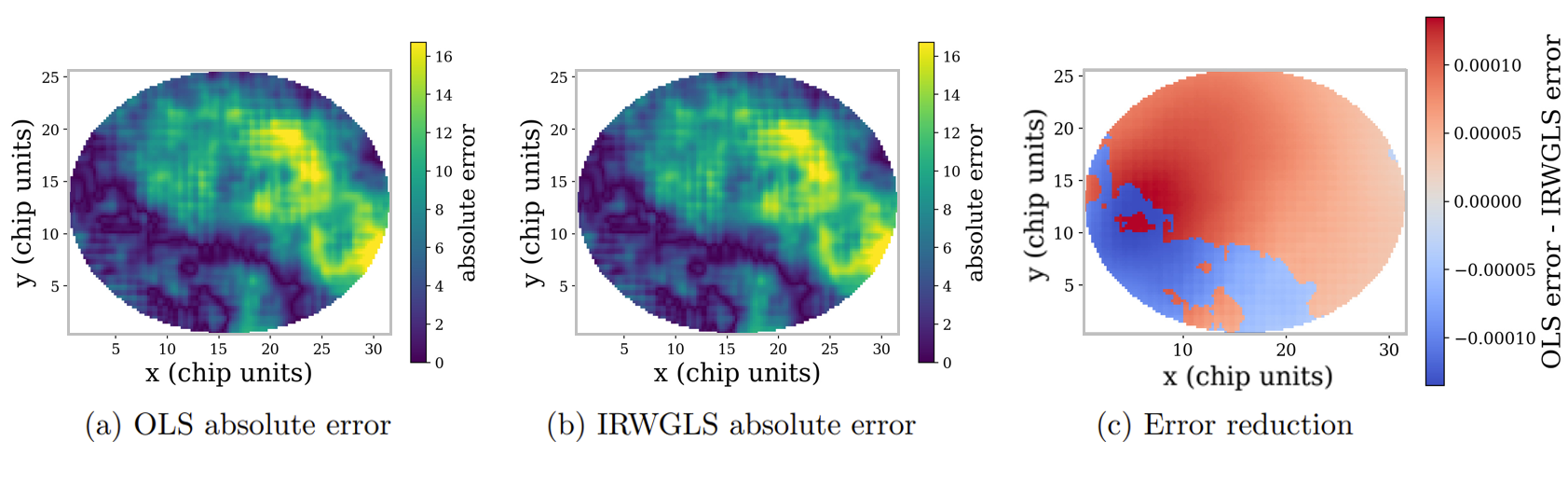}
    \caption{Absolute error of the estimated mean respectively using the OLS and IRWGLS.}
    \label{fig:multi_ols_irwgls_error_maps}
\end{figure}

Figure~\ref{fig:multi_ols_irwgls_error_maps} and Table~\ref{tab:multi_ols_irwgls_metrics} compare the estimated mean warpage using the OLS and IRWGLS, respectively. {\color{black} The OLS and IRWGLS estimates yield nearly identical RMSE and MAE with respect to the true population-mean surface.}

\vspace{0.3\baselineskip}
\begin{table}[!ht]
\centering
\caption{Comparison of the absolute error of the estimated mean between OLS and IRWGLS}
\label{tab:multi_ols_irwgls_metrics}
\small
\begin{tabular}{lccccc}
\toprule
Estimator & \makecell{True mean\\RMSE} & \makecell{True mean\\MAE} & \makecell{True mean\\bias} & \makecell{Raw residual\\RMSE} & \makecell{Raw residual\\MAE} \\
\midrule
OLS & 8.1062 & 6.8265 & 5.2034 & 0.0073 & 0.0057 \\
IRWGLS & 8.1061 & 6.8265 & 5.2033 & 0.0073 & 0.0057 \\
\bottomrule
\end{tabular}
\end{table}

\begin{figure}[htbp]
    \centering
    \begin{subfigure}{0.32\linewidth}
        \centering
        \includegraphics[width=\linewidth]{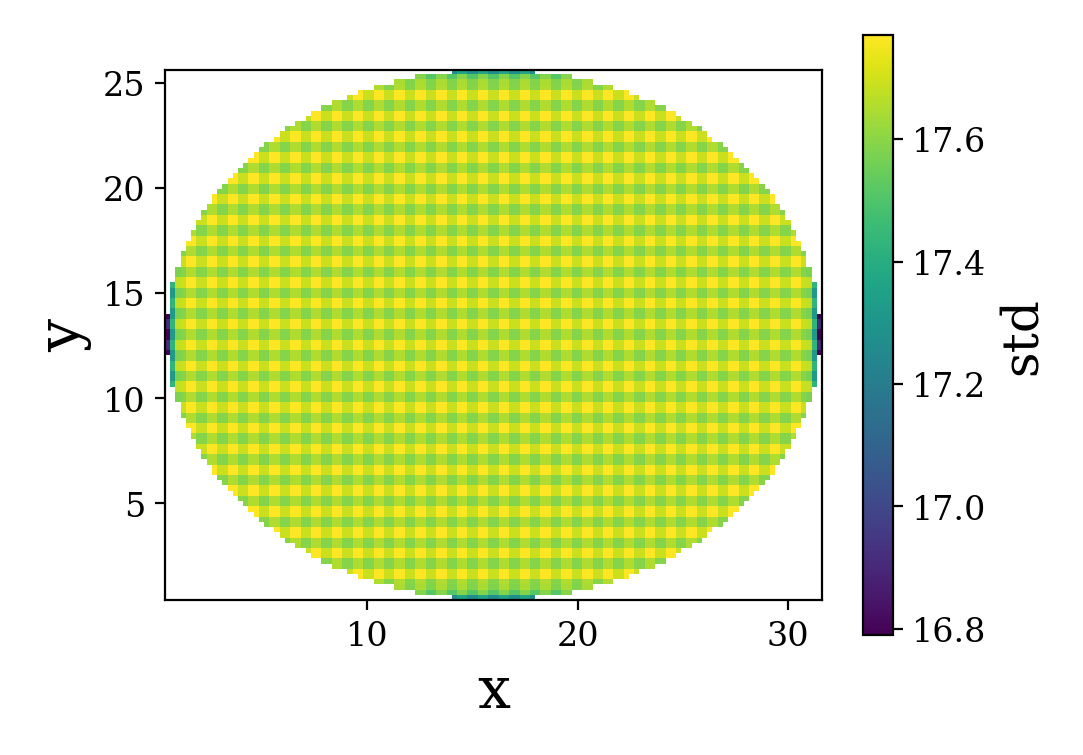}
        \caption{True standard deviation}
    \end{subfigure}
    \hfill
    \begin{subfigure}{0.32\linewidth}
        \centering
        \includegraphics[width=\linewidth]{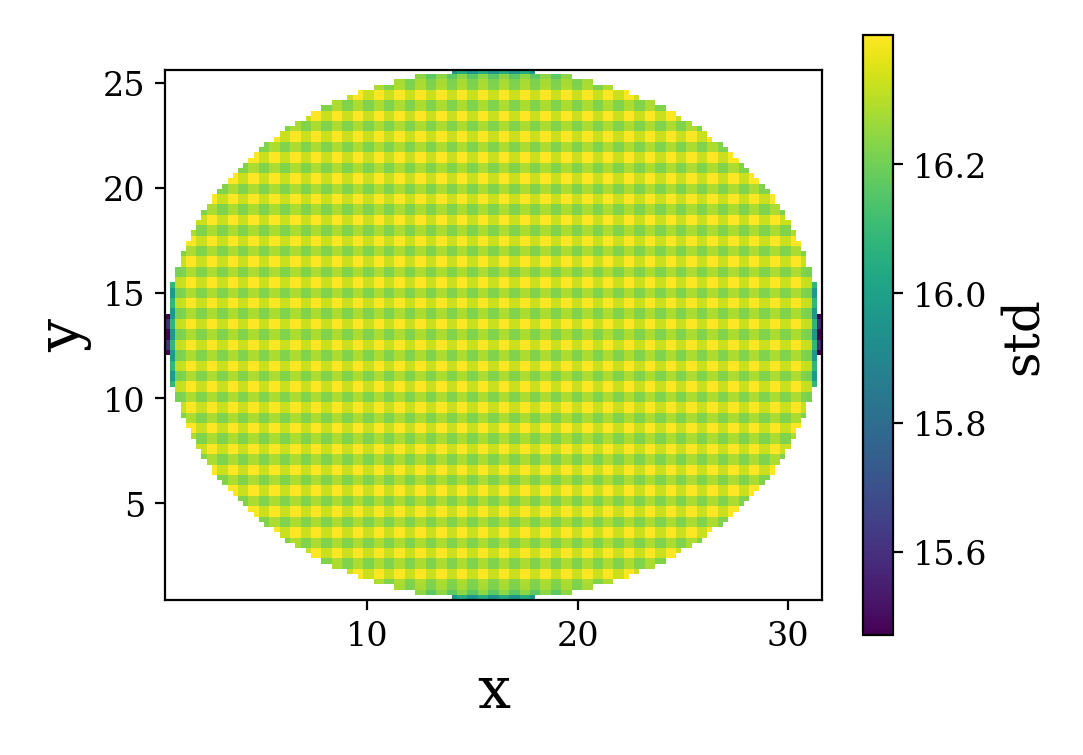}
        \caption{Estimated standard deviation}
    \end{subfigure}
    \hfill
    \begin{subfigure}{0.32\linewidth}
        \centering
        \includegraphics[width=\linewidth]{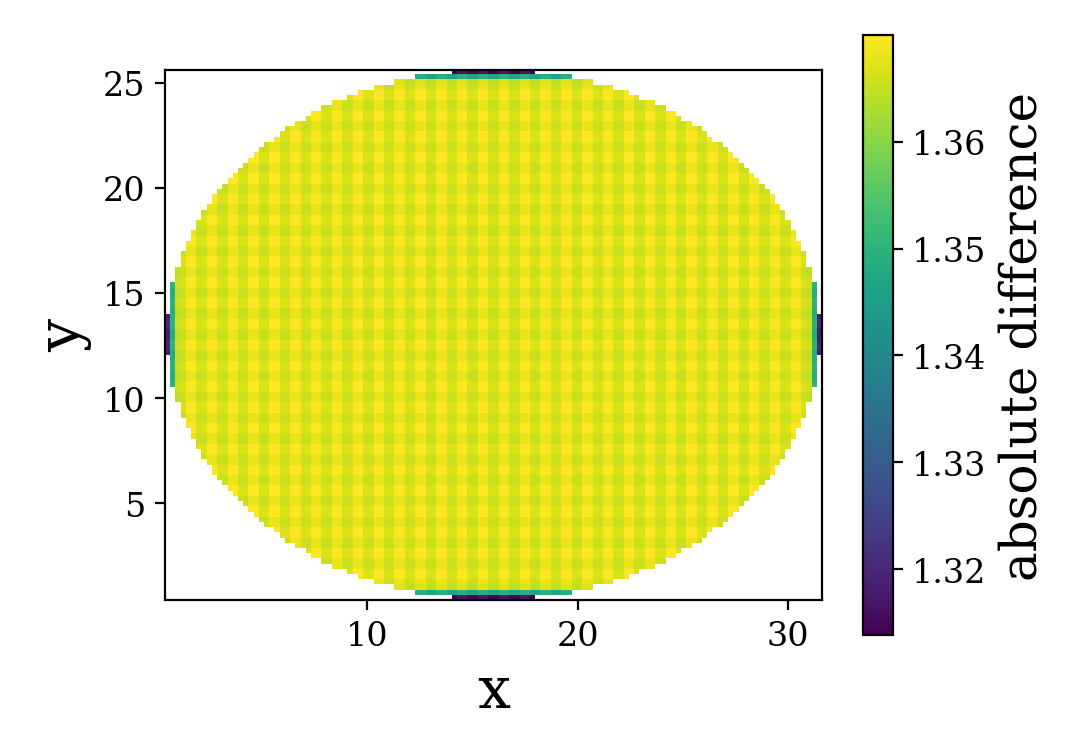}
        \caption{Absolute difference}
    \end{subfigure}
    \caption{True and estimated pointwise standard-deviation maps.}
    \label{fig:multi_std_recovery}
\end{figure}
\begin{figure}[!ht]
    \centering
    \begin{subfigure}{0.48\linewidth}
        \centering
        \includegraphics[width=\linewidth]{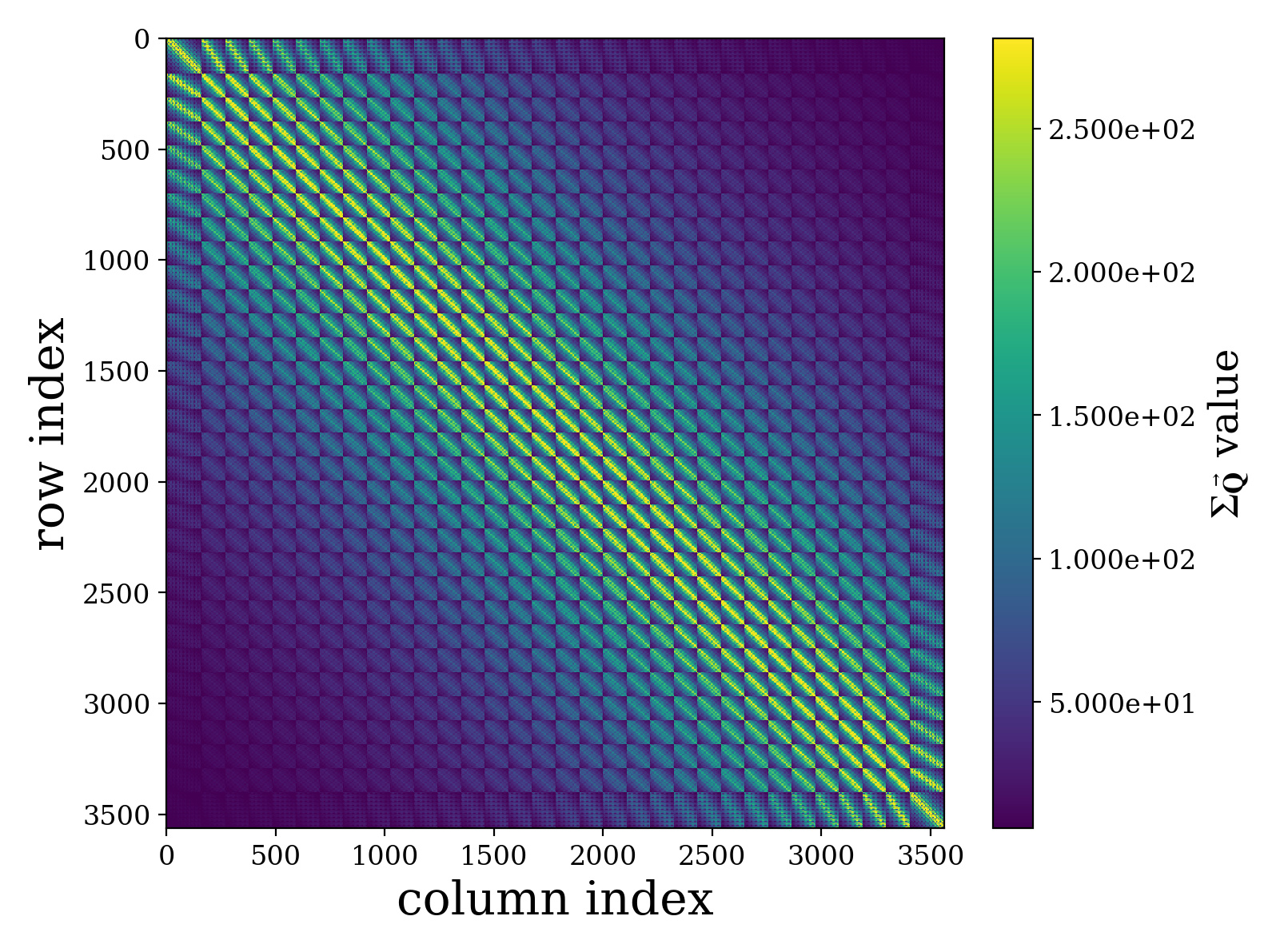}
        \caption{True $\bm{\Sigma}_{\vec{\bm{Q}}}$}
    \end{subfigure}
    \hfill
    \begin{subfigure}{0.48\linewidth}
        \centering
        \includegraphics[width=\linewidth]{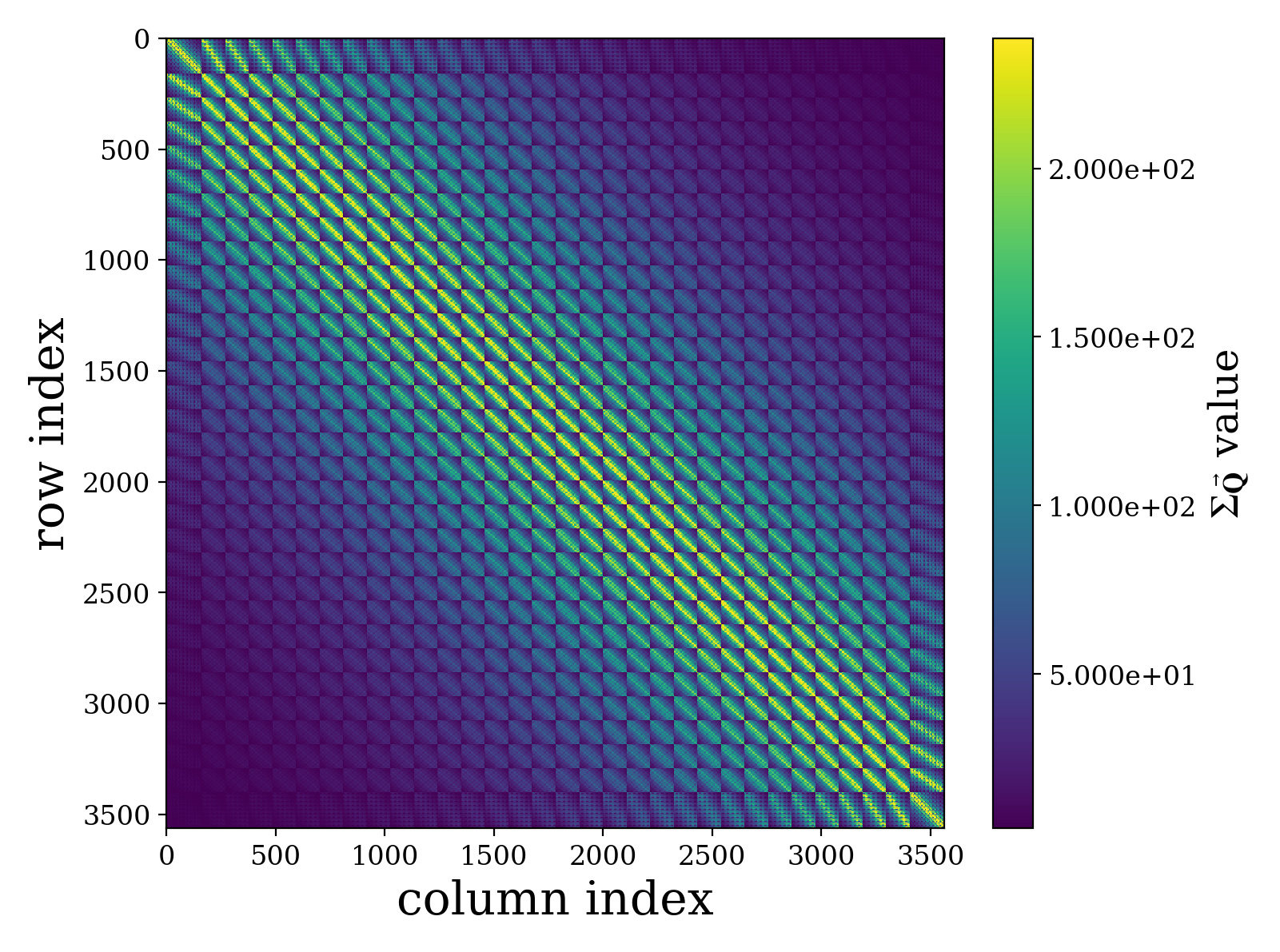}
        \caption{Estimated $\bm{\Sigma}_{\vec{\bm{Q}}}$}
    \end{subfigure}
    \caption{True and estimated covariance matrices in the free-control-point space.}
    \label{fig:multi_sigma_q}
\end{figure}
\begin{figure}[!ht]
    \centering
    \begin{subfigure}{0.48\linewidth}
        \centering
        \includegraphics[width=\linewidth]{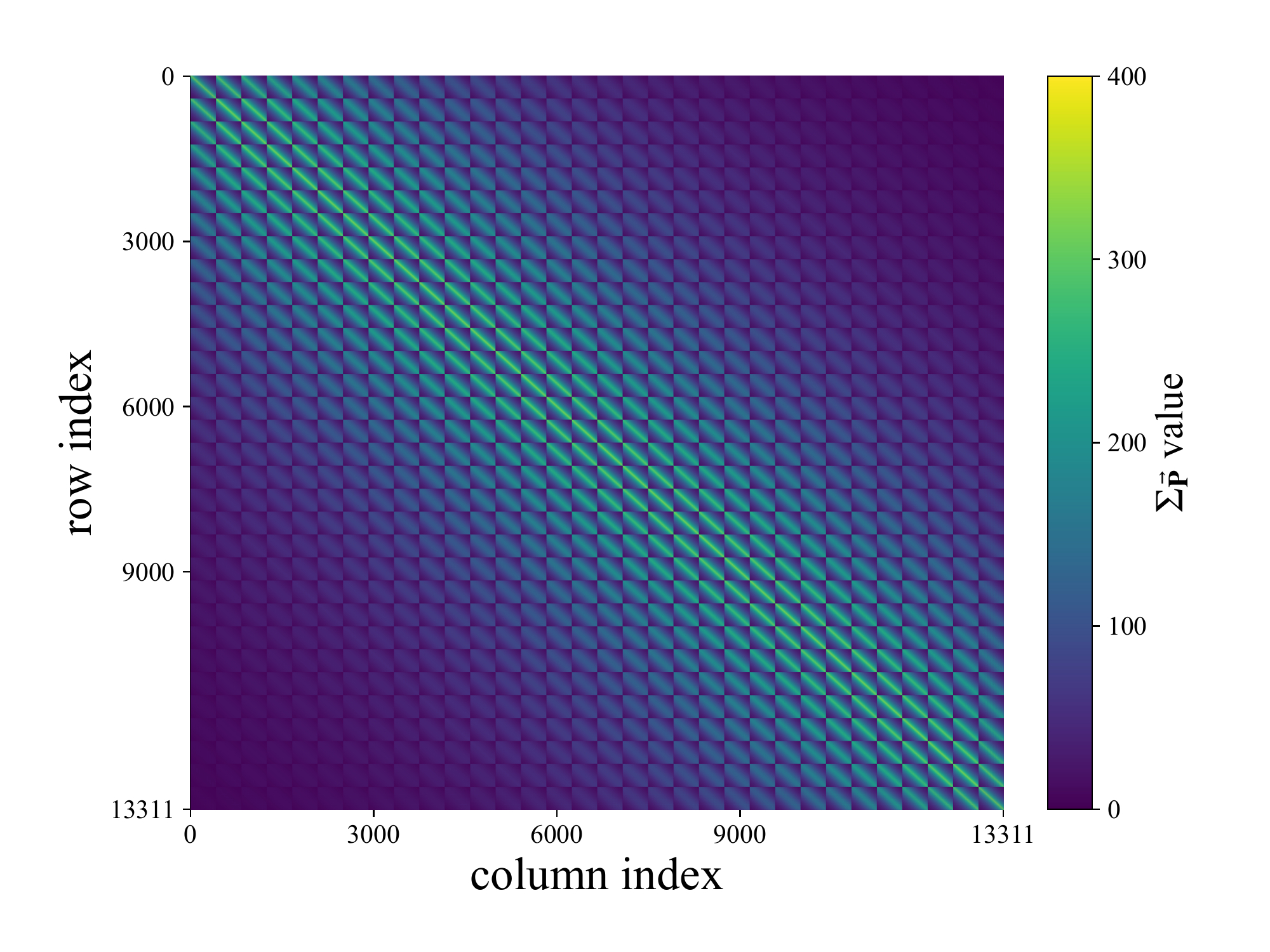}
        \caption{True $\bm{\Sigma}_{\vec{\bm{P}}}$}
    \end{subfigure}
    \hfill
    \begin{subfigure}{0.48\linewidth}
        \centering
        \includegraphics[width=\linewidth]{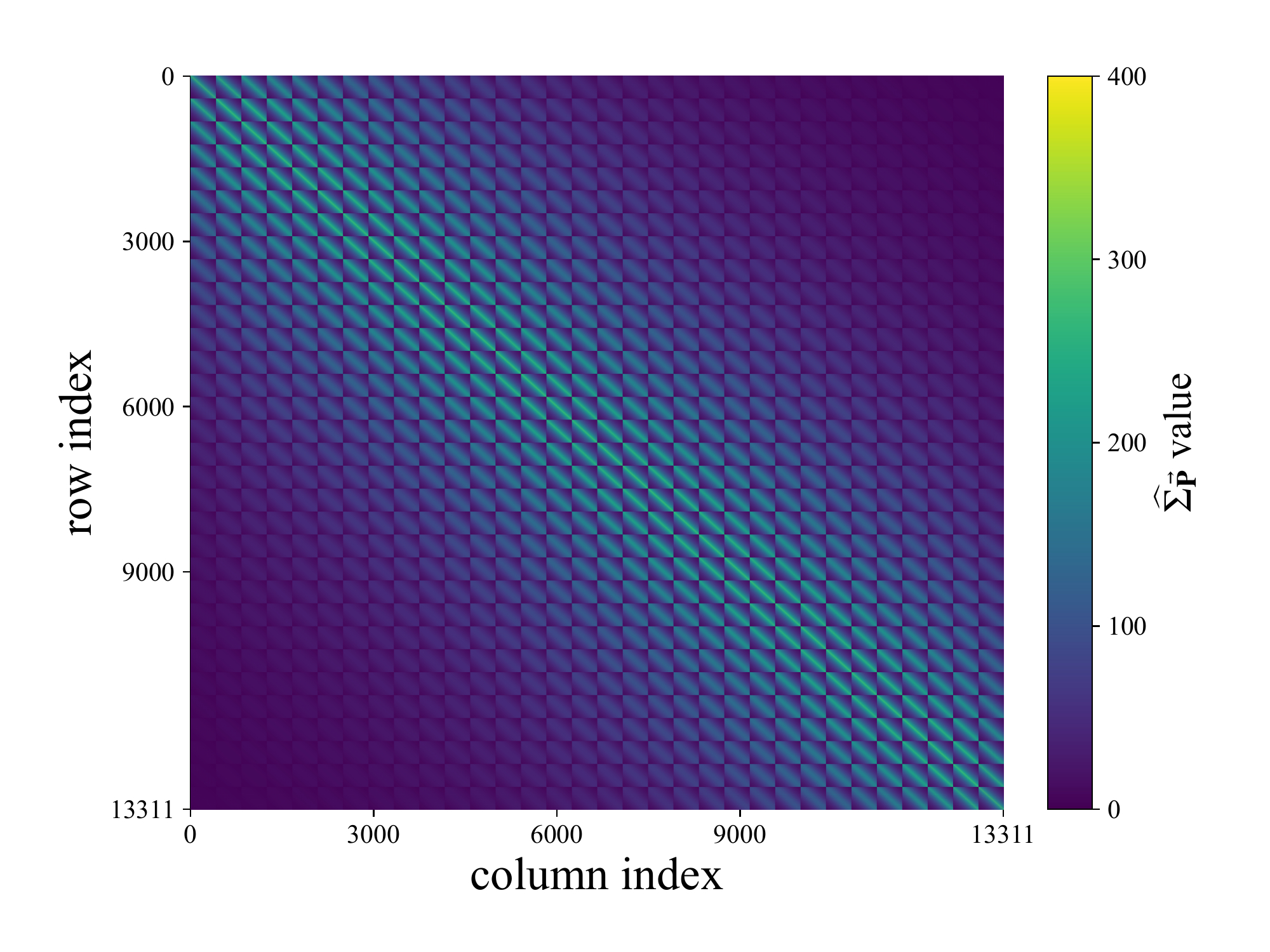}
        \caption{Estimated $\bm{\Sigma}_{\vec{\bm{P}}}$}
    \end{subfigure}
    \caption{True and estimated full-control-point covariance matrices}
    \label{fig:multi_sigma_p}
\end{figure}

Because the true covariance is known in this experiment when generating the synthetic data, the estimated uncertainty structure can be evaluated 
by comparing the true and estimated uncertainty maps; see Figure~\ref{fig:multi_std_recovery}. As shown in Table \ref{tab:validation_metrics}, the standard-deviation-map RMSE is 1.364908 and the MAE is 1.364901. Figure~\ref{fig:multi_sigma_q} and Figure~\ref{fig:multi_sigma_p} compare, respectively, the true and estimated covariance matrices $\bm{\Sigma}_{\vec{\bm{Q}}}$ and $\bm{\Sigma}_{\vec{\bm{P}}}$. It is seen that we are able to accurately estimate the between- and within-wafer covariance.

\begin{figure}[htbp]
    \centering
    \includegraphics[width=0.60\linewidth]{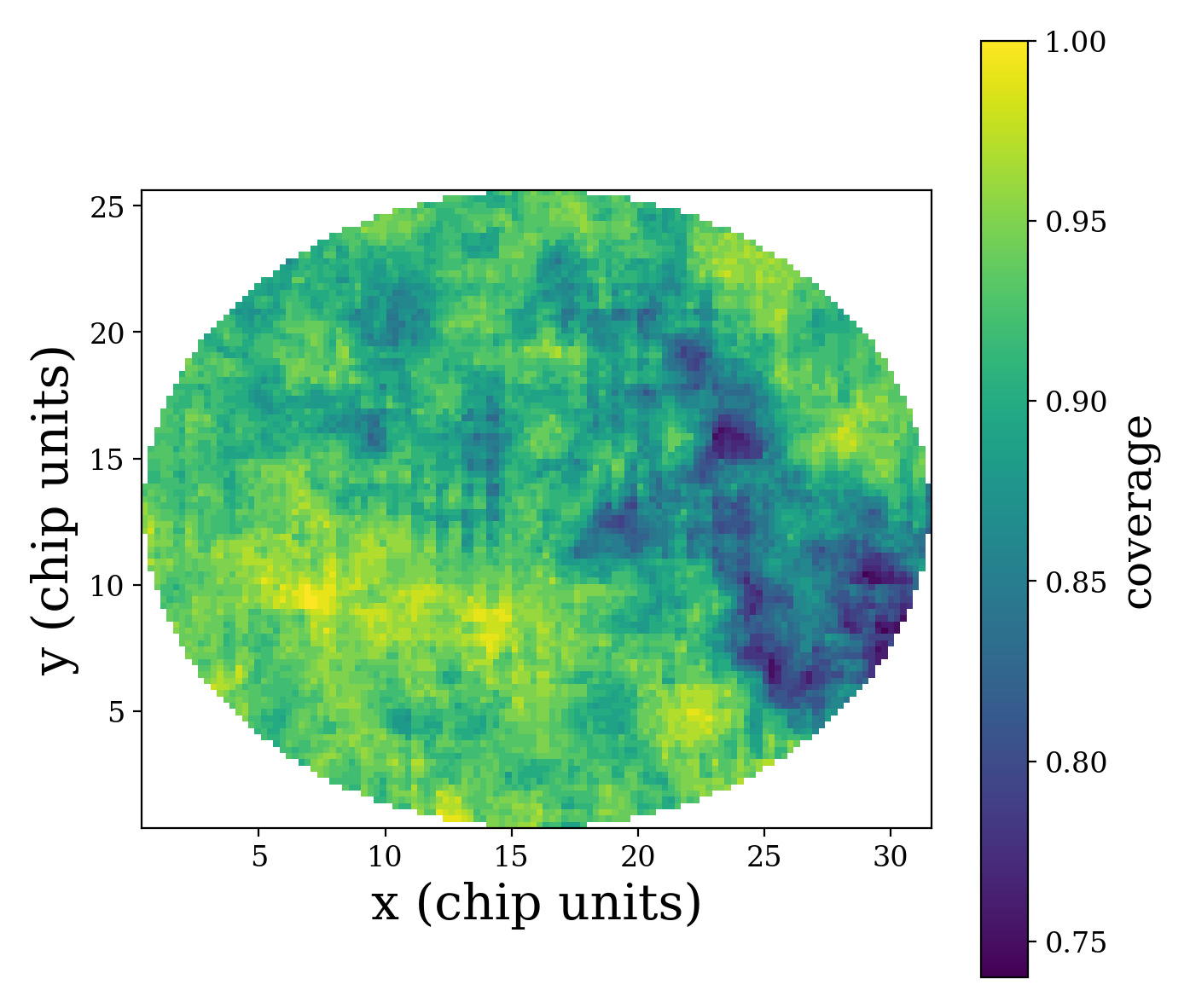}
    \caption{Held-out pointwise coverage map of the estimated 95\% uncertainty band.}
    \label{fig:multi_coverage}
\end{figure}

\begin{figure}[!ht]
    \centering
    \begin{subfigure}[t]{0.49\linewidth}
        \centering
        \includegraphics[width=\linewidth,height=0.33\textheight,keepaspectratio]{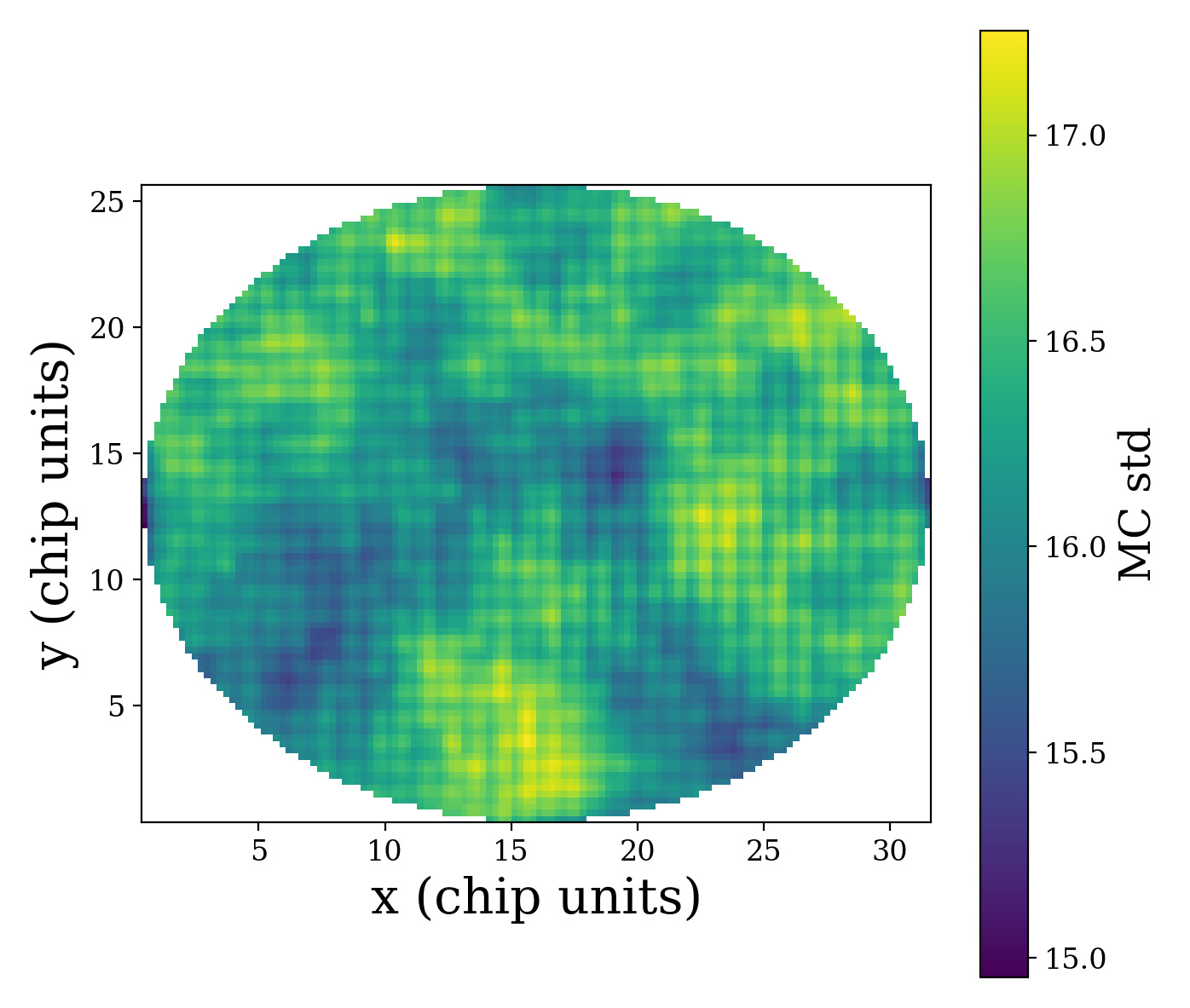}
        \caption{MC standard deviation}
    \end{subfigure}
    \hfill
    \begin{subfigure}[t]{0.49\linewidth}
        \centering
        \includegraphics[width=\linewidth,height=0.33\textheight,keepaspectratio]{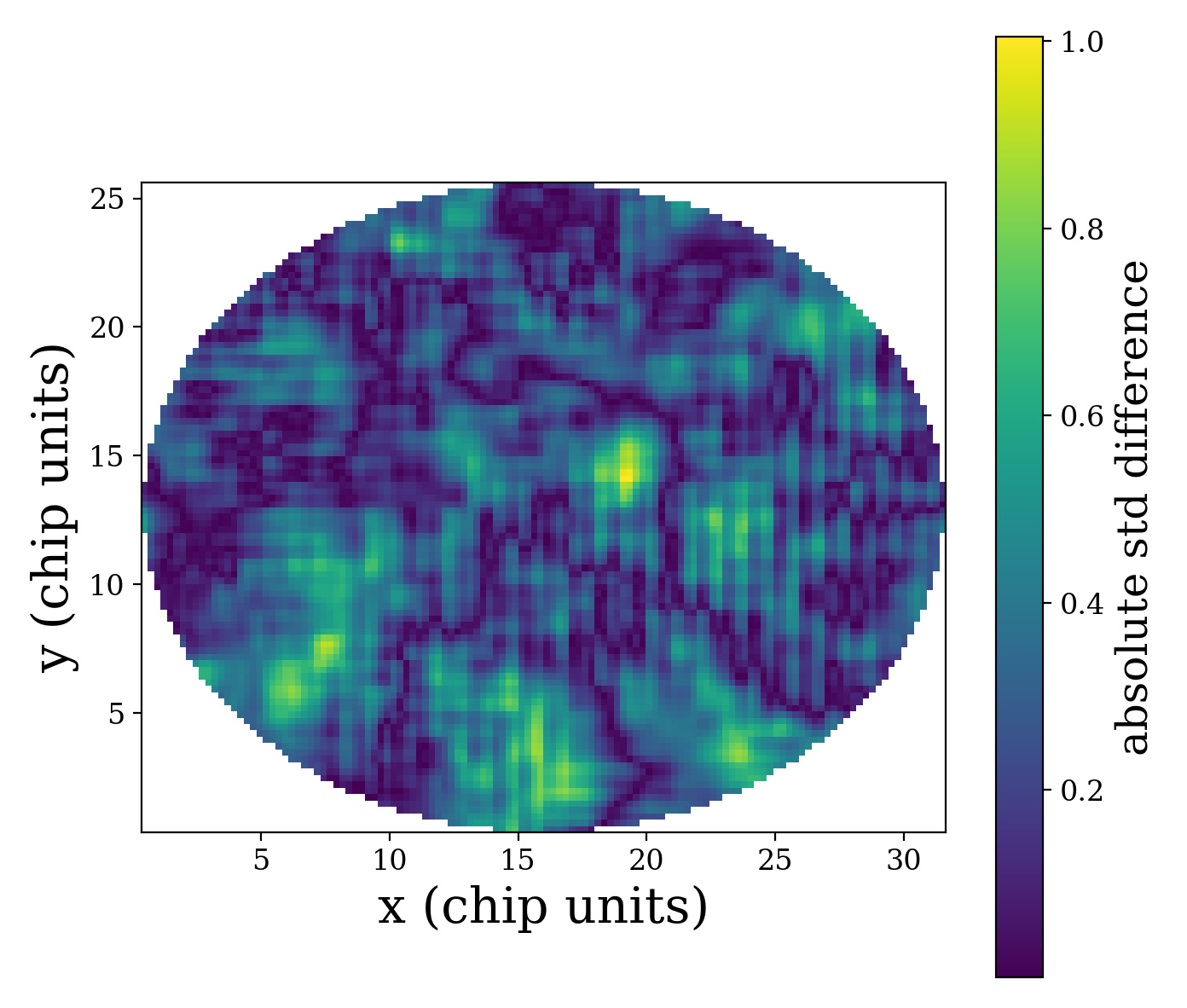}
        \caption{Absolute difference from analytic SD}
    \end{subfigure}
    \caption{Monte Carlo pointwise standard-deviation check for the multi-wafer covariance model.}
    \label{fig:multi_mc_std_check}
\end{figure}

\textbf{Out-of-Sample Evaluation and Uncertainty Quantification.} 
Held-out calibration is evaluated using 100 additional synthetic wafers generated from the same true model but not used during estimation. Figure~\ref{fig:multi_coverage} shows the pointwise coverage map used to evaluate calibration of the estimated 95\% uncertainty band with held-out synthetic wafers. Figure~\ref{fig:multi_mc_std_check} compares the pointwise standard deviation computed from Monte Carlo sampling with the analytic standard deviation obtained by covariance propagation. This comparison is a diagnostic for checking the internal consistency of uncertainty propagation and is not an additional population-recovery metric. The empirical pointwise coverage of the estimated 95\% band is 0.9082. This value is below the nominal level, which is consistent with the underestimation of the global scale parameter \(\xi_1\).

\vspace{0.3\baselineskip}
\begin{table}[!ht]
\centering
\caption{Multi-wafer validation metrics.}
\label{tab:validation_metrics}
\begin{tabular}{lr}
\toprule
Metric & Value \\
\midrule
Mean surface RMSE & 8.1061 \\
Mean surface MAE & 6.8265 \\
Std. map RMSE & 1.364908 \\
Std. map MAE & 1.364901 \\
Held-out 95\% band coverage & 0.9082 \\
\bottomrule
\end{tabular}
\end{table}

In summary, the two experiments demonstrate the performance of the proposed model. The single-wafer study shows that the chips-on-wafer B\'ezier representation can fit an image-derived wafer-shape point cloud, restrict the fitted surface to the real-chip domain, and provide chip-level surface representations and model-implied uncertainty diagnostics. The multi-wafer synthetic study is the main statistical validation of this paper and shows that, under a known data-generating process, the proposed Gaussian random-effect model can recover the population mean shape, covariance/noise hyperparameters, latent covariance structure, and pointwise uncertainty level. {\color{black} The OLS and covariance-weighted IRWGLS estimates yield nearly identical RMSE and MAE with respect to the true population-mean surface.} The held-out coverage result further evaluates calibration of the estimated uncertainty band, and the runtime diagnostics show that covariance hyperparameter estimation is the dominant computational cost while the free-control-point-space implementation makes the large chips-on-wafer problem computationally feasible.

\vspace{-6pt}
\section{Conclusions}
\label{sec:conclusion} 
\vspace{-6pt}
This paper proposed a B\'ezier-surface-based GP for chips-on-wafer warpage modeling. 
Systematic frameworks that explicitly represent chip-level geometry within wafer-level manufacturing flows as precise surface models have not been sufficiently investigated. This work addressed the gap by linking wafer-level metrology data to chip-level surface representation. Hence, it can serve as a starting point for analyzing chip warpage in mass-production environments, where this issue is becoming increasingly important in the semiconductor industry. For the B\'ezier-GP model, the geometric component represents each chip as a tensor-product B\'ezier patch and uses sparse operators to impose chip-to-chip continuity across the wafer map. This makes it possible to construct both a continuous wafer-level surface and chip-level patch representations over the chips-on-wafer domain defined from an ASC wafer map. The statistical component, i.e., the use of GP, models the free control points as a spatially correlated Gaussian random vector, enabling uncertainty propagation from latent chip-level shape parameters to wafer-level height and uncertainty maps. Experiments using an ASC wafer map and image-derived wafer-shape data evaluated both the geometric modeling capability and the statistical recovery capability of the proposed framework. 

Future work can extend the framework in several directions: (i) material-property-informed regularization can be introduced to suppress spike artifacts that may arise in polynomial-based fitting and to improve the physical plausibility of reconstructed surfaces, and (ii) the framework can be extended to an optimal sampling problem that replaces dense measurement recipes with uncertainty-aware sampling strategies. This extension can reduce the number of measurement points while maintaining warpage reconstruction accuracy, thereby supporting improved metrology throughput and practical recipe optimization.


\bibliography{references}

\appendix

\section*{Appendix \\ Point Cloud Data Measurement and Processing}
\begin{figure}[!ht]
    \centering
    \begin{subfigure}[t]{0.35\linewidth}
        \centering
        \includegraphics[width=0.8\linewidth,height=0.23\textheight,keepaspectratio]{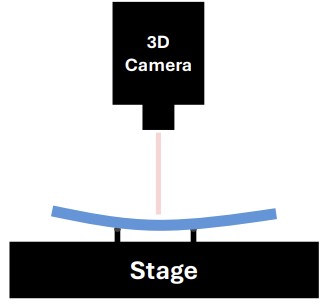}
        \caption{Measurement stage}
    \end{subfigure}
    \hfill
    \begin{subfigure}[t]{0.58\linewidth}
        \centering
        \includegraphics[width=0.8\linewidth,height=0.23\textheight,keepaspectratio]{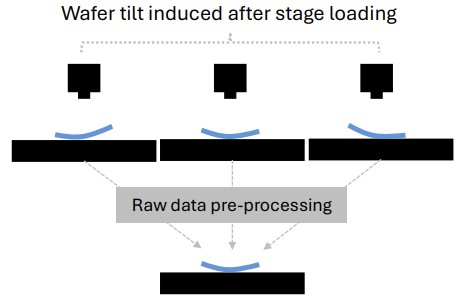}
        \caption{Stage-loading-induced wafer tilt}
    \end{subfigure}
    
    \vspace{6pt}
    
    \caption{Wafer-warpage measurement stage and stage-loading-induced wafer tilt.}
    \label{fig:measurement_process}
\end{figure}

\begin{figure}[!ht]
    \centering

    \begin{subfigure}[t]{0.33\linewidth}
        \centering
        \begin{minipage}[c][0.25\textheight][c]{\linewidth}
            \centering
            \includegraphics[width=\linewidth,height=0.25\textheight]
            {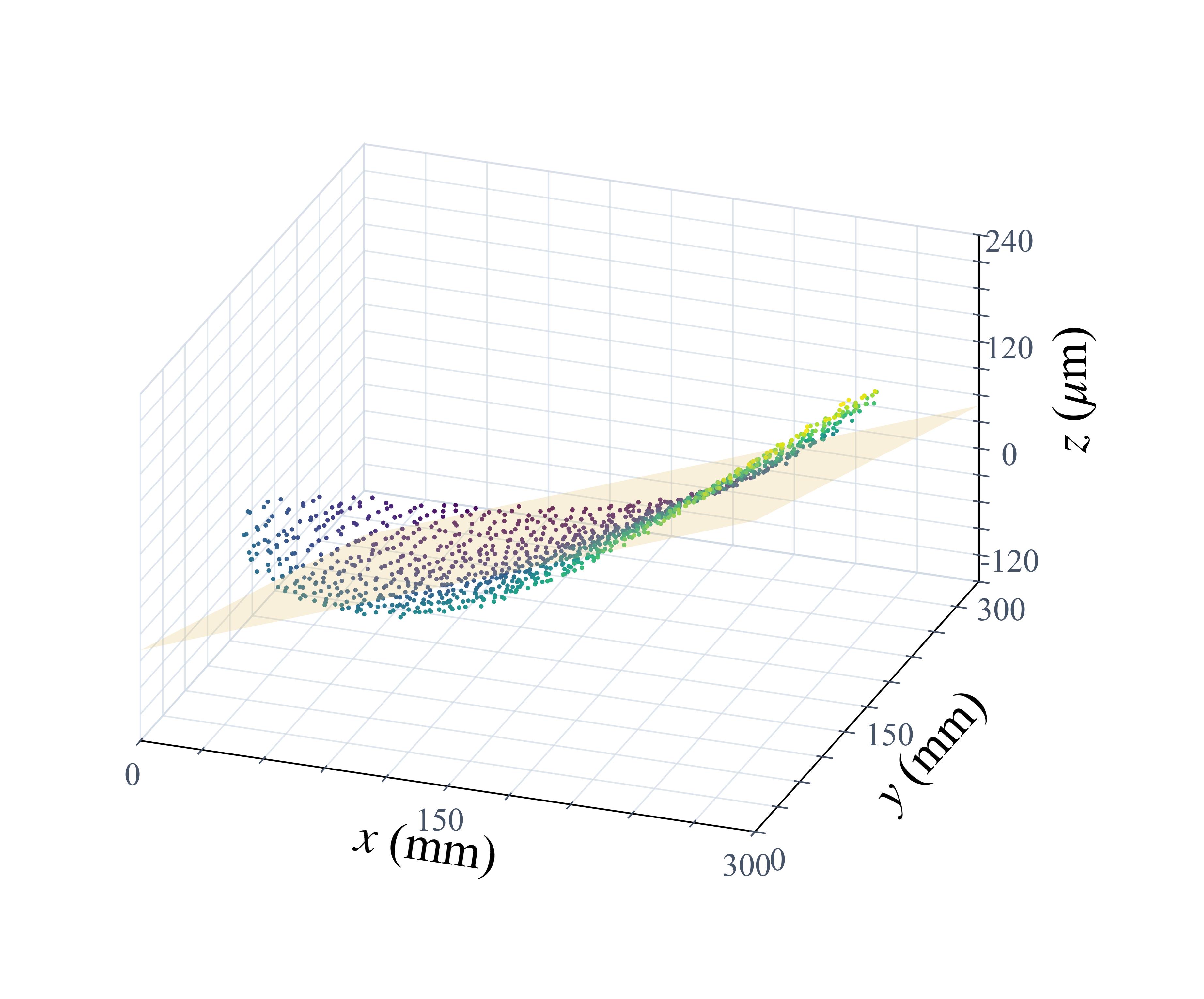}
        \end{minipage}
        \caption{Raw 3D point-cloud data}
    \end{subfigure}
    \hfill
    \begin{subfigure}[t]{0.33\linewidth}
        \centering
        \begin{minipage}[c][0.25\textheight][c]{\linewidth}
            \centering
            \includegraphics[width=\linewidth,height=0.25\textheight]
            {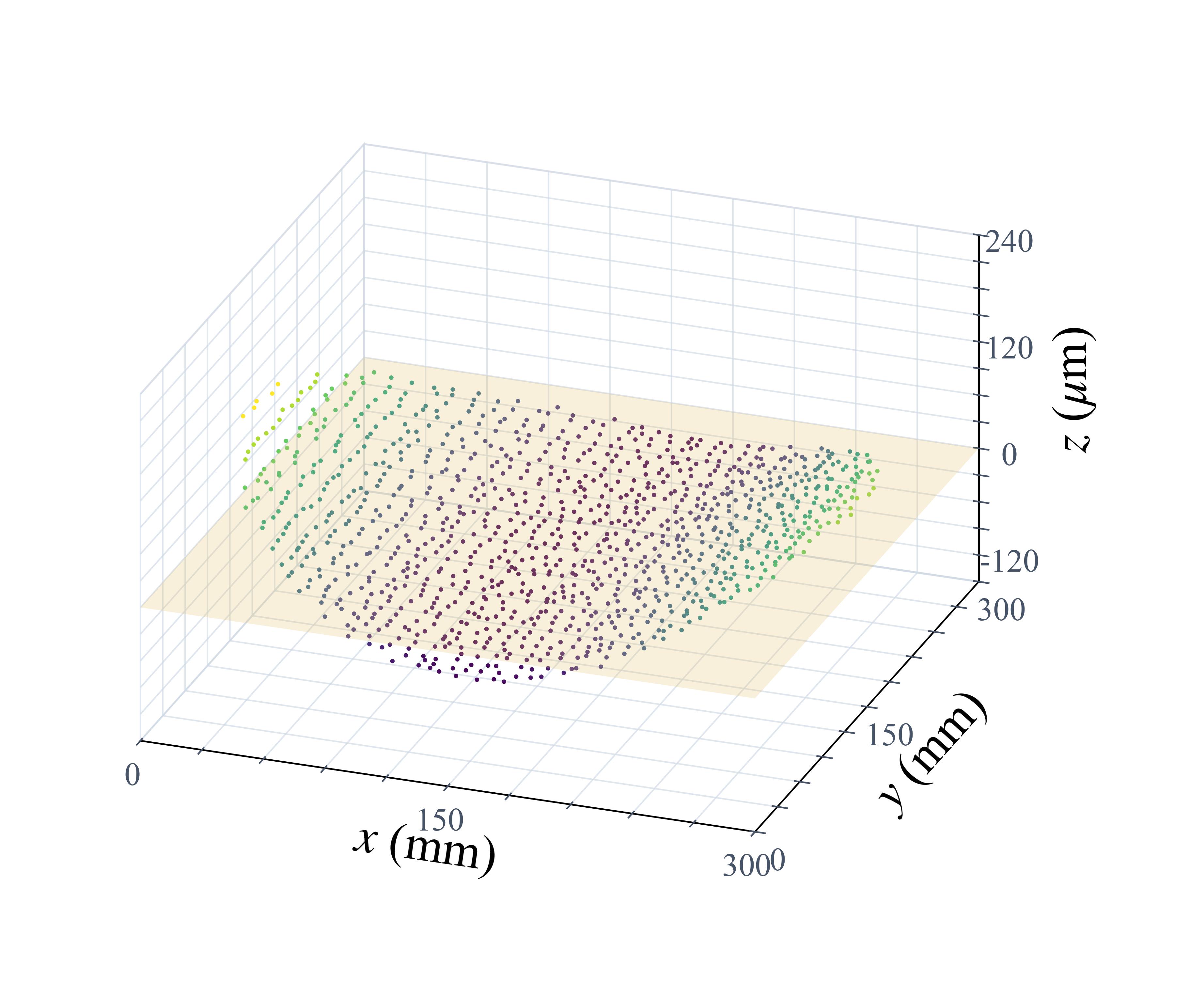}
        \end{minipage}
        \caption{Tilt-corrected data}
    \end{subfigure}
    \hfill
    \begin{subfigure}[t]{0.31\linewidth}
        \centering
        \begin{minipage}[c][0.25\textheight][c]{\linewidth}
            \centering
            \includegraphics[width=\linewidth,height=0.20\textheight]
            {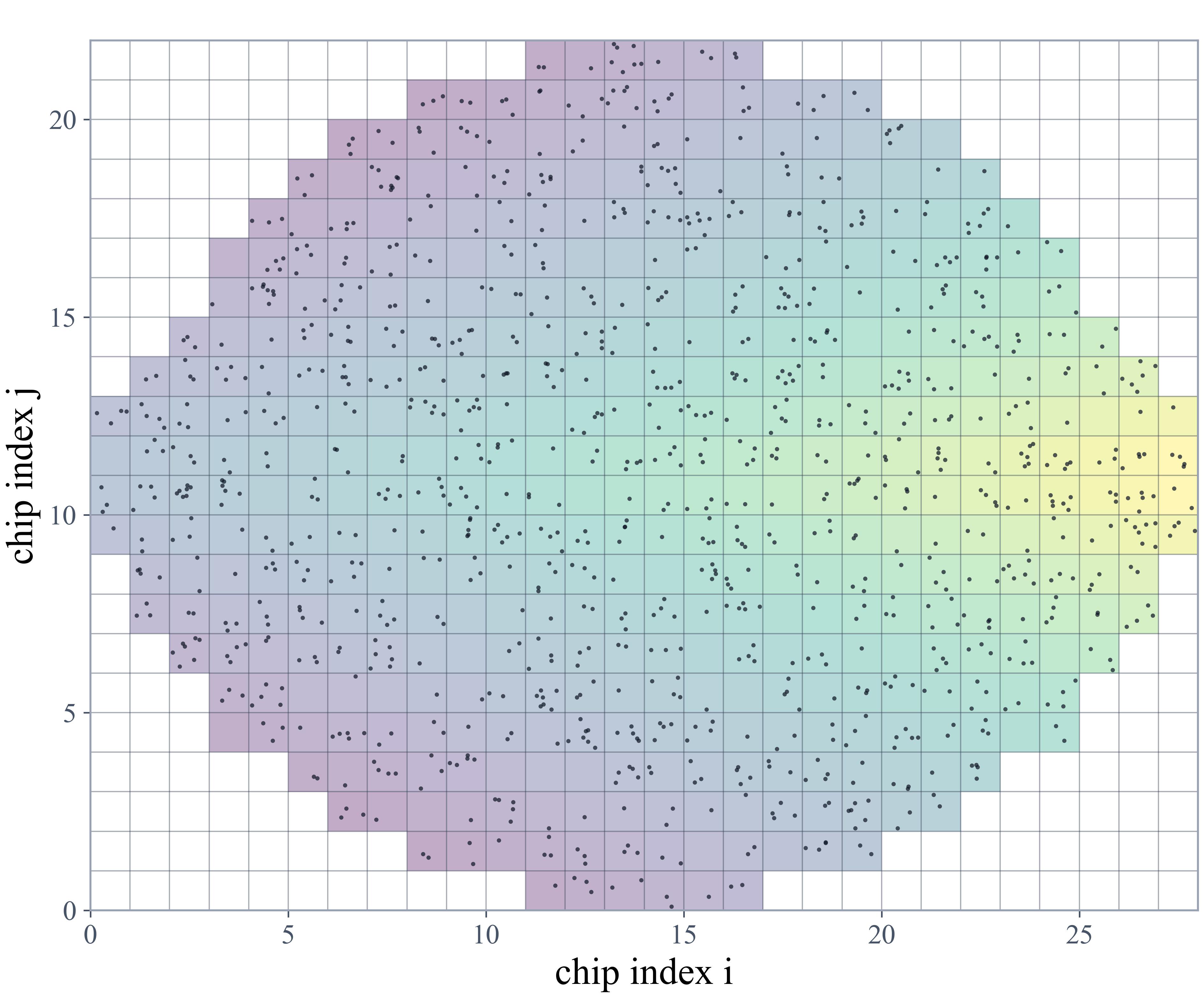}
        \end{minipage}
        \caption{Chip-wise organization}
    \end{subfigure}
    \caption{An illustration of point-cloud data processing.}
    \label{fig:pointcloud_preprocessing}
\end{figure}

\textcolor{black}{
This subsection describes the 3D point-cloud data obtained from wafer-level metrology systems, with the goal of extracting and normalizing chip-level warpage information for subsequent modeling. Figures~\ref{fig:measurement_process} and~\ref{fig:pointcloud_preprocessing} firstly show the measurement-stage configuration, stage-loading-induced wafer tilt, and the point-cloud data processing workflow considered in this study. In semiconductor manufacturing, wafer-shape (warpage/bow/saddle) data can be acquired using optical metrology techniques such as interferometry, 3D laser scanning, and Moir\'e-based methods.} From the perspective of stage design, metrology tools may employ flat stages, lifter-pin stages, or vertical-grabber stages. In practice, vertical-grabber configurations can face adoption barriers (e.g., patents and technical constraints), and thus many commercial suppliers predominantly provide tools based on horizontal (flat) stages.

In addition, the measurement sampling pattern may be implemented as line scan, one-shot scan, or point-by-point sampling. However, even when equipment is capable of dense acquisition (e.g., line scan or one-shot scan), production tools often record measurements only at specific pre-defined locations determined by a measurement recipe. As a result, both the number and spatial distribution of sampled points can vary across recipe versions and operating conditions, and may be irregular. In this work, we define the measurement data in a generic form so that the proposed framework can accommodate point clouds collected under any of the above modalities.

{\color{black}
Let the raw measured point cloud in tool coordinates be denoted by
$\{
\bm{z}^{\mathrm{raw}}_k
=
\left(
x^{\mathrm{raw}}_k,\,
y^{\mathrm{raw}}_k,\,
z^{\mathrm{raw}}_k
\right)^{\top}
\}_{k=1}^{N}$,
with $N$ being the total number of measurement points. 
}
During wafer handling in an EFEM (Equipment Front End Module), a wafer-handling robot transfers the wafer using an end-effector and loads it onto the measurement stage. When the wafer settles on the stage, it may exhibit a small gravitational tilt depending on the landing condition, which introduces a global tilt component unrelated to intrinsic warpage. To estimate this global tilt and define a leveled measurement frame, we fit a regression plane of the form
\[
z = a x + b y + c
\]
to the raw point cloud \(\{(x^{\mathrm{raw}}_k,y^{\mathrm{raw}}_k,z^{\mathrm{raw}}_k)\}_{k=1}^N\) via least squares.
Since the unit normal vector of the fitted plane is proportional to \([ -a,\,-b,\,1]^T\), we define
\begin{equation*}
\bm{n}
=
\frac{1}{\sqrt{a^2+b^2+1}}[-a,-b,1]^{\top}.    
\end{equation*}

Let the target normal be \(\bm{z}_{\mathrm{axis}}=[0,0,1]^T\), and define $\bm{v}_{\mathrm{axis}}=\bm{n}\times \bm{z}_{\mathrm{axis}}$, $
s=\|\bm{v}_{\mathrm{axis}}\|$, and $
\gamma=\bm{n}\cdot \bm{z}_{\mathrm{axis}}$. Then, 
it follows from the Rodrigues' formula that,
\[
\bm{R}
=
\bm{I}+[\bm{v}_{\mathrm{axis}}]_\times
+\frac{1-\gamma}{s^2}\left([\bm{v}_{\mathrm{axis}}]_\times\right)^2,
\]
\noindent where $[\bm{v}_{\mathrm{axis}}]_\times$ is the skew-symmetric matrix associated with $\bm{v}_{\mathrm{axis}}$, and $\bm{R}$ is a rotation matrix that maps \(\bm{n}\) to \(\bm{z}_{\mathrm{axis}}\). If \(s\approx 0\), we treat the data as already aligned and set \(\bm{R}\approx \bm{I}\).




{\color{black}
Let $R_{\mathrm w}$ denote the physical wafer radius.
In the leveled top-view frame, the origin is the lower-left
corner of the square enclosing the nominal wafer disk,
so that the wafer center is $(R_{\mathrm w},R_{\mathrm w})$.
For a 300 mm wafer, this center is $(150,150)$ mm.
This physical-coordinate origin is distinct from the
lower-left corner of the chip layout patch grid.

The wafer center position and notch orientation are assumed
to be available from the wafer-alignment procedure and
expressed in the measurement tool coordinate system.
Using this information and the known orientation of the chip
layout relative to the notch, we assume that the leveled
measurement axes are aligned with the chip-layout axes.
Let $(x_{\mathrm{ref}},y_{\mathrm{ref}})$ denote the wafer
center location in the original tool coordinates.
The corresponding reference point on the fitted plane is
\[
\bm p_{\mathrm{ref}}
=
[x_{\mathrm{ref}},y_{\mathrm{ref}},
a x_{\mathrm{ref}}+b y_{\mathrm{ref}}+c]^{\top}.
\]
The corrected coordinates are
\[
\begin{bmatrix}
x^{\mathrm{rot}}_k\\
y^{\mathrm{rot}}_k\\
z^{\mathrm{rot}}_k
\end{bmatrix}
=
\bm R
\left(
\begin{bmatrix}
x^{\mathrm{raw}}_k\\
y^{\mathrm{raw}}_k\\
z^{\mathrm{raw}}_k
\end{bmatrix}
-\bm p_{\mathrm{ref}}
\right)
+
\begin{bmatrix}
R_{\mathrm w}\\
R_{\mathrm w}\\
0
\end{bmatrix},
\qquad k=1,\ldots,N.
\]
The row and column conventions of the chip layout are
converted consistently so that the horizontal index increases
to the right and the vertical index increases upward.
A reversed indexing convention is handled in this index
conversion, rather than interpreted as a physical displacement.
}

Finally, to enforce a consistent reference height level, we apply a constant offset \(M\) to the leveled height:
\[
z^{\mathrm{rot}}_k \leftarrow z^{\mathrm{rot}}_k + M,\qquad k=1,\ldots,N.
\]

\noindent Hence, this tilt-removal process can be used as a preprocessing step at the metrology-tool level to correct raw height measurements and obtain warpage data with respect to a consistent reference frame. Height-correction functionality based on the same principle has also been implemented in commercial wafer-warpage measurement equipment.

{\color{black}
The physical position of the chip layout grid is specified by the patch pitches, \(\mathrm{pitch}_x\) and \(\mathrm{pitch}_y\), and by its registration relative to the wafer center. These pitches are the patch sizes already defined above, including the die-to-die spacing. Where available, product-layout information specifies a reference die, its reference point (such as the patch center or a specified corner), and its physical location relative to the wafer center. These data determine the grid origin; a reference-die index alone does not determine the physical registration.

When only the layout and patch pitches are available, we adopt a centered-grid placement as an explicit geometric assumption. Let $(\Delta_x,\Delta_y)$ denote the physical offset of the rectangular grid center from the wafer center. The lower-left grid origin is
\[
\begin{aligned}
x_{\mathrm g}&=R_{\mathrm w}+\Delta_x-\frac{n_x\,\mathrm{pitch}_x}{2},\\
y_{\mathrm g}&=R_{\mathrm w}+\Delta_y-\frac{n_y\,\mathrm{pitch}_y}{2}.
\end{aligned}
\]
The centered placement uses $\Delta_x=\Delta_y=0$; known product-specific offsets are used otherwise. For a centered grid, odd--odd column and row counts place the wafer center at the central patch center; even--odd counts place it on the central vertical patch boundary and central row centerline; odd--even counts give the opposite arrangement; and even--even counts place it at the intersection of the central patch boundaries. These are pitch-cell boundaries, not necessarily physical die edges. The parity rule describes the centered grid and does not by itself establish the physical registration of an arbitrarily cropped chip layout grid.

Any dummy padding or removal of empty rows and columns must preserve the physical positions of the original grid cells by updating the grid origin and indices consistently. If additional patches are needed to cover the selected measurement domain, this extension does not recenter the original die layout.
After registration, the measurement coordinates are normalized relative to the grid origin:
\[
\tilde{x}_k
=\frac{x^{\mathrm{rot}}_k-x_{\mathrm g}}{\mathrm{pitch}_x},
\qquad
\tilde{y}_k
=\frac{y^{\mathrm{rot}}_k-y_{\mathrm g}}{\mathrm{pitch}_y},
\qquad k=1,\ldots,N,
\]
with $z_k=z^{\mathrm{rot}}_k$. For points in the interior of the rectangular modeling domain, the patch indices are
\[
i_k=\lfloor\tilde{x}_k\rfloor,
\qquad j_k=\lfloor\tilde{y}_k\rfloor.
\]
An internal boundary point is assigned once using this half-open-cell convention. At the outer boundaries $\tilde{x}_k=n_x$ or $\tilde{y}_k=n_y$, the corresponding last patch is used and its local coordinate is set to one. Points outside the modeling grid are not assigned by taking a fractional part alone.

The points assigned to $c_{i,j}$ are re-indexed by $d=1,\ldots,D_{i,j}$, where $D_{i,j}$ may vary across chips. Their local coordinates are
\[
\tilde{u}_{i,j}(d)=\tilde{x}_{i,j}(d)-i,
\qquad
\tilde{v}_{i,j}(d)=\tilde{y}_{i,j}(d)-j,
\qquad d=1,\ldots,D_{i,j}.
\]
Thus, each observation is associated with its patch index, local coordinates, and processed height. The chip layout mask distinguishes actual-chip cells from dummy cells independently of this coordinate assignment. Measurement-domain and edge-exclusion filters are applied in the physical frame. The resulting local coordinates enter the existing B\'ezier basis evaluation and the construction of $\bm J$; the continuity operators and probabilistic model retain their existing form.
}

The data processing pipeline converts the raw data into a processed data set that is aligned with \(\Omega\) and represented in terms of (i) global coordinates, (ii) local coordinates within each chip, and (iii) the associated height measurements, which together serve as the input to the subsequent chip-level surface modeling. \textcolor{black}{In fact, the processing procedure described is consistent with some commercialized horizontal-stage automated wafer-level 3D metrology systems. In such non-vacuum wafer-loading measurement processes, tilt correction and coordinate normalization are effective for improving measurement consistency and achieving the required Gauge R\&R performance.}


\end{document}